\documentclass[5p]{elsarticle}
\usepackage{graphicx}
\graphicspath{{./Figures/}}

\usepackage[tight,footnotesize]{subfigure}
\usepackage{balance}
\usepackage{mathtools}
\usepackage{enumerate}
\usepackage{comment}
\usepackage{mdwlist}
\usepackage{float}
\usepackage{balance}
\usepackage{subfigure}
\usepackage{enumerate}
\usepackage{multirow}

\usepackage{color,soul}

\journal{Computer Communications}

\begin{document}

\begin{frontmatter}
\title{Analytical modeling and analysis of interleaving on correlated wireless channels}

\author[address1,address3]{Dmitri Moltchanov\corref{cor1}}
\cortext[cor1]{Corresponding Author.} \ead{dmitri.moltchanov@tut.fi}

\author[address2]{Pavel Kustarev}
\ead{kustarev@lmt.ifmo.ru}

\author[address1,address4]{Yevgeni Kucharyavy}
\ead{yk@cs.tut.fi}

\address[address1]{Department of Electronics and Communications Engineering, \\ 
Tampere University of Technology, Tampere, Finland\vspace{0.3cm}}

\address[address2]{Department of Embedded Systems,\\
ITMO University, St.-Petersburg, Russia\vspace{0.3cm}}

\address[address3]{Peoples’ Friendship University of Russia (RUDN University)\\ 6 Miklukho-Maklaya St, Moscow, 117198, Russian Federation\vspace{0.3cm}}

\address[address4]{School of Business Informatics\\Faculty of business and management\\National Research University Higher School of Economics, Moscow, Russia}

\begin{abstract}
Interleaving is a mechanism universally used in wireless access technologies to alleviate the effect of channel correlation. In spite of its wide adoption, to the best of our knowledge, there are no analytical models proposed so far. In this paper we fill this void proposing three different models of interleaving. Two of these models are based on numerical algorithms while one of them allows for closed-form expression for packet error probability. Although we use block codes with hard decoding to specify the models our modeling principles are applicable to all forward error correction codes as long as there exists a functional relationship (possibly, probabilistic) between the number of incorrectly received bits in a codeword and the codeword error probability. We evaluate accuracy of our models showing that the worst case prediction is limited by 50\% across a wide range of input parameters. Finally, we study the effect of interleaving in detail demonstrating how it varies with channel correlation, bit error rate and error correction capability. Numerical  results reported in this paper allows to identify the optimal value of the interleaving depth that need to be used for a channel with a given degree of correlation. The reference implementations of the models are available \cite{1}.
\end{abstract}


\end{frontmatter}

\section{Introduction}\label{sect_00}

In spite of significant progress in hardware design and associated signal processing
algorithms made over the last decades wireless channels still remain prone to
transmission errors. The reason is that reliability and hardware complexity is often traded for additional capacity at the air interface. Forward error correction (FEC) codes  along with retransmission techniques are supposed to be responsible for concealing residual errors. FEC codes are nowadays used in most modern wireless access technologies.

It is well documented that FEC codes show their best performance when bit errors happen at random without any sort of dependence between them. On the other hand, wireless channels are known to exhibit a high degree of correlation
manifesting itself in clipping of bit errors. Although there are codes that may tolerate a certain degree of error clipping (e.g. Reed-Solomon codes) correlated channel statistics lead to their sub-optimal performance.

In spite of universal usage of interleaving, to the best of our knowledge there are no
analytical models capturing its performance. The reason is twofold. First, most studies of wireless channel performance have been carried out assuming that the correlation in the bit error process is effectively removed using interleaving. In those investigations, where the correlation in the bit error process has been explicitly assumed, no interleaving functionality was considered. System level simulators that are widely used nowadays to evaluate performance of wireless access technologies may include interleaving as a basic block at the physical layer. However, interleaving is still implemented in rather unguided manner, i.e. the interleaving depth is set to some default value. Due to these reasons, interleaving is regarded as one of the features of wireless channels prohibiting their detailed cross-layer analytical studies. In this paper we will fill this void proposing three analytical models for interleaving having different degree of complexity and accuracy. Our models are suitable for any FEC codes with hard decoding that can correct $k$ bits (symbols) in a codeword of length $n$. This is the case for block codes such as BCH and RS codes. Minor modifications are required for codes with soft decoding, e.g. RS or turbo codes. Nevertheless, the models are applicable as long as there is relationship, possibly probabilistic, between the number of incorrectly received bits in a codeword and the codeword error probability.

%turbo codes, convolutional and low-density parity check (LDPC) codes. %Finally, our model can also be used to estimated performance of OFDM systems that inherently ''interleaves'' symbols over a predefined set of channels.

The rest of the paper is organized as follows. In Section \ref{sect_01} we highlight the importance of accurate modeling of interleaving process and introduce the system model we work with. Further, in Section \ref{sect_03}, three models for interleaving having different degree of complexity and accuracy are introduced. We assess accuracy of the models in Section \ref{sect_04}. The effect of interleaving is analyzed in detail in Section \ref{sect_05}. Conclusions are given in the last section.

\section{System Model}\label{sect_01}

The notation used in the paper is provided in Table \ref{tab:notation}.

\begin{table}[!h]\footnotesize
\caption{Notation used in the paper.}
\label{tab:notation}
\begin{center}
\begin{tabular}{p{0.15\columnwidth}p{0.75\columnwidth}}
\hline
\textbf{Parameter}           & \textbf{Definition}   \\
\hline\hline
$\Delta{t}$                     & Time to transmit a single bit\\
\hline
$I$                    		  	& Number of codewords\\
\hline
$n$                    		  	& Length of the codewords in bits\\
\hline
$k$                    		  	& Number of data bits in a codewords\\
\hline
$l$                    		  	& Number of data bits that can be corrected\\
\hline
$M$                    		  	& Number of interleaved codeblocks in a packet\\
\hline
$c_j^{(i)}$                     & Bit $j$ of codeword $i$\\
\hline
$\{S(u)\}$  & Bit error process\\
\hline
$\alpha,\beta$                 & Transition probabilities of bit error process\\
\hline
$c$                 			& Lag-1 NACF of the bit error process\\
\hline
$p_E$                 			& Bit error probability\\
\hline
$p_I$                 			& Interleaved codeblock error probability\\
\hline
$p$                     		& Packet error probability\\
\hline
$p_C$                     		& Codeword error probability\\
\hline
$c_C$                     		& Lag-1 NACF of the codeword error process\\
\hline
$\nu_{00},\nu_{11}$        & Probability of correct/incorrect first two codewords \\
\hline
$f_1(1),f_2(1)$                &Bit error probability is states 1 and 2\\
\hline
$D$                &Transition matrix of the bit error process\\
\hline
$D(1),D(0)$                &Transition matrices with and without error\\
\hline
$A(0),A(1)$                &Supplementary matrices, $A(0)+A(1)=1_I$\\
\hline
$1_I$                &Identity matrix\\
\hline
$D(i,n)$                & $nI$-steps matrix with $i$ incorrect bits\\
\hline
$D(i,j,n)$                &$nI$-steps matrix with $i$ and $j$ incorrect bits\\
\hline
$\vec{h}$                &Initial state probability vector\\
\hline
$\vec{v}$                &Transitions rates from transient to absorbing state\\
\hline
$P$                			&Rate matrix of the absorbing chain\\
\hline
$Q$                		&Rate matrix of transient states\\
\hline
$q_{ij}$                		&Transition rate between state $i$ and $j$\\
\hline
$\vec{0}$                		&Zero vector of appropriate size\\
\hline
$T$                		&Time till absorption\\
\hline
$F_T(k)$                		&CDF of time till absorption in $k$ steps\\
\hline
$\vec{\pi}_C$                		&Stationary distribution of the codeword process\\
\hline
$D_C(0)$                & Transition matrix of codeword process with error\\
\hline
$\alpha_C,\beta_C$                 & Transition probabilities of codeword error process\\
\hline
$\Phi(x)$                  & Error function\\
\hline
$\mu,\sigma$            & Mean and variance of $p_I$ estimate\\
\hline
$\hat{p}$                 & Estimate of the packet error probability\\
\hline
$N$                 & Number of experiments\\
\hline
$\gamma$                 & Confidence probability\\
\hline
$I_L$                 & Indicator of the packet loss event\\
\hline
\end{tabular}
\end{center}\vspace{-0.5cm}
\end{table}\normalsize

\subsection{Wireless channel model}

In this study, to make the model universally applicable we abstract the channel organization mechanisms assuming the bit error statistics as the input to the model.

We consider the bit error process as a covariance stationary binary process, where $1$ and $0$ denote incorrect and correct bit reception, respectively. Consider a discrete-time environment with constant slot duration $\Delta{t}$ corresponding to the amount of time required to transmit a single bit over a wireless channel. We model the bit error process using the discrete-time Markov modulated process with irreducible aperiodic Markov chain $\{S(u),u=0,1,\dots\}$, $S(u)\in\{0,1\}$. When at most single event is allowed to occur in a slot this process is known as switched Bernoulli
process (SBP). To parameterize a covariance stationary binary process only mean value and lag-1 normalized autocorrelation (NACF) coefficient have to be captured. It was shown in \cite{3} that there is a special case of SBP called interrupted Bernoulli process (IBP) exactly matching mean and lag-1 NACF value. It is given by
\begin{align}\label{eqn:005}
\begin{cases}
\alpha = (1 - c)p_E\\
\beta  = (1 - c)(1 - p_E)\\
\end{cases}
\begin{cases}
f_{1}(1) = 0\\
f_{2}(1) = 1\\
\end{cases},
\end{align}
where $f_{1}(1)$ and $f_{2}(1)$ are bit error probabilities in states 1 and 2,
respectively, $\alpha$ and $\beta$ are transition probabilities from state 1 to state 2
and from state 2 to state 1, respectively, $c$ is the lag-1 NACF value of bit error
observations, $p_E$ is the BER. Letting $D$ be transition probability matrix and
defining the following matrices
\begin{align}\label{eqn:006}
A(0)=
\begin{pmatrix}
1 & 0\\
0 & 0\\
\end{pmatrix},\qquad{}
A(1)=
\begin{pmatrix}
0 & 0\\
0 & 1\\
\end{pmatrix}
\end{align}
the model is described by two matrices $D(0)=D\times{}A(0)$ and $D(1)=D\times{}A(1)$
describing transitions between states of the model with correct and incorrect reception
of a bit. Since $A(0)+A(1)=1_I$, where $1_I$ is the identity matrix.

Note that the channel model introduced above may not exactly capture correlation
properties of various propagation environments. However, in practical situations, exact
behavior of NACF is not known. The proposed model provide the so-called
"first-order" approximation of correlated channel behavior. A special advantage of the model is that
it can be easily tuned to explore the qualitative and quantitative effects of correlation. Still the extension of the model to the case of general finite-state Markov
chain (FSMC, \cite{4}) is straightforward. Indeed, irrespective of the number of states
used to represent the bit error process and probabilities of bit errors in each state we
can always partition $D$ into $D(0)$ and $D(1)$ such that $D=D(0)+D(1)$. Finally, if wireless channels conditions exhibit piecewise stationary behavior as
was reported in a number of studies (see e.g. \cite{5,6}), this model may represent statistical characteristics of covariance stationary parts. In this case, (\ref{eqn:005}) is
interpreted as a model for a limited duration of time during which mean value and NACF of bit error observations remain constant.

\subsection{Interleaving process}

Assume that data are encoded into $I$, $I>0$, codewords of the same length $n$. We denote these codewords by $c^{(i)}$, $i=1,2,\dots$ where bit $j$ of $i$th codeword is $c_{j}^{(i)}$. Without interleaving these bits would have been sent as
\begin{align}\label{eqn:001}
c_{1}^{(1)},c_{2}^{(1)},\dots,c_{n}^{(1)},\dots,c_{1}^{(I)},\dots,c_{n}^{(I)},
\end{align}
and in case of a deep channel fade we may incorrectly receive a significant amount of bits belonging to the same codeword. This may eventually lead to inability to decode this codeword even when the overall ''average'' channel quality is acceptable. 

Let us now introduce interleaving of depth $I$, $I>1$. According to the concept, we first combine $I$ codewords into a matrix of size $n\times{}I$, where each codeword represents a row, i.e.
\begin{align}\label{eqn:002}
\begin{pmatrix}
c_{1}^{(1)} & c_{2}^{(1)} & c_{3}^{(1)} & \dots & c_{n}^{(1)}\\
c_{1}^{(2)} & c_{2}^{(2)} & c_{3}^{(2)} & \dots & c_{n}^{(2)}\\
\vdots      & \vdots      & \vdots      & \ddots& \vdots     \\
c_{1}^{(I)} & c_{2}^{(I)} & c_{3}^{(2)} & \dots & c_{n}^{(I)}\\
\end{pmatrix}
\end{align}

Given the matrix (\ref{eqn:002}), we perform its column-wise transmission, i.e. column $n$ is
transmitted first with $c_{1}^{(1)}$ being the first bit sent, then column $n-1$
starting with $c_{2}^{(1)}$, etc. The transmitted sequence of bits looks as
\begin{align}\label{eqn:003}
c_{1}^{(1)},c_{1}^{(2)},\dots,c_{1}^{(I)},\dots,c_{n}^{(1)},c_{n}^{(2)},\dots,c_{n}^{(I)}.
\end{align}

Observing (\ref{eqn:003}) we see that those bits that go back-to-back in a codeword are
transmitted $I$ bits apart. The interleaving procedure tries to ensure that adjacent
bits in a codeword are not similarly affected by the current channel conditions of a
channel. Thus, interleaving tries to reduce the memory of a channel. The parameter $I$ is called the interleaving depth. The reverse operation is performed at the receiving end. 

When interleaving of depth $I>1$ is used the length of a packet is $MIn$ bits, where $M$ is the number of interleaving blocks. When the product $In$ is not small (i.e., more than few tens) a good approximation of the packet error probability is obtained by
\begin{align}\label{eqn:004}
p=1-(1-p_{I})^{M},
\end{align}
where $p_{I}$ is the probability of incorrect reception of $I$ interleaved codeblock. Thus, it is sufficient to estimate the probability of correct reception of an interleaving block consisting of $I$ codewords. Applying (\ref{eqn:004}) we shrink the memory of the channel model at $In$. Since $In$ is often large, this effect is negligible in practical applications.

\begin{comment}

Interleaving is highly effective and brings only slight processing overhead. The only operation needed is the permutation of bits of $I$ codewords. Still extremely high values of $I$ are often avoided as it brings no additional performance gains as we will see later. Since a packet to be transmitted usually contains more than $nI$ bits we need to implement a number of interleaving procedures, $v$, for a single packet. A part of a packet interleaved in one such procedure is called an interleaved codeblock. 

When interleaving of depth $I>1$ is used the length of a packet is $vIn$ bits, where $v$ is the number of interleaving blocks. When the product $In$ is not small (i.e., more than few tens) a good approximation of the packet error probability is obtained by
\begin{align}\label{eqn:004}
p_{P}=1-(1-p_{I})^{v},
\end{align}
where $p_{I}$ is the probability of incorrect reception of $I$ interleaved codeblock.
Thus, it is sufficient to estimate the probability of correct reception of an
interleaving block consisting of $I$ codewords. Applying (\ref{eqn:004}) we
shrink the memory of the channel model at $In$. Since $In$ is often large, this effect is negligible in practical applications.

\end{comment}

In this paper we study interleaving using packet error probability as the main metric of interest. We define the packet error probability as the probability that at least one codeword is incorrectly decoded. Although our models are developed using Bose-Hocquenghem-Chaudhuri (BCH) FEC codes as an example, they can be extended to the case of any type of FEC code. For those codes having no closed-form expression for codeword error probability (e.g. convolutional, turbo or low-density parity check codes) it can be obtained via simulation studies. A BCH code is represented by a triplet $(n,k,l)$, where $n$ is the size of the codeword, $k$ is the number of data bits in a codeword, $l$ is the number of incorrectly received bits that can be corrected. A codeword is correctly decoded when the number of bit errors is less or equal to $l$. 

\subsection{Related work}

% No models for interleaving: NB! add references

There have been several attempts to model the interleaving mechanism in the past, see e.g., \cite{dunscombe1989optimal,chen2004improving,el2012performance,kiyani2008performance, kang2008probabilistic,kang2010hybrid} for a brief account of studies. The authors in \cite{dunscombe1989optimal} the authors develop an optimal interleaving scheme for convolutional codes when the error burst size, known in advance, is greater than the allowed interleaving depth. In \cite{chen2004improving} the authors proposed to combine the functionality of FEC codes and interleaver into a single module. The modification allows to dynamically change the interleaving depth and strength of the FEC to dynamically adapt to changing wireless channel conditions. The performance of the proposed scheme has been evaluated using the computer simulations. An optimal interleaving scheme for transmission of audio information over the wireless channels has been developed in \cite{el2012performance}. The simulation approach has been chosen for performance assessment. The study in \cite{kiyani2008performance} analyze interleaving as a part of the signal space diversity concept for wireless channel, where they derived a closed form expression for the upper bound of bit error rate $M$-ary phase shift keying in Rayleigh fading channel. A comprehensive analytical model for delay analysis of FEC codes with interleaving over fading channels has been proposed in \cite{kang2008probabilistic,kang2010hybrid}. The bit error process was assumed to follow FSMC. As the major emphasis of the work was on delay induced by the interleaving procedure no results on the interleaved codeblock and packet error probabilities have been provided.

% Short summary

Summarizing the related work we conclude that the interleaving studies performed so far introduced significant simplifications to the wireless channel behavior or relied on the simulation studies to derive performance metrics of interest or concentrated on simple bounds on optimal interleaving depth. Is spite of the universal use of the interleaving mechanism, the simple yet accurate interleaving model is still missing. %and the researchers analyzing performance of wireless technologies do not include interleaving into the modeling procedure leading to overestimation of the wireless channel performance. The interleaving can be explicitly taken into account in link- or system-level simulations. However, the parameters of the interleaving scheme are often set to to their default values that may result in non-optimal performance of wireless channels. Thus, having a simple yet accurate model of interleaving is of special importance. 

\section{Interleaving Model}\label{sect_03}

\subsection{Approach at the glance}

The core idea of the models specified below consists in translating the memory of the bit error model to the interleaved codeblock error probability. such that the Markov structure of the model is preserved. This is done by formulating a new absorbing Markov process describing the dynamics of the codeword transmission process by explicitly tracking the number of incorrectly received bits in successive codewords. To determine the interleaved codeblock error probability we need to estimate the joint distribution of the number of incorrectly received bit in first two codewords parameterizing the model and then estimate the absorption time. As both steps are computationally intensive, we further formulate two simplified models. The first simplified model address the problem of absorption time estimation by defining a two-state Markov chain whose parameter are directly derived from the joint distribution of the number of incorrectly received bits in the first two codewords. For this model the interleaved codeword error probability is provided in closed-form. The last model assumes that the correlation between codewords is the same as correlation between individual bits removing the need for calculating the joint distribution. 

\subsection{Absorbing Markov chain model}

We represent the process of transmission of codewords in an
interleaved codeblock using an absorbing Markov chain $\{S(u'),u'=0,1,\dots\}$
with the state-space $S(u')\in\{0, 1,\dots, l, l+1\}$, where states $0,1,\dots,l$
corresponds to the correct reception of a codeword, while state $l+1$ aggregates those
states resulting in incorrect reception of a codeword. We start with the initial state
distribution $\vec{h}=(h_{0}, h_{1}, . . . , h_{l+1})$ and observe the process for
exactly $I$ steps. As one may observe, the probability of absorption in no more than $I$ steps
is the interleaved codeblock error probability, $p_{I}$.

The canonical form of the absorbing Markov chain is \cite{kemeny1960finite}
\begin{align}\label{eqn:012}
P=
\begin{pmatrix}
Q           & \vec{v}   \\
\vec{0}^{T} & 1         \\
\end{pmatrix}
\end{align}
where $Q$ is $l\times{}l$ matrix whose elements, $q_{ij}$, $i,j=0,1,\dots,l$, define
transition probabilities between transient states of the absorbing Markov chain,
$\vec{0}^{T}$ is the vector of zeros, and $\vec{v}=(v_{0},v_{1},\dots,v_{l})$ is the
vector containing transitions between transient and absorbing states. Introducing $\vec{e}$ as
a vector of ones of size $l$, and noticing that the sum of all rows in (\ref{eqn:012})
must be $1$ we have that
\begin{align}\label{eqn:014}
\vec{v}+Q\vec{e}=\vec{e},
\end{align}
implying that the model is completely characterized by $Q$.

For a Markov chain defined in (\ref{eqn:012})-(\ref{eqn:014}) we are interested in the probability of absorption in no more than $I$ steps. We compute it as follows. Recall, that the time
till absorption is a discrete distribution of phase-type over the set of positive
integers \cite{o1999phase}. This distribution is also the first passage time distribution to the state
$l+1$. Let $T$ denote the random variable describing the time till absorption. The
cumulative distribution function (CDF) of the number of transitions till absorption is given by
\begin{align}\label{eqn:015}
F_{T}(k)=Pr\{T\leq{}k\}=1-\vec{h}Q^{k}\vec{e},
\end{align}
where $\vec{h}$ is the initial state probability vector. Letting
$k=I$ in (\ref{eqn:015}) gives us the probability we are looking for.

\begin{figure}[h!]
\centering
\includegraphics[width=\columnwidth]{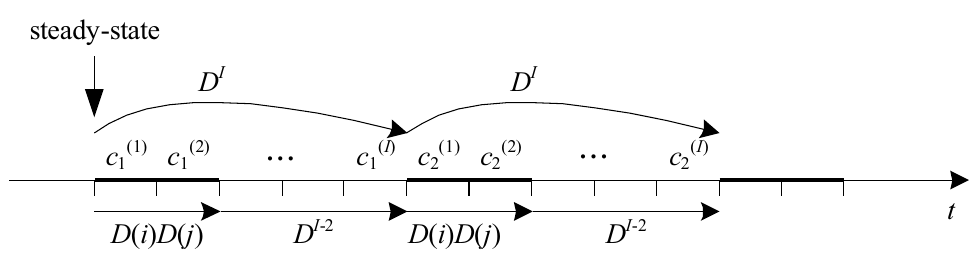}
\caption{Estimating two-dimensional distribution.}
\label{fig:twoDim}
\end{figure}

We need three parameters to specify the model: matrix $Q$, describing transition probabilities between transient states, $\vec{v}$, describing transitions between transient and absorbing states and the initial state probability vector, $\vec{h}$. Both $Q$ and $\vec{v}$ can be obtained by estimating two-dimensional distribution of the number of incorrectly received bits received in the first and second codewords. The idea of the numerical algorithm is that bits of the first and second codewords are transmitted back-to-back over the wireless channel and these pairs of bits are separated by $I-2$ transmission slots as shown in Fig. \ref{fig:twoDim}.

%Let $D(i,j)$ be matrices containing two-steps transition probabilities
between states of the bit error model associated with $i$ and $j$, $i,j=0,1$ incorrectly
received bit in the first and second slots, respectively. We have $D(i,j)=D(i)D(j)$. 

Let the set of matrices $D(i,j,n)$ contain $nI$-step conditional transition probabilities of the bit error process associated with with $i$, $j$, $i,j=0,1,\dots,n$, incorrectly received bits in the first and second codeword, respectively. We get $D(i,j,n)$ recursively as
\begin{align}\label{eqn:017}
D(i,j,n)=\sum_{k=0}^{1}\sum_{k=0}^{1}D(i-k,j-l,k-1)D(i)D(j).
\end{align}

Now, we can estimate probabilities of transitions as
\begin{align}\label{eqn:020}
&q_{ij}=\vec{\pi}D(i,j,n)\vec{e},\,i,j=0,1,\dots,l.\nonumber\\
&v_{i}=\sum_{j=l+1}^{n}\vec{\pi}D(i,j,n)\vec{e},\,i=0,1,\dots,l.
\end{align}
where $\vec{\pi}$ is the steady-state distribution of the bit error process, $\vec{e}$ is the vector of ones of appropriate size.

The only thing left is to compute the elements of the initial state probability vector
$\vec{h}$. According to the definition $h_{i}$, $i=0,1,\dots,l$, is the probability that
there are exactly $i$ incorrectly received bits in a first codeword.
To get elements $\vec{h}$ it is sufficient to know one-dimensional distribution of
the number of incorrectly received bits in a codeword. This distribution can be obtained
from two-dimensional distribution $q_{ij}$ by summing up over
all $j$.

\subsection{Two-state Markov chain model}

The model introduced in the previous section consists of two successive steps. We first estimate two-dimensional distribution of the number of errors in first two codewords and then apply an absorbing Markov chain to model the process of successive reception of $I$ codewords. Further, we need to compute the interleaved codeblock error probability, $p_{I}$, by taking an $l\times{}l$ matrix $Q$ into a power of $I$. When the error correction capability $l$ and interleaving depth $I$ are both rather large this incurs an additional computational overhead.

Once two-dimensional distribution of incorrectly received bits in first two codewords is obtained we can
define a new discrete-time two-step Markov chain $\{S_{C}(u'),u'=0,1,\dots\}$,
$S_{C}(u')\in\{0,1\}$ modeling the process of codeword reception, where $0$ denotes a
correctly received codeword while $1$ implies incorrect reception. Note that a new
interval duration is exactly $n$ times that of the bit error process, where $n$ is the
length of a codeword. Having two-dimensional
distribution $q_{ij}$, $i,j=0,1,\dots,n$ one can estimate both the codeword error
probability and lag-1 NACF value as
\begin{align}\label{eqn:021}
&p_{C}=\sum_{i=0}^{l}\sum_{j=0}^{n}q_{ij},\nonumber\\
&c_{C}=\frac{\sum_{i=0}^{1}\sum_{j=0}^{1}(i-p_{C})(j-p_{c})\nu_{ij}}{p_{C}-p_{C}^{2}},
\end{align}
where in the denominator of the latter expression we used a well-known property of binary
processes. Here, $\nu_{ij}$, $i=0,1$, is the probability of outcome of first two codewords
transmissions that can be estimated using $q_{ij}$ as
\begin{align}\label{eqn:022}
\nu_{00}=\sum_{i=0}^{l}\sum_{j=0}^{l}q_{ij},\,\nu_{11}=\sum_{i=l+1}^{n}\sum_{j=l+1}^{n}q_{ij},
\end{align}
and $\nu_{01}=1-\nu_{00}$, $\nu_{10}=1-\nu_{11}$.

The interleaved codeblock error probability, $p_{I}$, is 
\begin{align}\label{eqn:023}
p_{I}=\vec{\pi}_{C}[D_{C}(0)]^{I}\vec{e},
\end{align}
where $\vec{\pi}_{C}$ is the steady-state probability vector of the codeword error process, $D_{C}(0)$ is the matrix having transitions between states with correct reception of a codeword. The transition matrix $D_C(0)$ can be found similarly $D(0)$ of the bit error process using (\ref{eqn:005}) and (\ref{eqn:006}).

Note that (\ref{eqn:023}) is easier to compute compared to finding the first passage time in absorbing Markov chain as steady-state probabilities of the model as well as eigenvectors and eigenvalues of the transition probability matrix are available in closed form as 
\begin{align}
\vec{\pi}_C=(\alpha_C/(\alpha_C+\beta_C),\beta_C/(\alpha_C+\beta_C)).
\end{align}

Replacing the absorbing Markov chain model, where we used transitions between states of the process defined over the number of incorrectly received errors in a codeword, by the two-state Markov chain we loose in accuracy of the model. Particularly, this procedure deteriorates the memory structure of the original model. Still, if the loss in accuracy is not drastic, the gain we potentially get avoiding estimation of first passage time distribution can be a deciding factor for choosing this model, especially for large values of $I$ and $l$.

\subsection{Simple two-state Markov chain model}

%While it is in principle computationally feasible we would still like to avoid this step.

Both models introduced above involve estimation of two-dimensional distribution. For large values of $n$, say $n>1000$, we need to estimate a matrix containing $10E6$ elements. For practical applications we would like to have an easy-to-compute approximation, preferably providing closed form expression, that is developed below. The major assumption is that for any $I>1$ the correlation between outcomes of codewords reception is likely similar to that between individual bit transmissions.

To parameterize the codeword error process we need the codeword error probability and the lag-1 NACF value. The latter is readily available. Although the codeword error probability can be obtained from two-dimensional distribution of the number of errors in two adjacent codewords we avoid this step estimating the number of errors in a single
codeword directly. Let $D(j,n)$, $j=0,1,\dots,n$, be the set of matrices describing transitions of a channel model with exactly $j$ errors in a bit pattern of length $n$. Starting from $D(j,1)=D^{I-1}(j)$, $j=0,1$, where $D^{I-1}(j)$ is defined as $D^{I-1}(j)=D(j)D^{I-1}$, $j=0,1$, we estimate these matrices recursively using
\begin{align}\label{eqn:024}
&D(j,i)=\sum_{k=0}^{1}D(j-k,i-1)D_{I-1}(k),\,i=1,2,\dots,n-1,\nonumber\\
&D(j,n)=\sum_{k=0}^{1}D(j-k,n-1)D(k),
\end{align}
leading to the following codeword error probability
\begin{align}\label{eqn:025}
p_{C}=1-\vec{\pi}_{C}\left(\sum_{j=0}^{l}D(j,n)\right)\vec{e}.
\end{align}

Once $p_{C}$ is found we use apply (\ref{eqn:023}) to
estimate $p_{I}$. Recall that the number of visits to a state of a two-state Markov chain can be found using closed-form as shown in \cite{7} reducing the complexity of the model further.

\subsection{Feasible extensions}

In this paper we use BCH codes as an example. However, our model can be extended to the case of other block codes with hard decoding. For example, consider Reed-Solomon (RS) codes. As opposed to BCH codes RS codes operates over symbols with each symbols consisting of 2d bits. An RS code is defined as $(n,k)$, where $n$ is the length of the codeword, $k$ is the number of error correcting bits, while the difference $(n-k)$ represents the number of data symbols. The correction capability of any RS code is $\lfloor{}(n-k)/2\rfloor{}$ symbols. A symbol is incorrectly received if there is at least one bit error in this symbol. It is easy to see that RS codes are inherently more resistant to correlation as it does not matter how many bits are incorrectly received in a single symbol. In practice $d=3$, resulting in the symbol length of eight bits is used.

Adapting the proposed approach to the case of RS code is intuitively simple. The only difference is that we need to define the symbol error process by firstly estimating matrices describing transitions between states of the channel model with $i$, $i=0,1,\dots,2^{d}$, incorrectly received bits and then relating them to the matrices describing correct and incorrect reception of a symbol. Then, we proceed similarly to
BCH codes. %Note that operating over symbols of length $2^{d}$ rather than on individual bits allows to define simpler model when channel correlation is weak. For example, assuming $d=3$ the correlation become negligible at the lag $2^{3}=8$ for any value of $\lambda$ less than $0.7$. Thus, for such values of lag-1 NACF we can assume that successive symbol transmissions are independent and use simple model for uncorrelated channel. This also highlights that operating over symbols of length $2^{d}$ is roughly similar to implementing interleaving of depth $I=2^{d}$.

The proposed approach can also be extended to the case of codes with soft decoding for which there is no deterministic relationship between the number of errors occurring in a codeword and the outcome of a codeword transmission (e.g. convolutional or turbo codes). For these codes the relationship takes probabilistic form and can be obtained via simulations for any finite code length. Knowing the probabilities of $i$, $i=0,1,\dots,n$, errors in a codeword of size $n$ and the probability of incorrect reception of a codeword conditioned on $i$ errors, we can estimate the probability of incorrect reception of a codeword. The difference compared to BCH or RS codes is that it might be non-negligible for any number of bit errors. Then, we proceed similarly to the BCH case described above.

\section{Comparison of models}\label{sect_04}

The purpose of this section is to evaluate the accuracy of the proposed model. We do it
by comparing the results obtained using our model with those derived with simulations of
the packet transmission process. We use the following short notation to
refer to the defined models:
\begin{itemize}
  \item{model 1: two-dimensional distribution, two-state MC;}
  \item{model 2: one-dimensional distribution, two-state MC;}
  \item{model 3: two-dimensional distribution, absorbing MC.}
\end{itemize}

\subsection{Simulations and data analysis}

The simulator is written in C and implements the following simple algorithm for
simulating interleaving of BCH codewords. To generate the bit error sequence with a
predefined bit error probability and lag-1 NACF value we use discrete autoregressive
model of order one, DAR(1). Note that DAR(1) and Markov model we used to represent the bit error process in this paper are stochastically equivalent. The reason for using DAR(1) is that the process is given by a recursion which is more convenient for simulation studies. In our simulations we count the number of bits and once it reaches $nI$ we estimate the number of errors in each codeword. Let $I_{L}$ be the loss indicator of the packet. If all the codewords are received correctly this interleaved codeblock is marked as received correctly. We then proceed with the next interleaved codeblock. If all $v$ codeblocks belonging to a packet are received correctly then this packet is marked as correctly received and associated with indicator value is $I_{L}=0$. Otherwise, it is incorrectly received and the associated value of loss indicator is set to $1$. In order to obtain reliable statistics in all our experiments we simulated for $N=100000$ packets. %The simulation time is several seconds using an average modern computer. However, as we will see, this number is not large enough to get accurate results for working ranges of BER making the proposed analytical models useful.

The metric of interest is the packet error probability, $p$, i.e. the probability that at
least one interleaved codeblock out of $IM$ is received incorrectly. Observe that the packet
error probability is actually the mean of the loss indicator of a packet $I_{L}$. Thus, we
can replace the problem of estimating the probability of an event by the problem of
estimation of the mean of the loss process represented by a binary sequence of 1s and 0s. The unbiased, consistent and effective point estimate of the packet loss
probability, $\hat{p}$, is obtained by averaging the values of all obtained indicators.
%Further, recall, that the common rule of thumb states that in order to obtain reliable estimate of the mean of the binary process one needs to have at least $30$ non-zero outcomes. This drastically limits the range of input parameters implying that the point estimates are not reliable if the packet loss probability is below $30/100000=3E-4$.

\begin{figure*}[t!]
\centerline{
  \subfigure[$c=0.0$]{
    \includegraphics[width=0.5\textwidth]{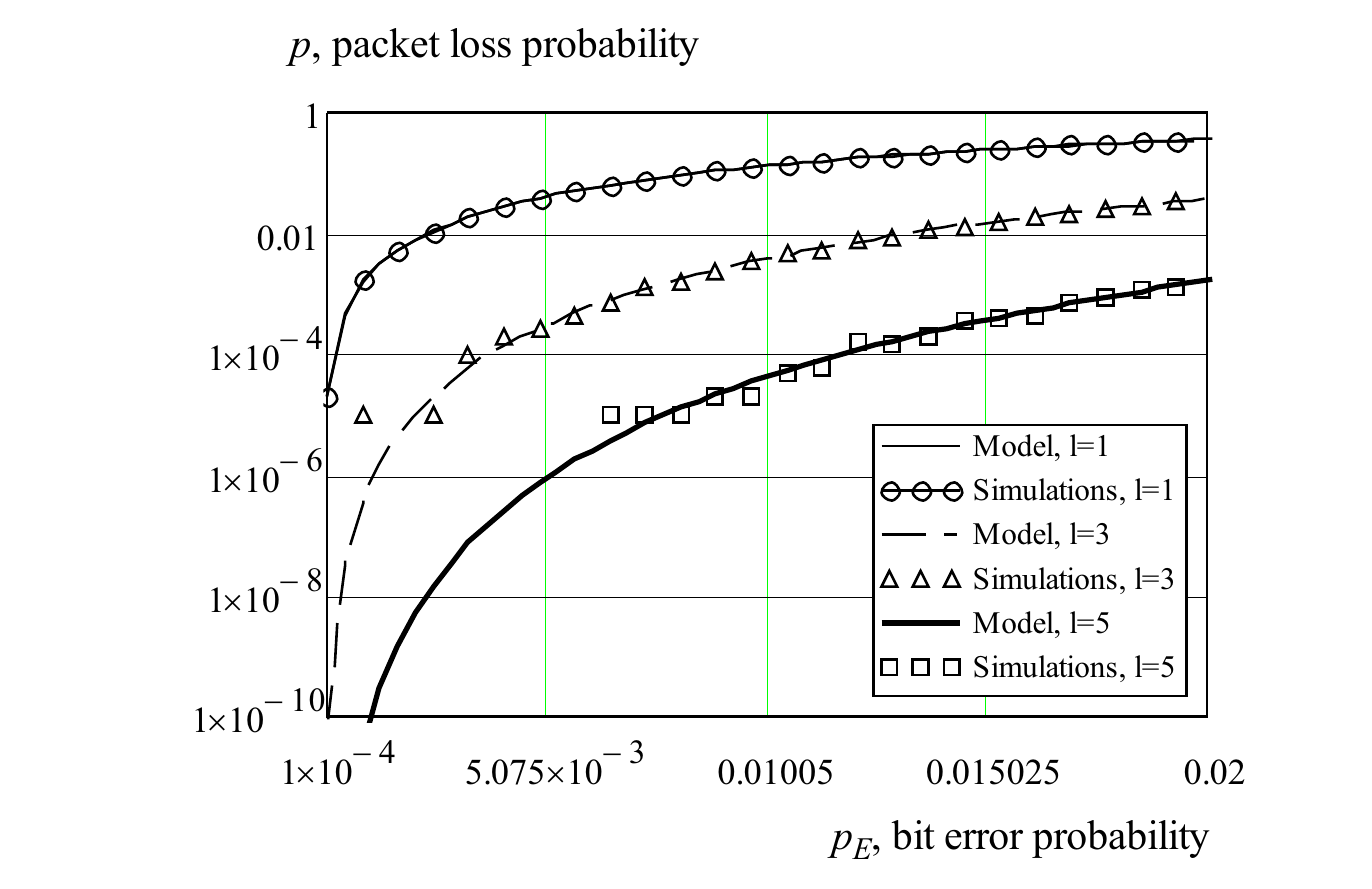}
    \label{fig:firstSingle_c_1}
  }
  \subfigure[$c=0.3$]{
    \includegraphics[width=0.5\textwidth]{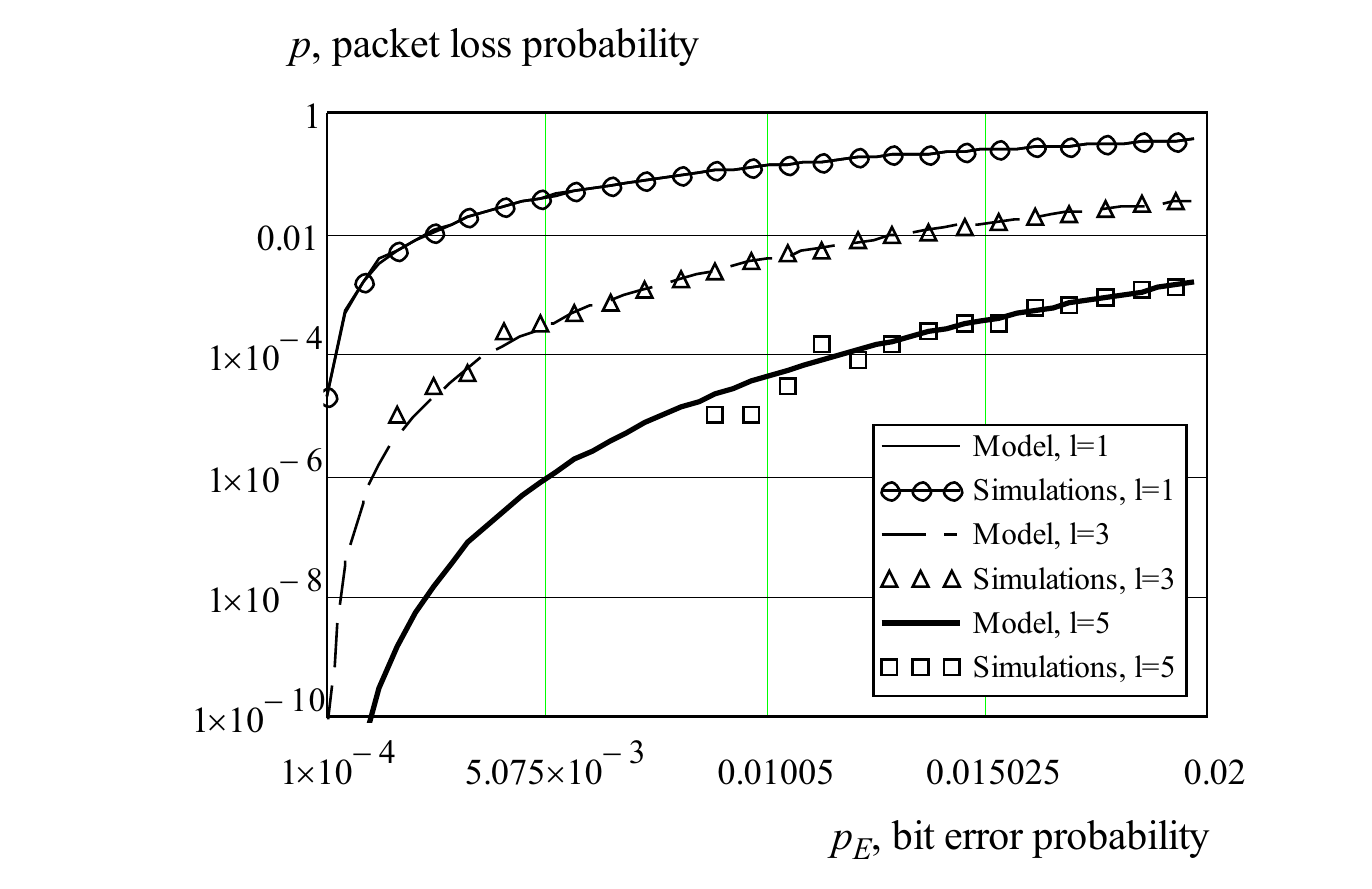}
    \label{fig:firstSingle_c_2}
  }
} \centerline{
  \subfigure[$c=0.6$]{
    \includegraphics[width=0.5\textwidth]{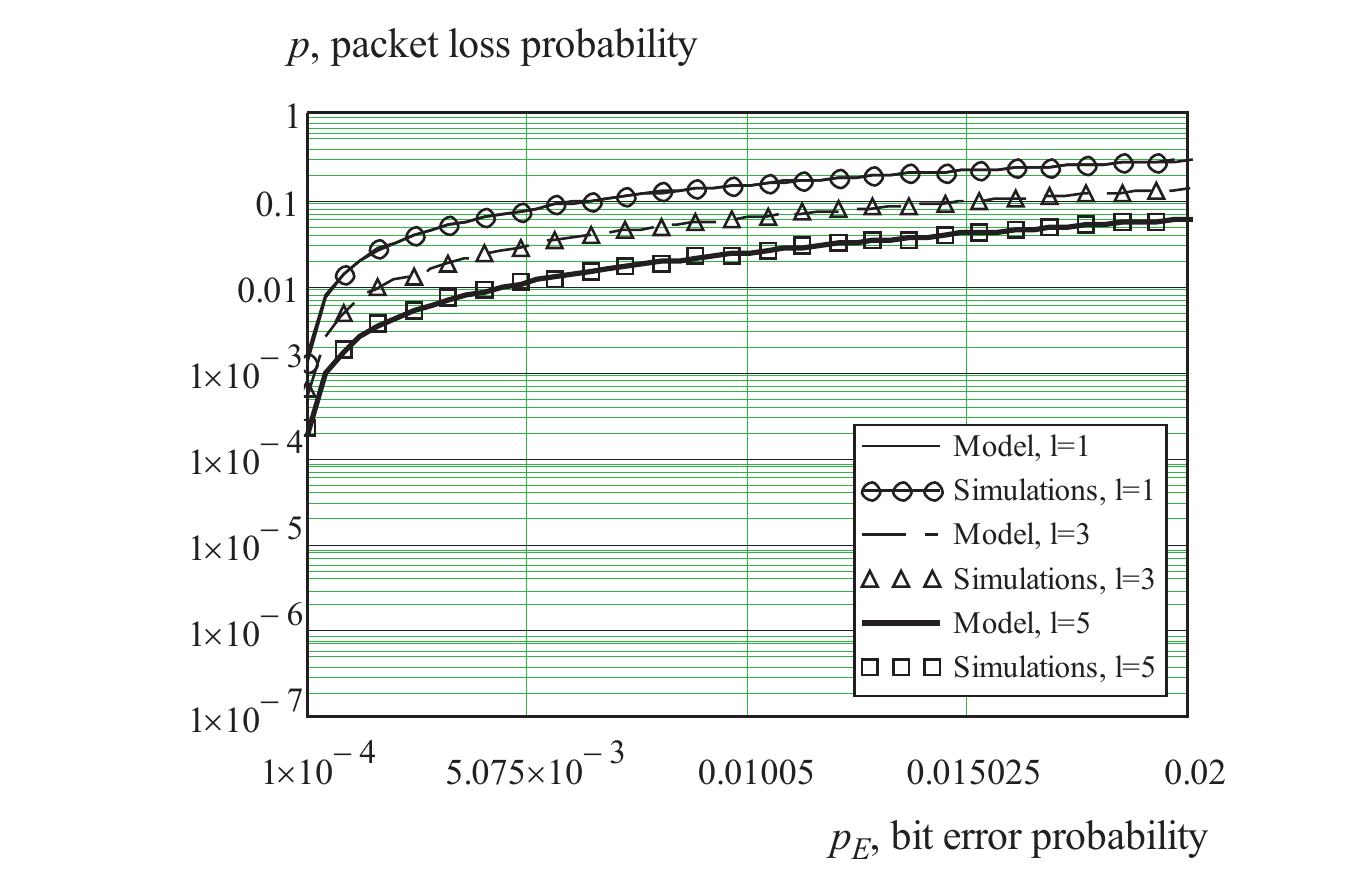}
    \label{fig:firstSingle_c_3}
  }
  \subfigure[$c=0.9$]{
    \includegraphics[width=0.5\textwidth]{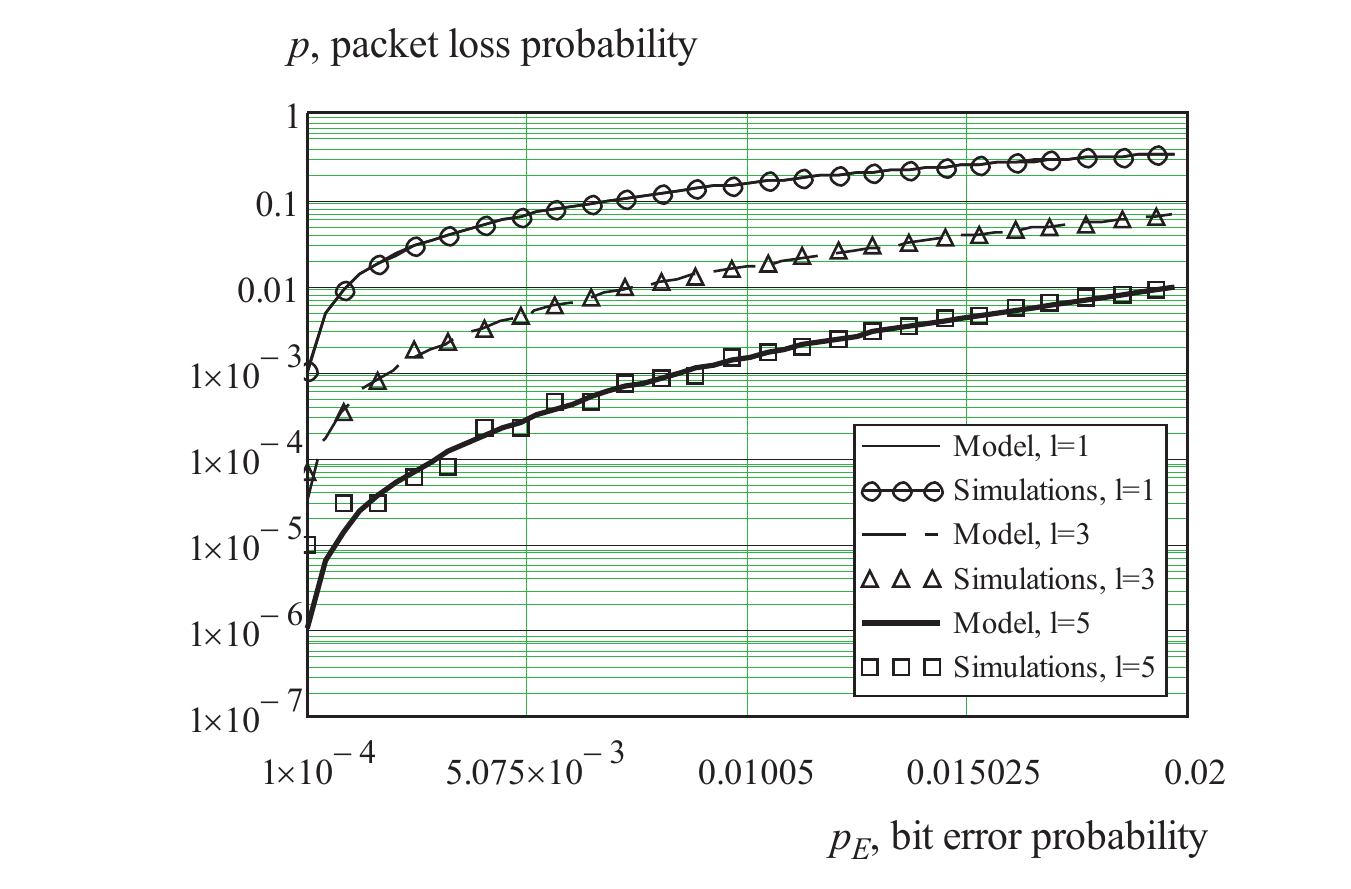}
    \label{fig:firstSingle_c_4}
  }
} \caption{First codeword error probability.}
  \label{fig:firstSingle_c}
\end{figure*}

In addition to the point estimate of the packet error probability we are interested in
interval estimates as they allow to make reliable conclusions about
deviations of our models from the actual performance. There are a number of ways how to
obtain the interval estimate for our data. Here, we briefly explain the approach we use.
When the number of experiments $N$ (simulated packets) is large while the point estimate
of the packet error probability is not extremely small or very close to $1$ the
distribution of the point estimate $\hat{p}$ is approximately Normal. The following holds for a Normal distribution with mean $\mu$ and standard deviation $\sigma$
\begin{align}\label{eqn:026}
Pr\{|X-\mu|<c\}=2\Phi(C/\sigma)=\gamma,
\end{align}
where $\Phi(\cdot)$ is the error function, $C$ is some positive constant, and $\gamma$ is
the confidence probability. Observing that $\mu=p$ and $\sigma=\sqrt{p(1-p)/N}$ from
$2\Phi(C/\sigma)$ we see that $t=\sigma\sqrt{N}/\sqrt{p(1-p)}$ leading to $c=t\sqrt{p(1-p)/N}$
and we may rewrite (\ref{eqn:026}) as
\begin{align}\label{eqn:027}
Pr\{|X-p|<t\sqrt{p(1-p)/N}\}=2\Phi(t)=\gamma.
\end{align}

Thus, the following holds with probability $\gamma$
\begin{align}\label{eqn:028}
|\hat{p}-p|<t\sqrt{p(1-p)/N},
\end{align}
whose solutions are
\begin{align}\label{eqn:029}
p_{1}=\hat{p}\pm{}t\sqrt{\hat{p}(1-\hat{p})/N},
\end{align}
where $t$ is found as $\Phi(t)=\gamma/2$.

%It is important to note that when $p$ is small the assumption of Normal distribution of the point estimate $\hat{p}$ does not hold. As we will see in the next section this would result in confidence intervals being exceptionally wide. Thus, only those estimates having values greater than $3E-4$ can be considered as reliable.

For comparison purposes we use BCH codes of length $63$ with correction capabilities
$l=1,3,5$, corresponding to $k=57,45,36$ data bits. The results presented here are
representative implying that our conclusions holds for BCH codes with other codeword
lengths but similar code rates $k/n$. The size of a packet is $nIv$, where $v$ is the
number of interleaved codeblocks. In order to demonstrate the effect of different
interleaving depths we need to keep the packet size constant. We keep the product $vI$
constant by choosing the following pairs of $(I,v)$: $(1,16)$, $(2,8)$, $(4,4)$, $(8,2)$,
$(16,1)$. With this choice of $(I,v)$ the whole packet size (including all headers) is
always $63\times{}16=1008$ bits ($126$ bytes). We exclude the cases $c=0$ and $(I,v)=(1,16)$ as they are trivial.

\subsection{Identifying the ranges of parameters of interest}

\subsubsection{First codeword error probability}

All the models proposed in this paper consist of two steps. In one of them (model 2) we
estimate the first codeword error probability at the first step. In the other two, the
first step consists in estimation of two-dimensional distribution of the number of
incorrectly received bits in the first two codewords. A widely used approach for
verification of complex models is to check their parts separately. This gives us
information as to which part of the model provides accurate approximation and/or which
part of the model is the source of bias. Here, we will evaluate accuracy of the first
codeword error probability for model 2, while in the next subsection we will take a look
at the probability of incorrect reception of the first two codewords.

Fig. \ref{fig:firstSingle_c} compares the modeling results with those obtained via
simulations for different values $l$ and $c$. We see that our approach allows to closely
approximate this metric implying that the first part of the model correctly predicts the
first codeword error probability. Some deviations from the simulation data are again seen
when the first codeword error probability approaches $10E-5$. However, these deviations are not systematic and rather randomly spread around the line representing the modeling results. They are attributed to the limited number of experiments set to $N=10^5$.

\begin{figure*}[t!]
\centerline{
  \subfigure[$l=3$]{
    \includegraphics[width=0.5\textwidth]{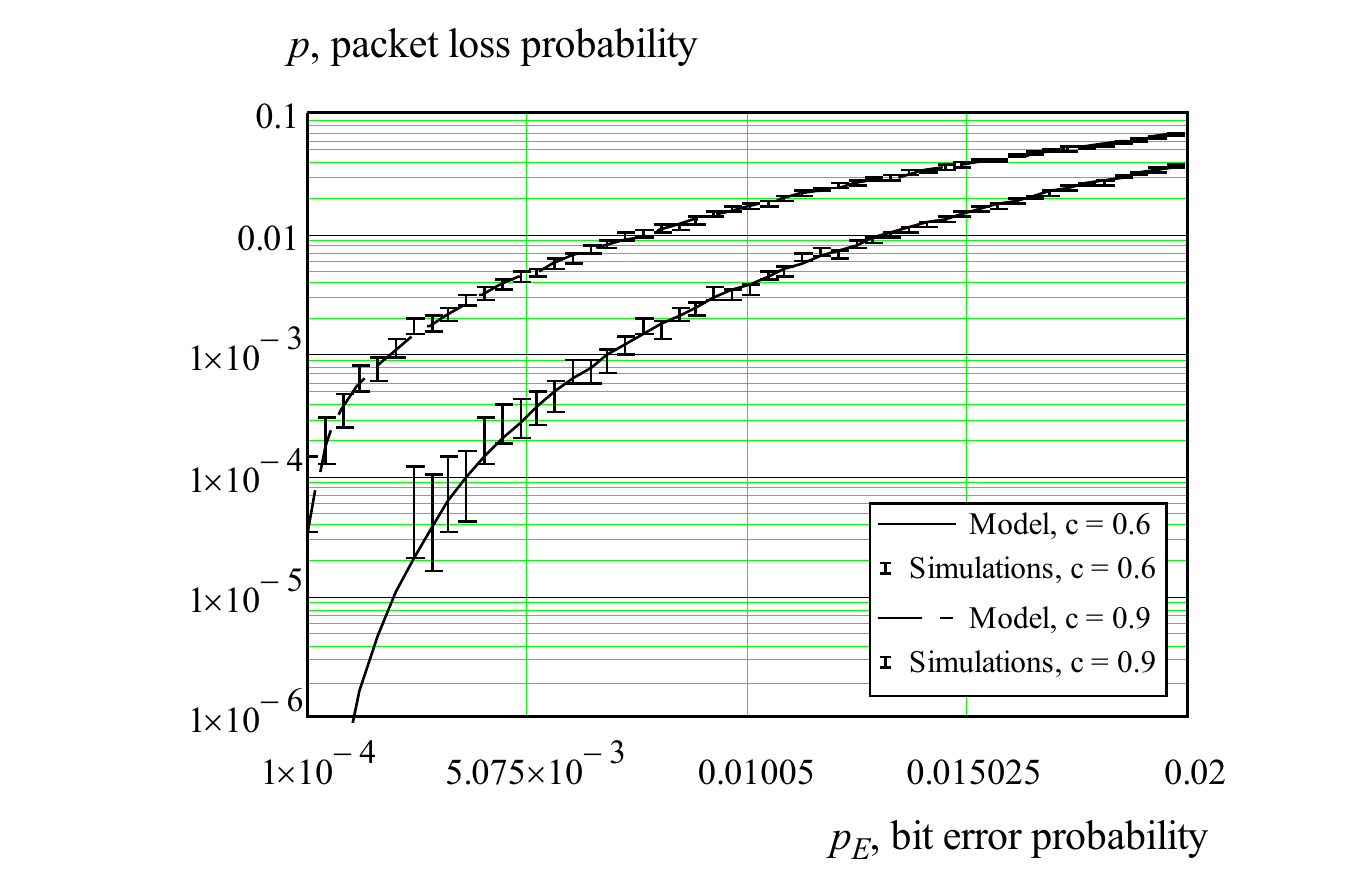}
    \label{fig:firstSingle_Int_c_1}
  }
  \subfigure[$l=5$]{
    \includegraphics[width=0.5\textwidth]{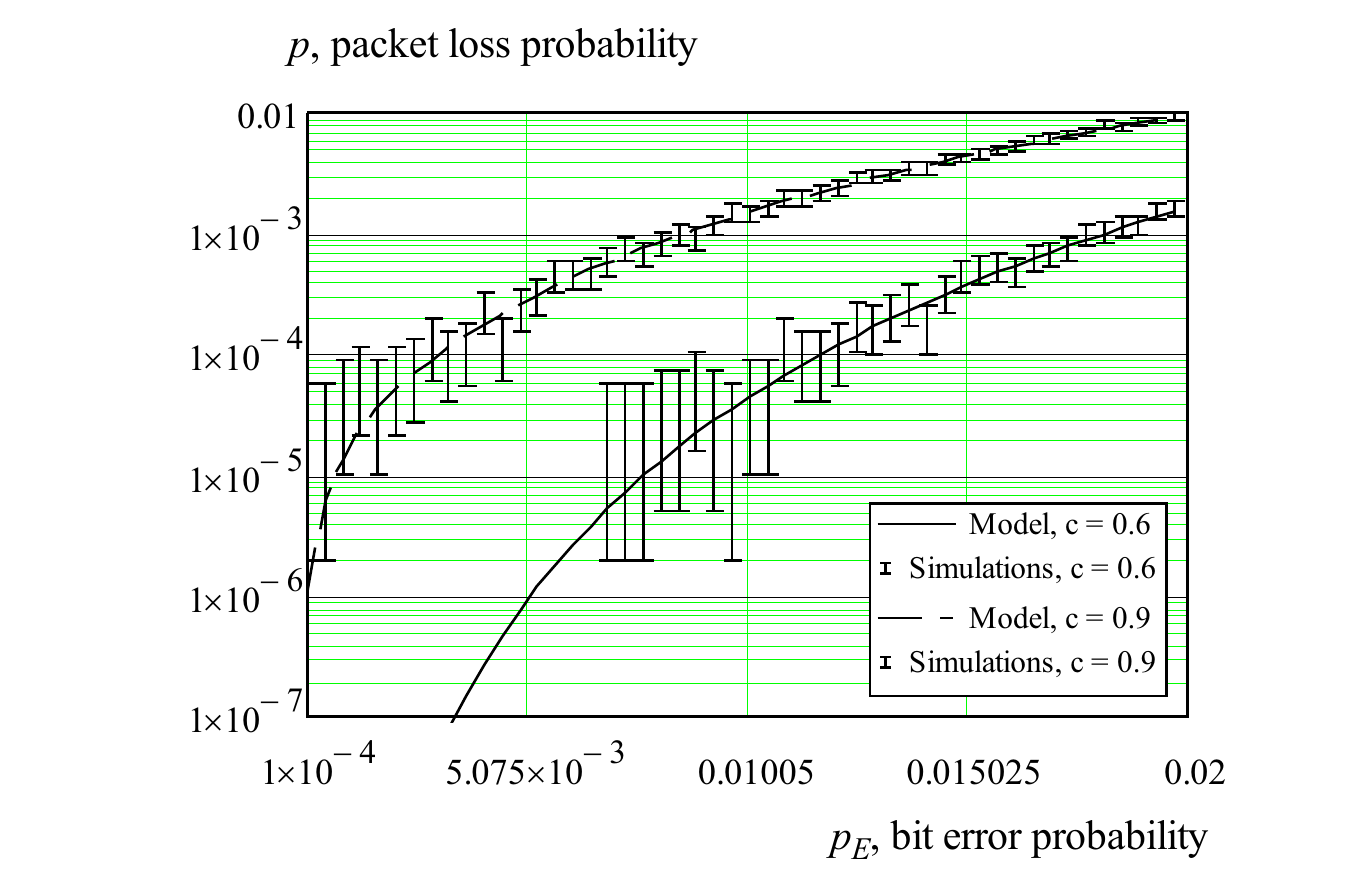}
    \label{fig:firstSingle_Int_c_2}
  }
} \caption{Interval estimations for first codeword error probability.}
  \label{fig:firstSingle_Int_c}
\end{figure*}

Comparison of the modeling results with interval estimations (the confidence level set to
$0.05$) for different values of $c$ and $l$ shown in Fig. \ref{fig:firstSingle_Int_c} approves our conclusions. All analytical results are in-between intervals obtained using simulations except for that region, where the packet error probability falls below $10E-4$. One may also observe that as the absolute value of the BER increases the confidence intervals gets shorter. As we already discussed above this is the inherent feature of our interval estimator. %For small values of BER we get only few incorrectly received packets implying that the packet loss probability becomes smaller while the estimations become biased.

\begin{figure*}[t!]
\centerline{
  \subfigure[$c=0.6$]{
    \includegraphics[width=0.5\textwidth]{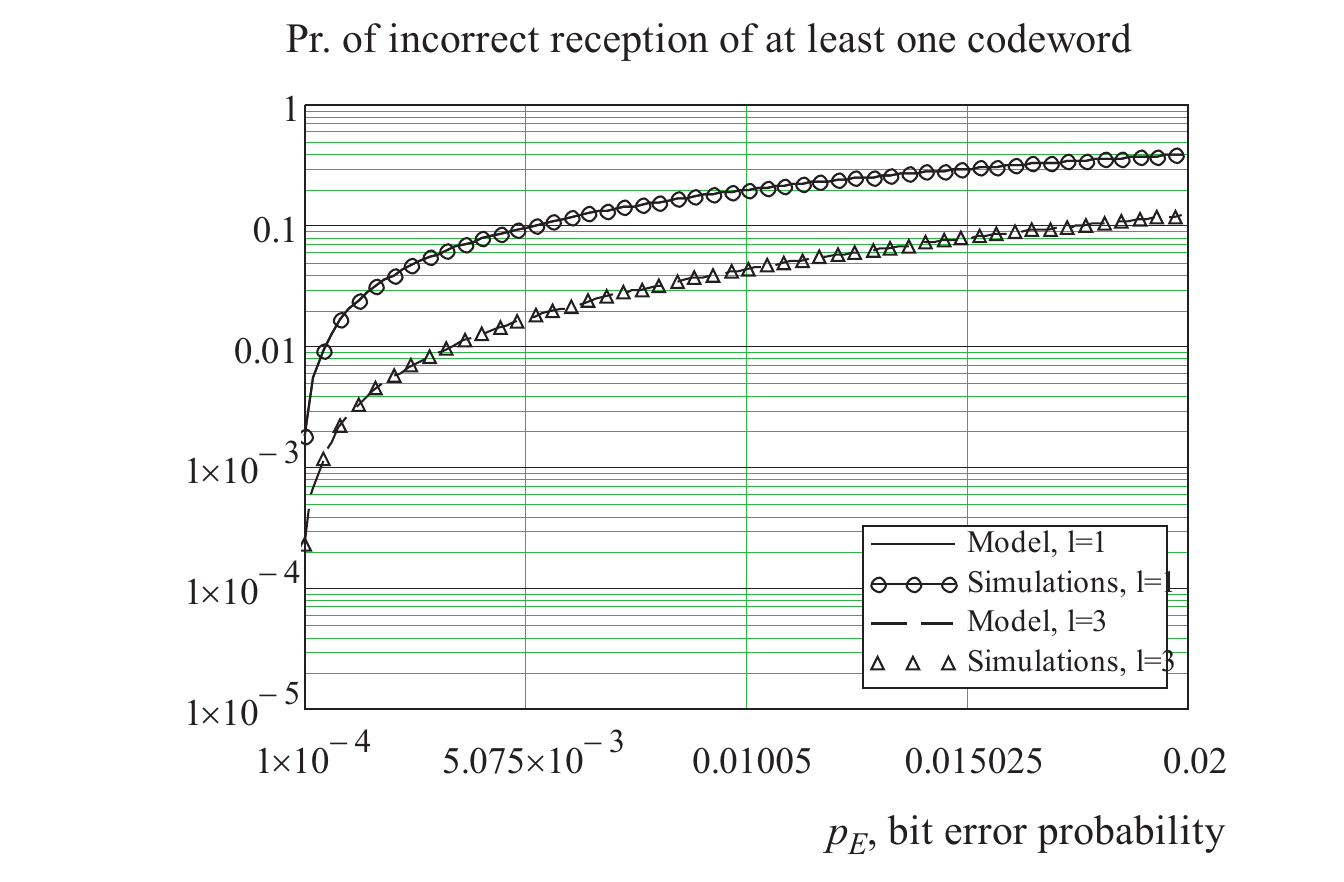}
    \label{fig:firstDouble_Int_c_1}
  }
  \subfigure[$c=0.9$]{
    \includegraphics[width=0.5\textwidth]{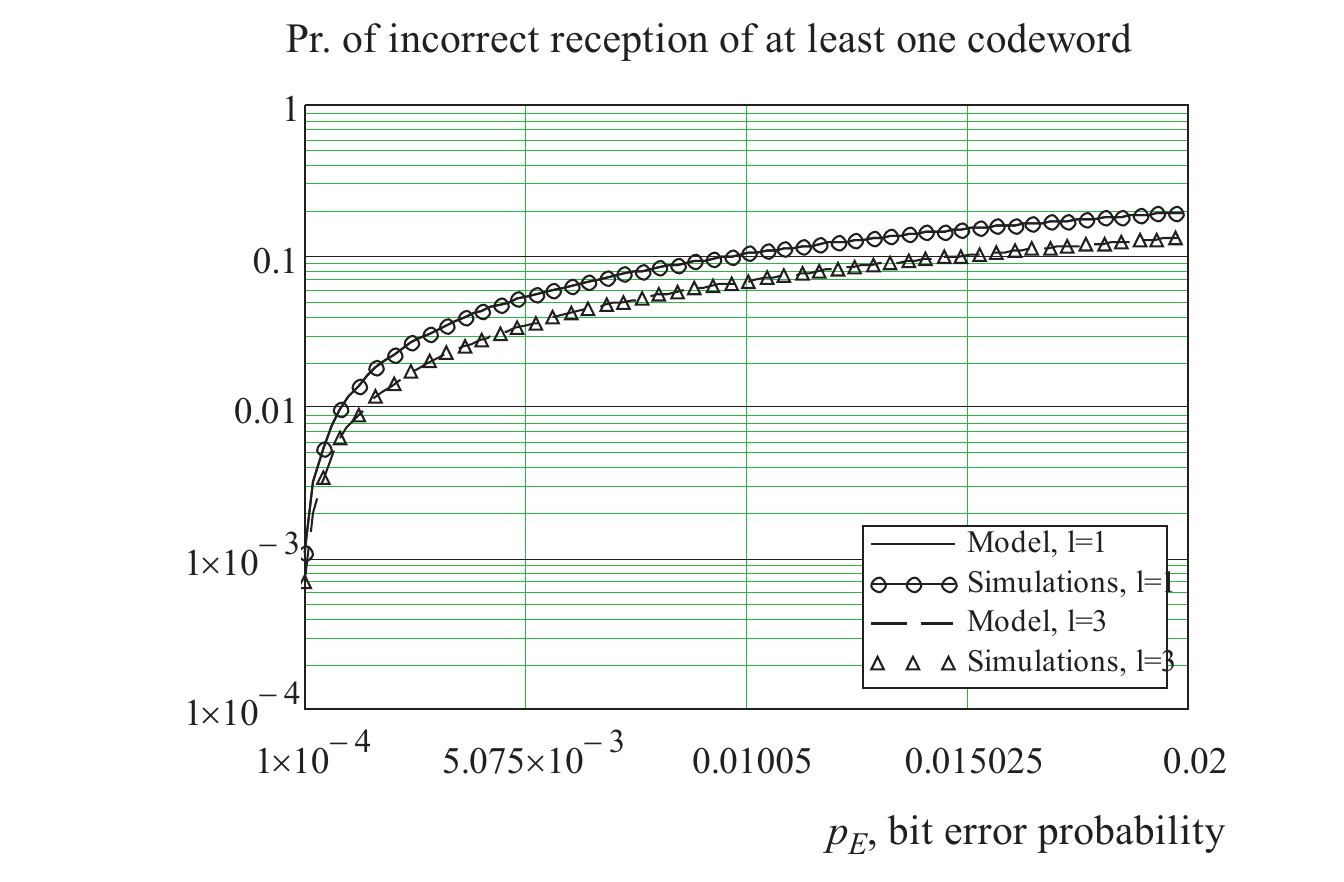}
    \label{fig:firstDouble_Int_c_2}
  }
} \caption{Incorrect reception of at least one out of two first codewords.}
  \label{fig:firstDouble_Int_c}
\end{figure*}

\subsubsection{Incorrect reception of first two codewords}

Estimation of two-dimensional distribution of the number of incorrectly received bits in
the first two codewords is the first step for models 1 and 3. We use it at the second
step of the modeling procedure to specify transition probabilities of the absorbing
Markov chain. To evaluate accuracy of this step it is sufficient to consider the
probability of incorrect reception of at least one of these codewords. This gives the
probability of the complement of the event consisting in correct reception of both
codewords. There is a strict functional relationship between this event and the joint
two-dimensional distribution of the number of incorrectly received bits in first two
codewords.

The comparison between the point estimates obtained via simulations and the modeling data for probability of incorrect reception of at least one codeword out of the first two is
shown in Fig. \ref{fig:firstDouble_Int_c}. We see that the modeling results closely match
the simulation ones implying that we make no mistakes at this step. It is important to
note that (as we will see in the next section) such a good approximation means that these two models correctly predict the probability of incorrect reception of a packet for any packet size and interleaving scheme $(I,v)=(2,8)$. %The reason is that we perfectly approximate the probability of incorrect reception of one interleaved codeblock while the effect of correlation between successive codeblocks is negligible.

% We still use old figures here

%Based on the results of this section we see that we are not interested in those cases when $c=0.0$ and/or $I=1$. For these parameters all the models provide accurate approximation of the packet loss probability. Furthermore, for the case $c=0.0$ there is a closed-form expression that is easier to use, see (\ref{eqn:007}). We also showed that at the first step of the modeling procedure we make no mistakes for all the models. Thus, all possible inaccuracies that we may notice further are incurred by the second step of the modeling procedure.

\subsection{Performance comparison of models}

\begin{figure*}[t!]
\centerline{
  \subfigure[$l=1,c=0.3$]{
    \includegraphics[width=0.33\textwidth]{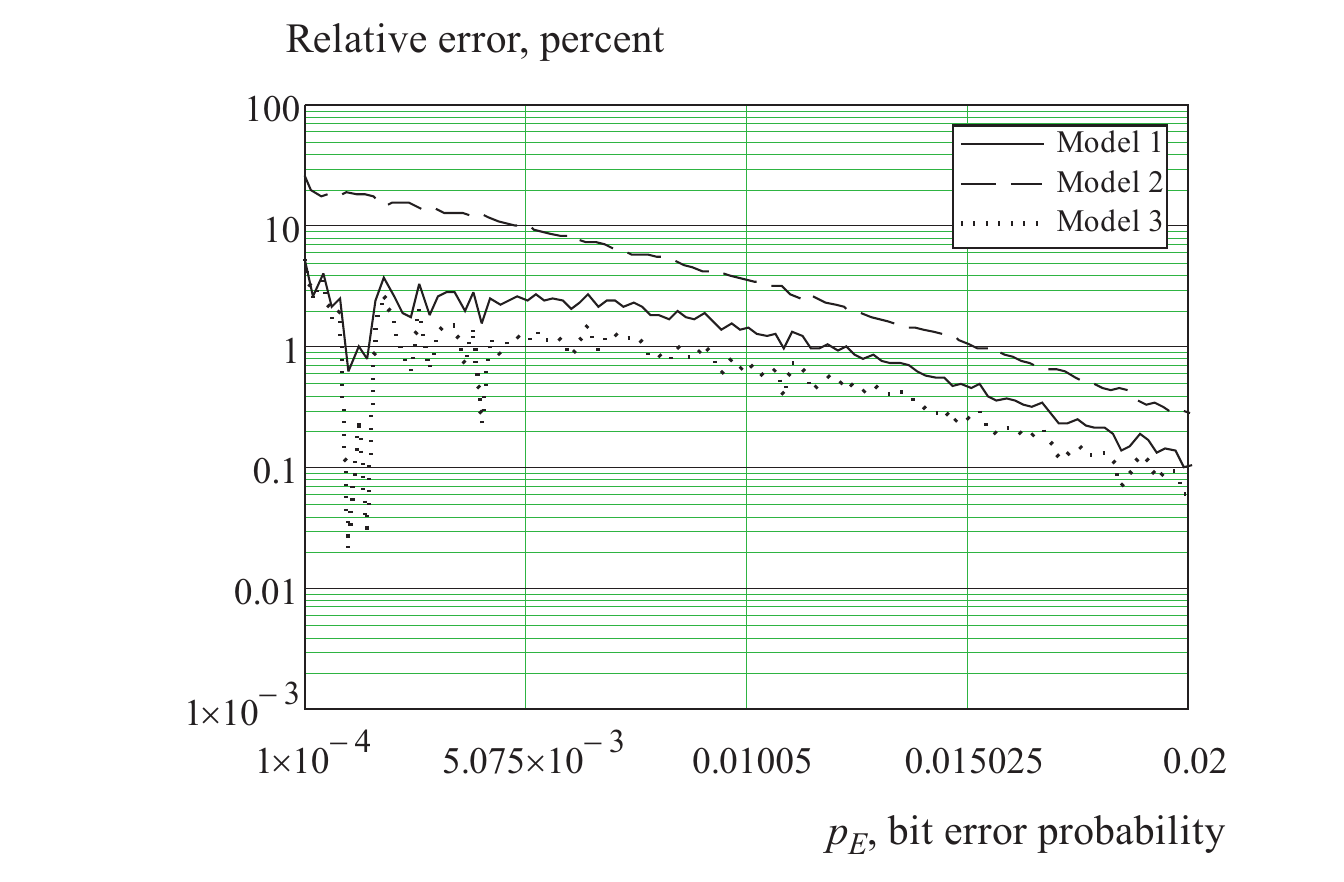}
    \label{fig:i16v1_l1c03}
  }
  \subfigure[$l=1,c=0.6$]{
    \includegraphics[width=0.33\textwidth]{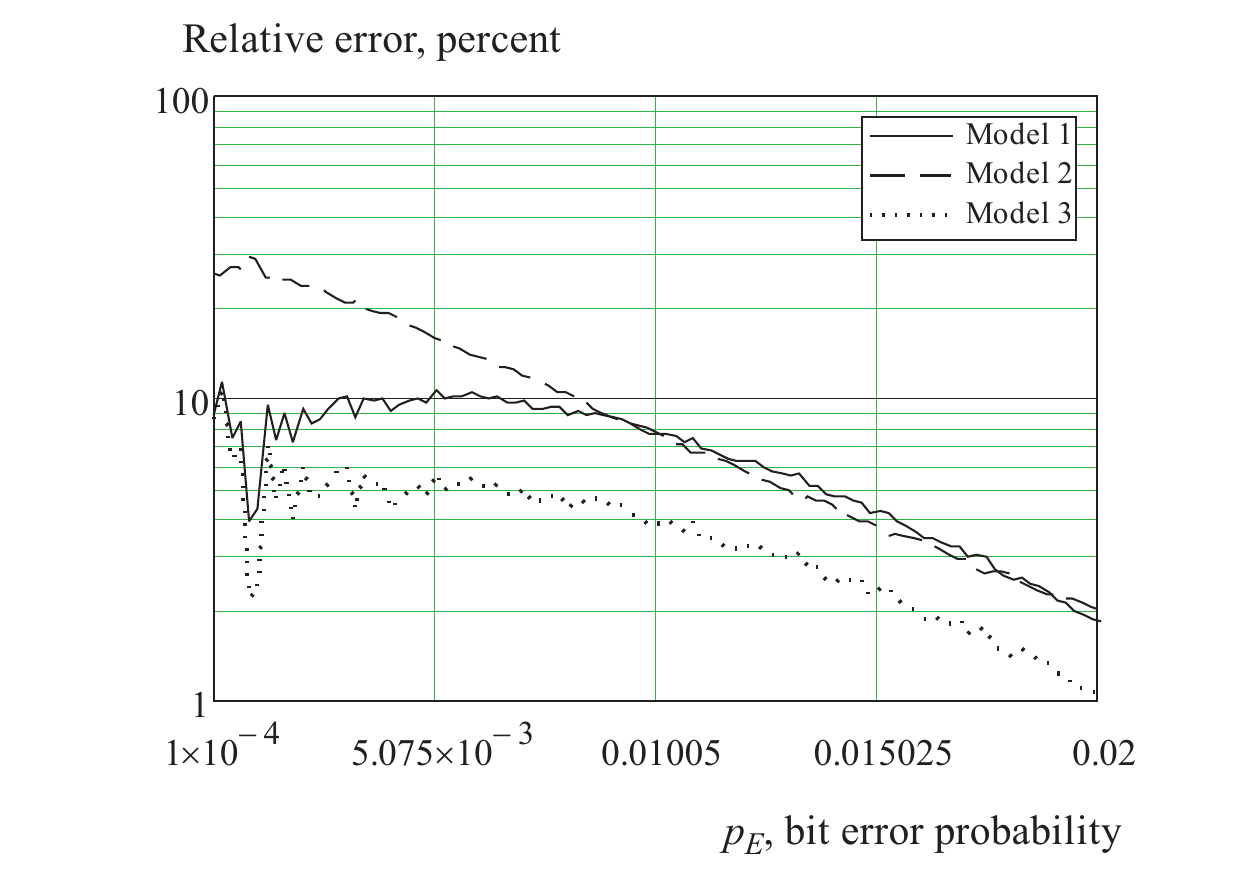}
    \label{fig:i16v1_l1c06}
  }
  \subfigure[$l=1,c=0.9$]{
    \includegraphics[width=0.33\textwidth]{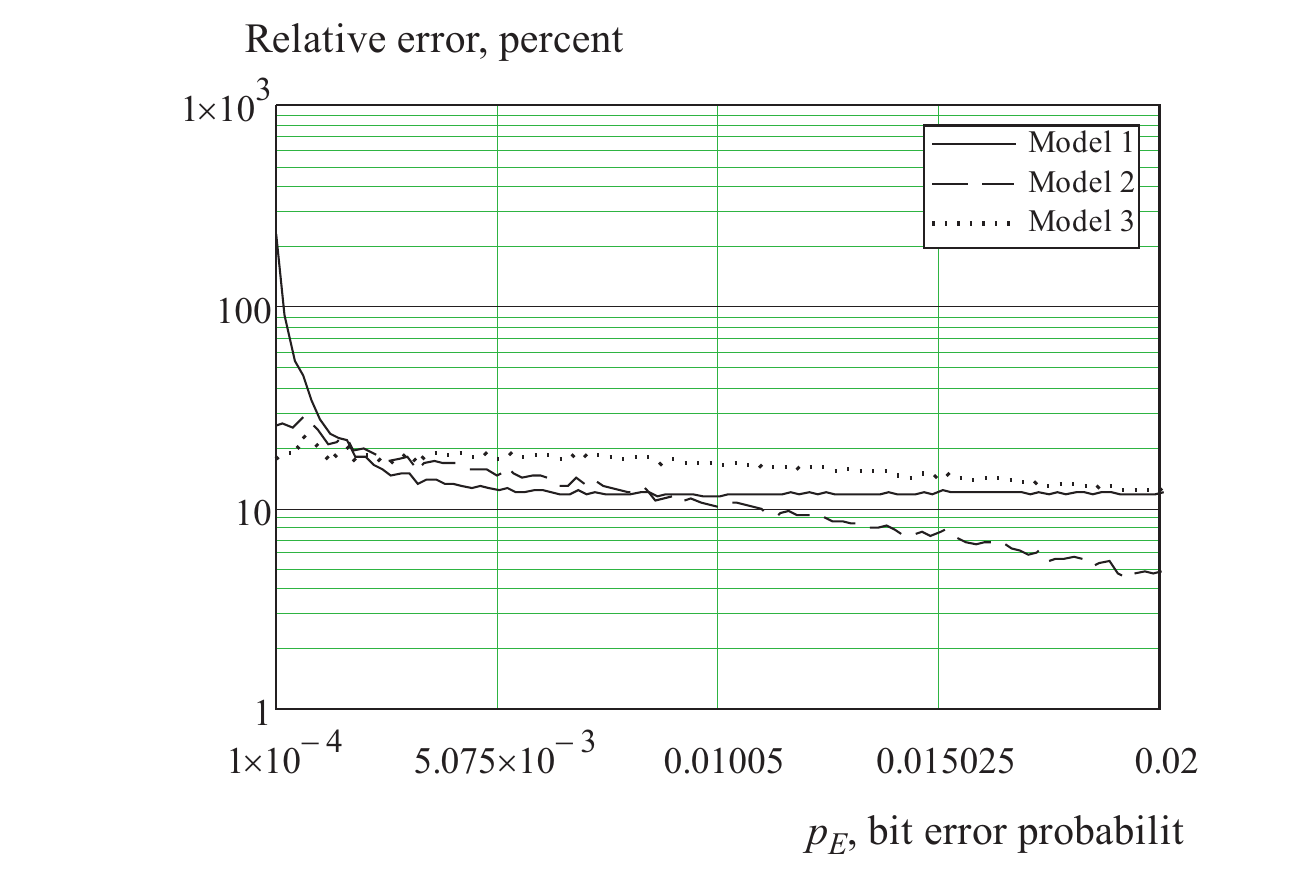}
    \label{fig:i16v1_l1c09}
  }
} \centerline{
  \subfigure[$l=3,c=0.3$]{
    \includegraphics[width=0.33\textwidth]{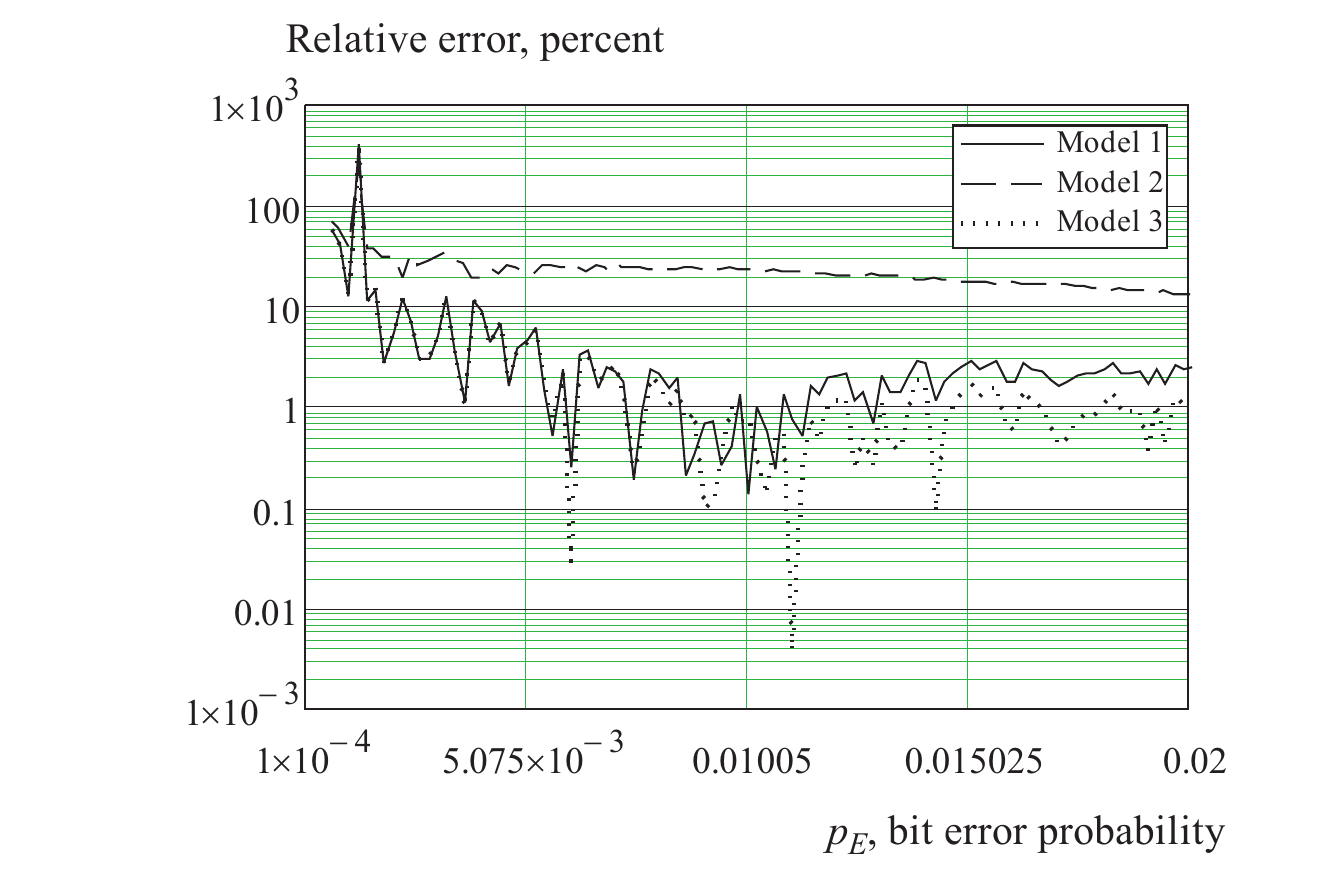}
    \label{fig:i16v1_l3c03}
  }
  \subfigure[$l=3,c=0.6$]{
    \includegraphics[width=0.33\textwidth]{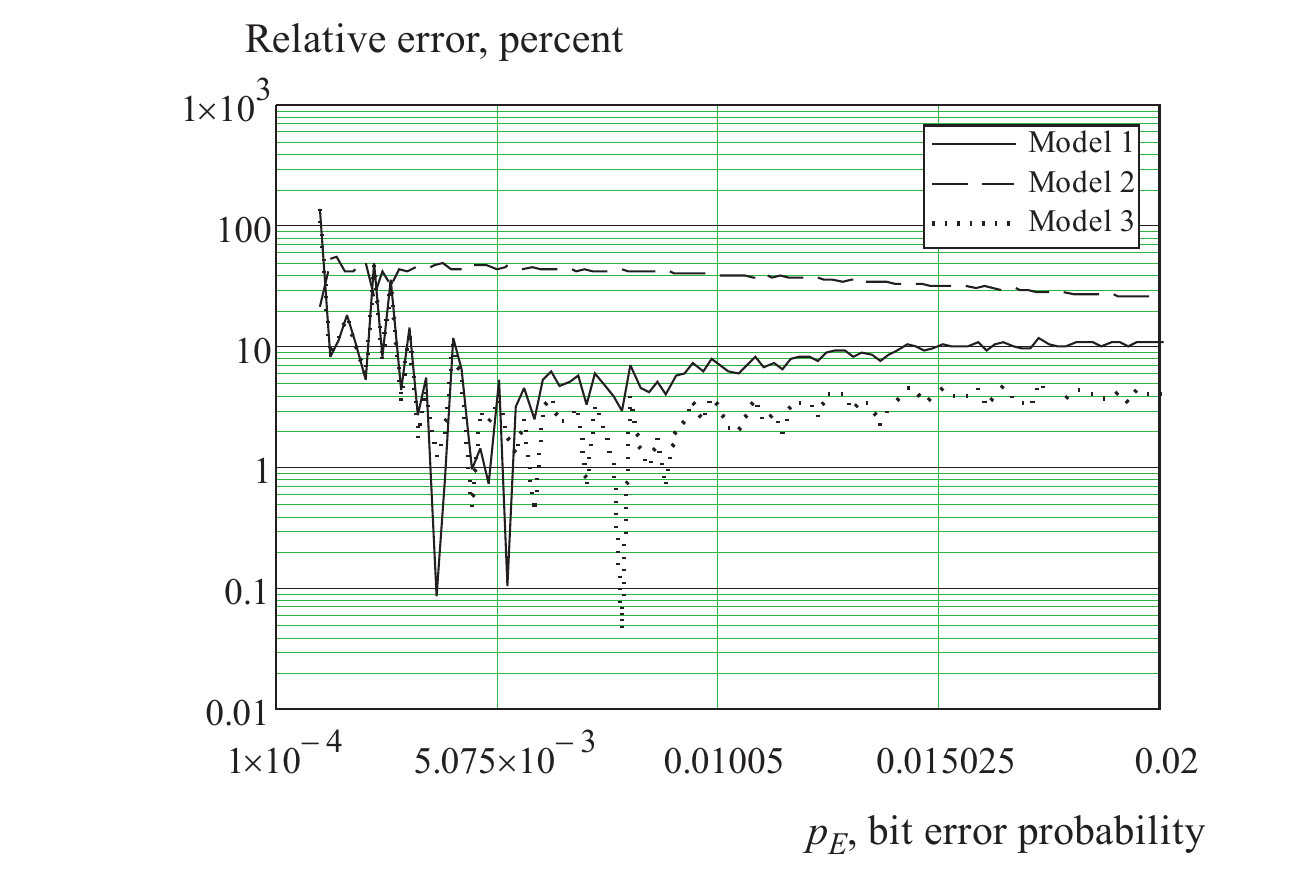}
    \label{fig:i16v1_l3c06}
  }
  \subfigure[$l=3,c=0.9$]{
    \includegraphics[width=0.33\textwidth]{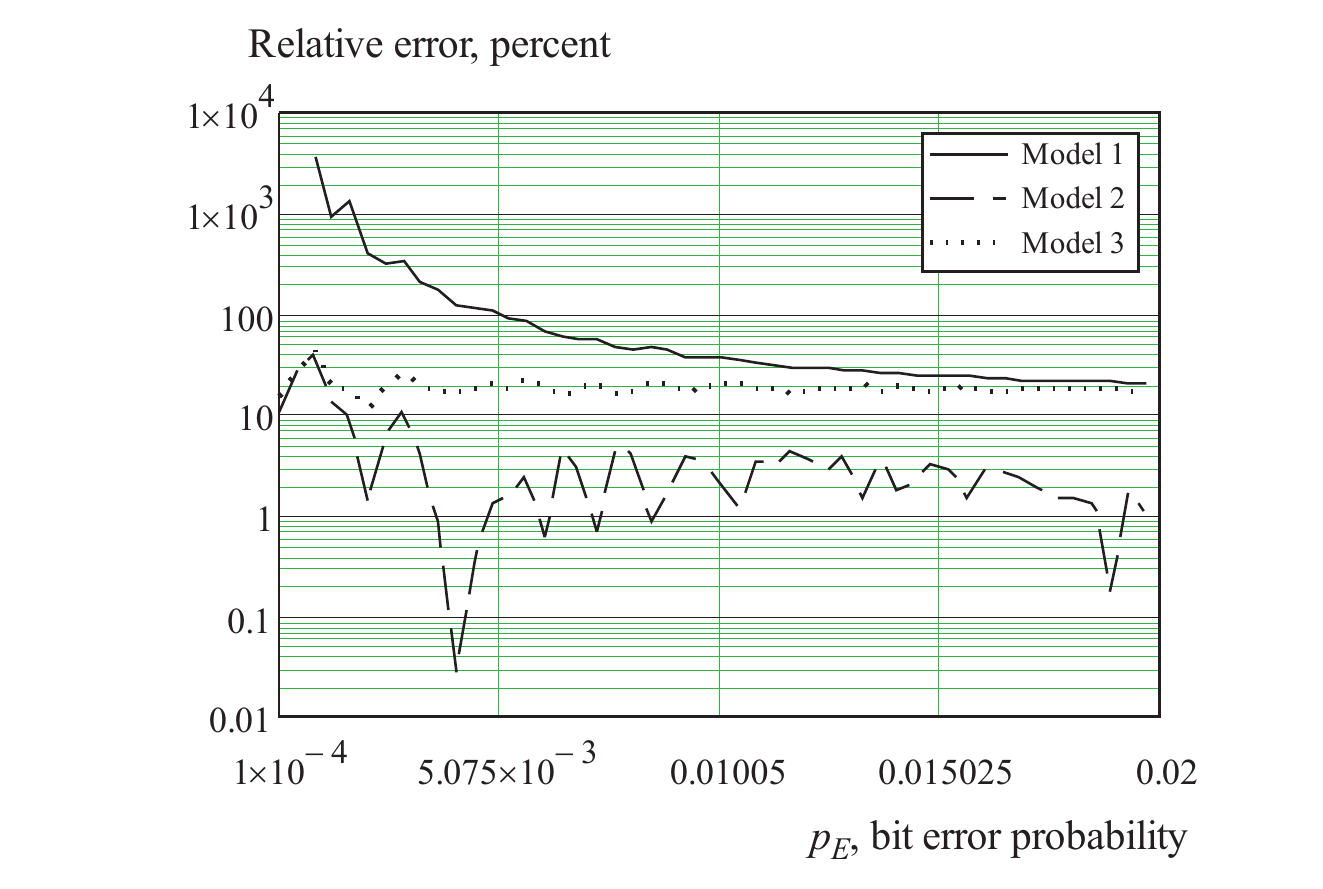}
    \label{fig:i16v1_l3c09}
  }
} \centerline{
  \subfigure[$l=5,c=0.3$]{
    \includegraphics[width=0.33\textwidth]{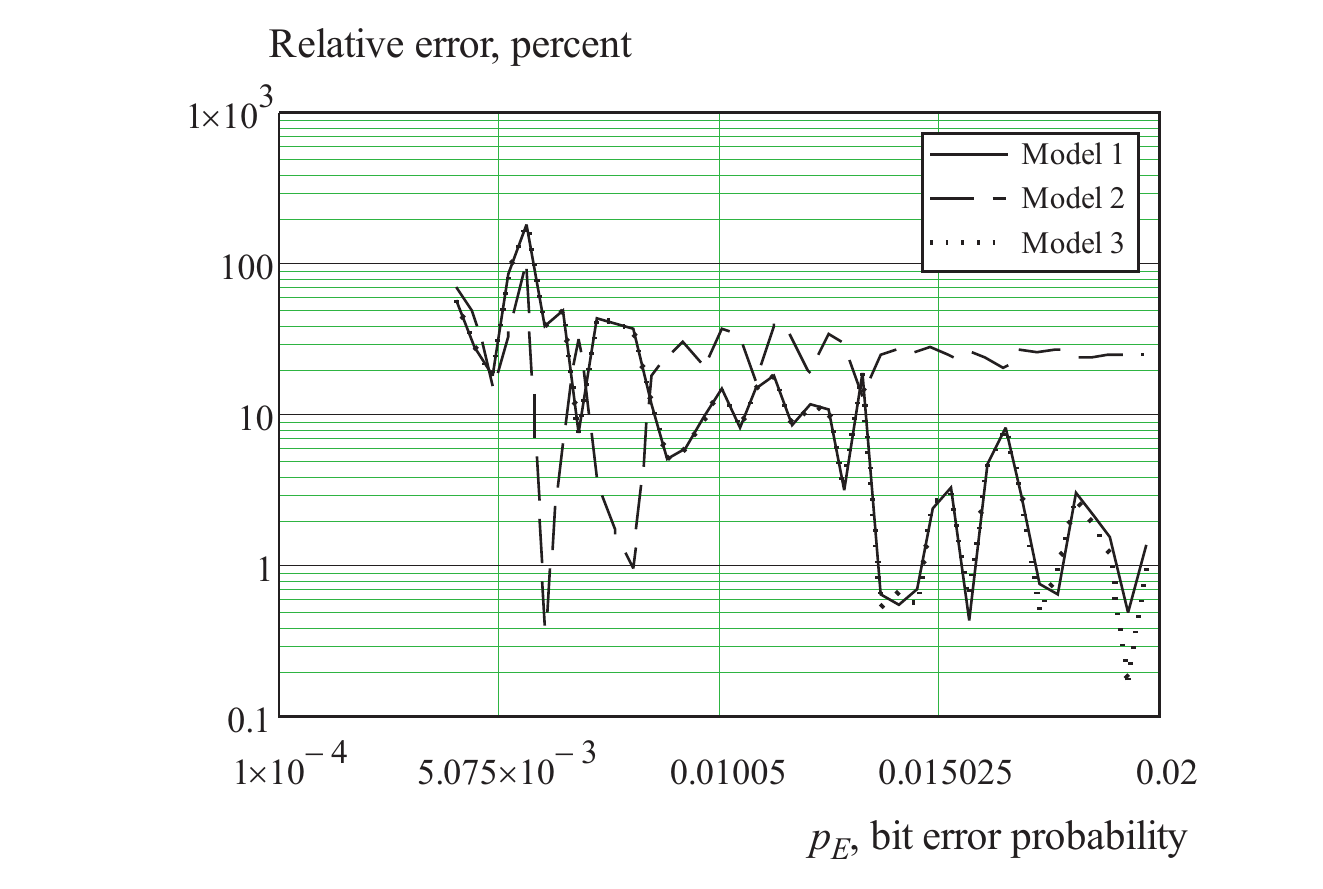}
    \label{fig:i16v1_l5c03}
  }
  \subfigure[$l=5,c=0.6$]{
    \includegraphics[width=0.33\textwidth]{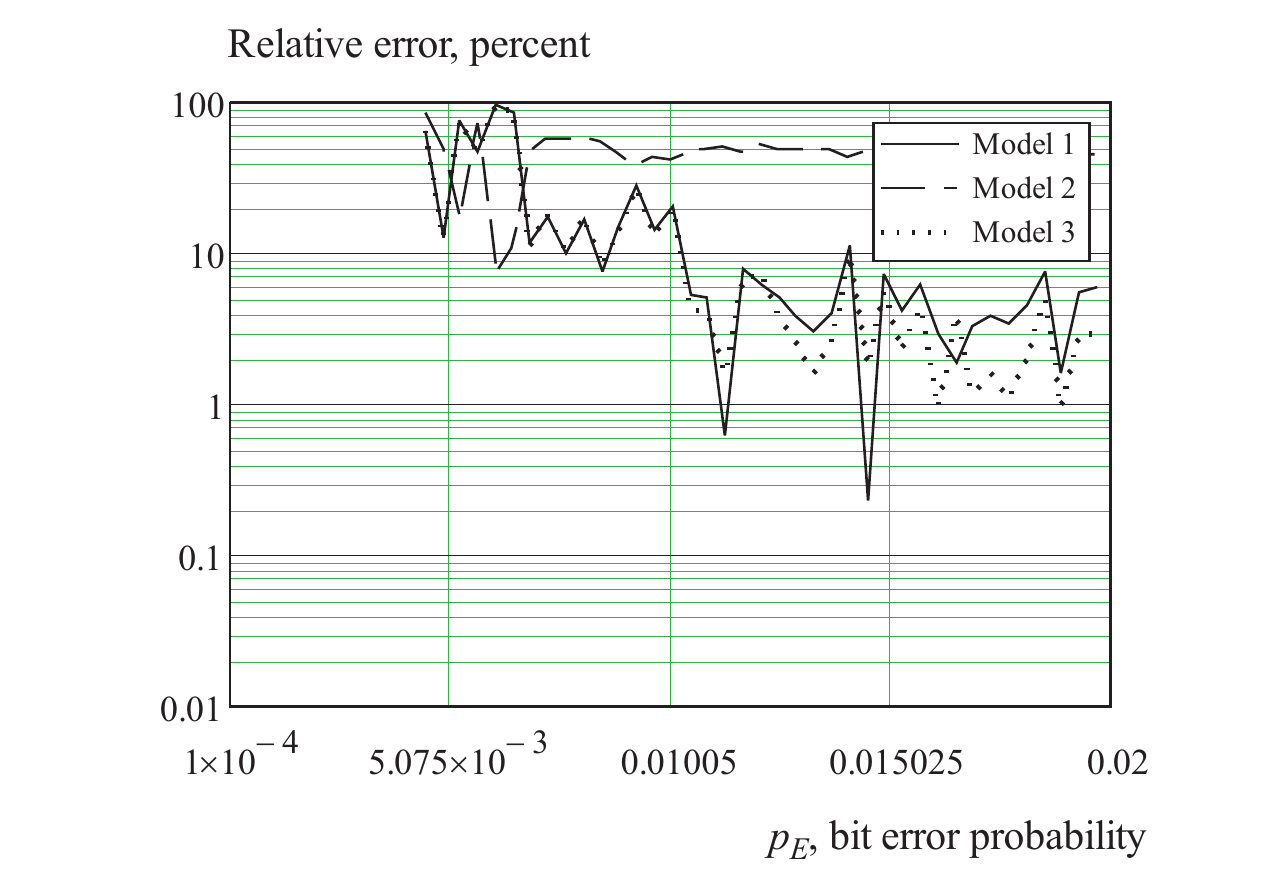}
    \label{fig:i16v1_l5c06}
  }
  \subfigure[$l=5,c=0.9$]{
    \includegraphics[width=0.33\textwidth]{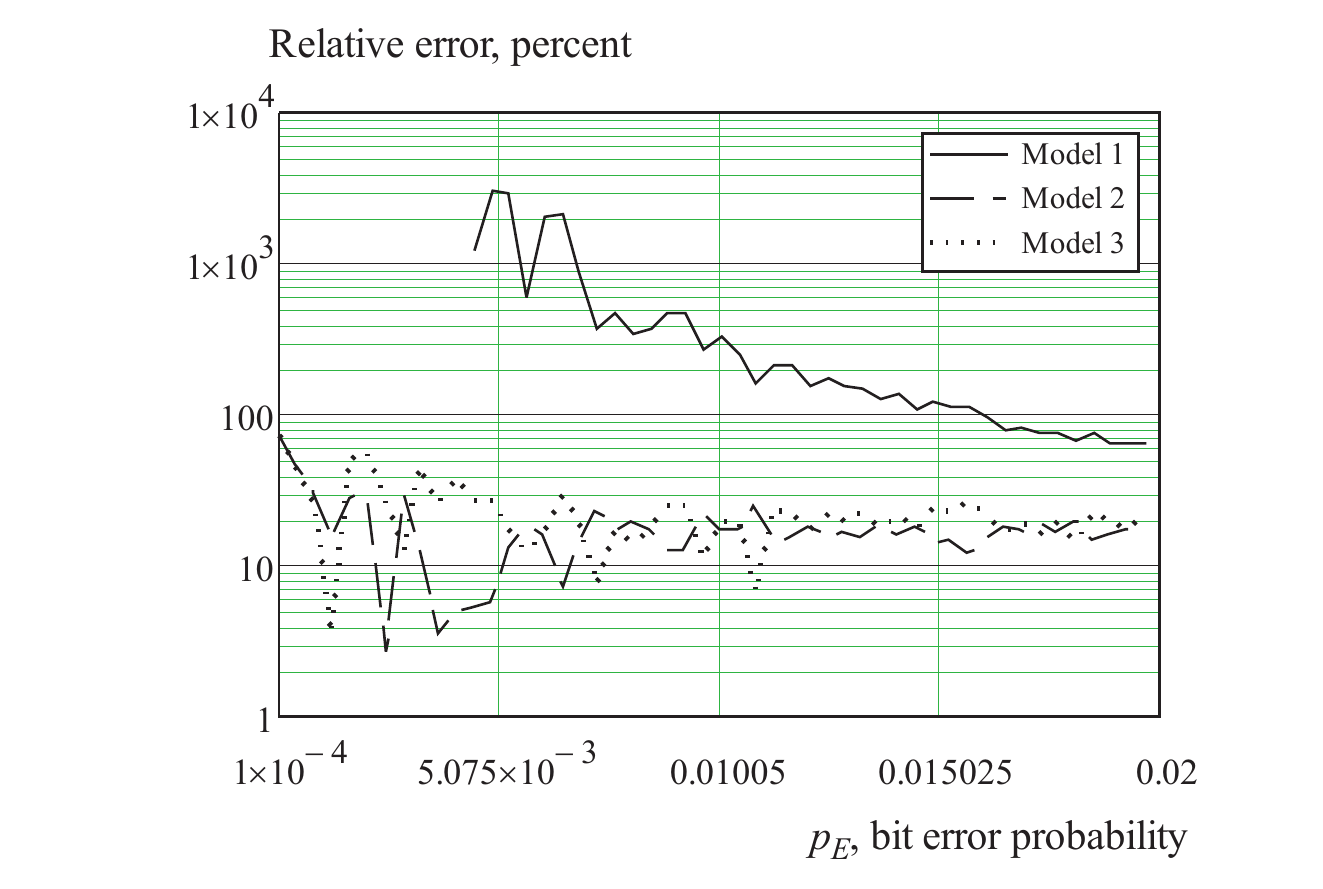}
    \label{fig:i16v1_l5c09}
  }
} \caption{Comparison of three models for $(I,v)=(16,1)$.}
  \label{fig:i16v1}
\end{figure*}

%\subsubsection{Other values of $(I,v)$}

%Once we learned the behavior of our models for $(I,v)=(2,8)$ we need to see what happens when $I$ increases. In order to keep the packet size constant the maximum possible choice of $I$ is $16$ implying the $(I,v)=(16,1)$ interleaving scheme. 

Fig. \ref{fig:i16v1} illustrates the relative error of approximation of the packet error probability for all three models and different values of $c$, $l$ and $(I,v)=(16,1)$ interleaving scheme. The comparative behavior of models is similar to the case $(I,v)=(2,8)$ for all considered values of $l$ and small to medium correlation ($c=0.3$ and $c=0.6$). For small to medium values of $c$ the model 2 is the worst one while the models 1 and 3 provide approximately similar accuracy. In this range of $c$ these two models deviate from empirical results at most by $10\%$ while the prediction according to the model 2 deviates by at most $30\%$. As the lag-1 NACF grows these conclusions are no longer valid. First of all, surprisingly, model 3 is not the best for $c=0.9$ and $l=1$ as both models 1 and 2 outperform it by few percents. As $l$ grows while $c$ remains set at $0.9$, the model 1 performs worse and worse with deviations reaching two orders of magnitude for small values of BER and $l=3,5$. The model 2 continues to outperform the model 1 for $l=3$ and is comparable to the model 3 for $l=5$. If one should chose a single model for all the values of BER, $c$, and $l$ the model 3 is recommended. Another advantage of model 3 is that it does not allow for extreme bias for the small values of BER. However, for all the input values there are simpler model that performs as good as (sometimes better than) model 3. For $c=0.3,0.6$, $l=1,3,5$, this is the model 1 while for $c=0.9$ this is the model 2. Anyway, for any choice of the input parameters there is a model allowing to get results deviating from the empirical data by at most $10\%$.

\begin{figure*}[t!]
\centerline{
  \subfigure[$(I,v)=(4,4),l=1$]{
    \includegraphics[width=0.33\textwidth]{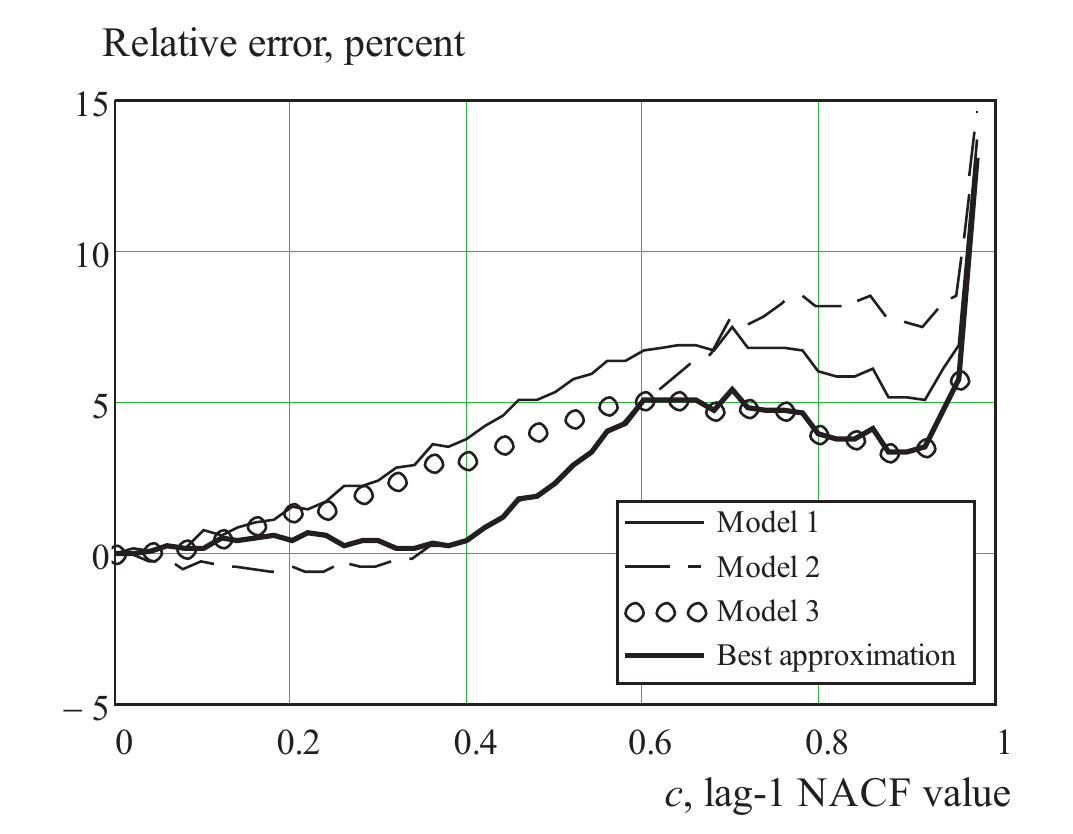}
    \label{fig:comp_corr_001_44l1}
  }
  \subfigure[$(I,v)=(8,2),l=1$]{
    \includegraphics[width=0.33\textwidth]{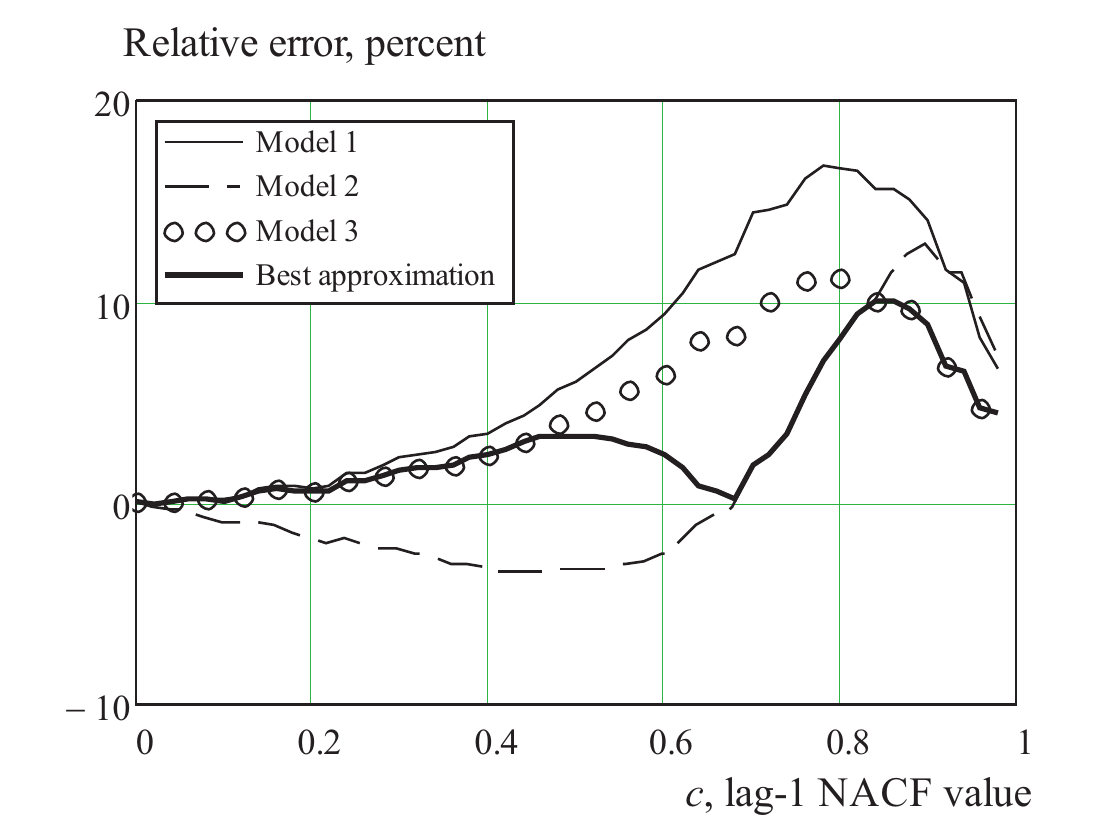}
    \label{fig:comp_corr_001_82l1}
  }
  \subfigure[$(I,v)=(16,1),l=1$]{
    \includegraphics[width=0.33\textwidth]{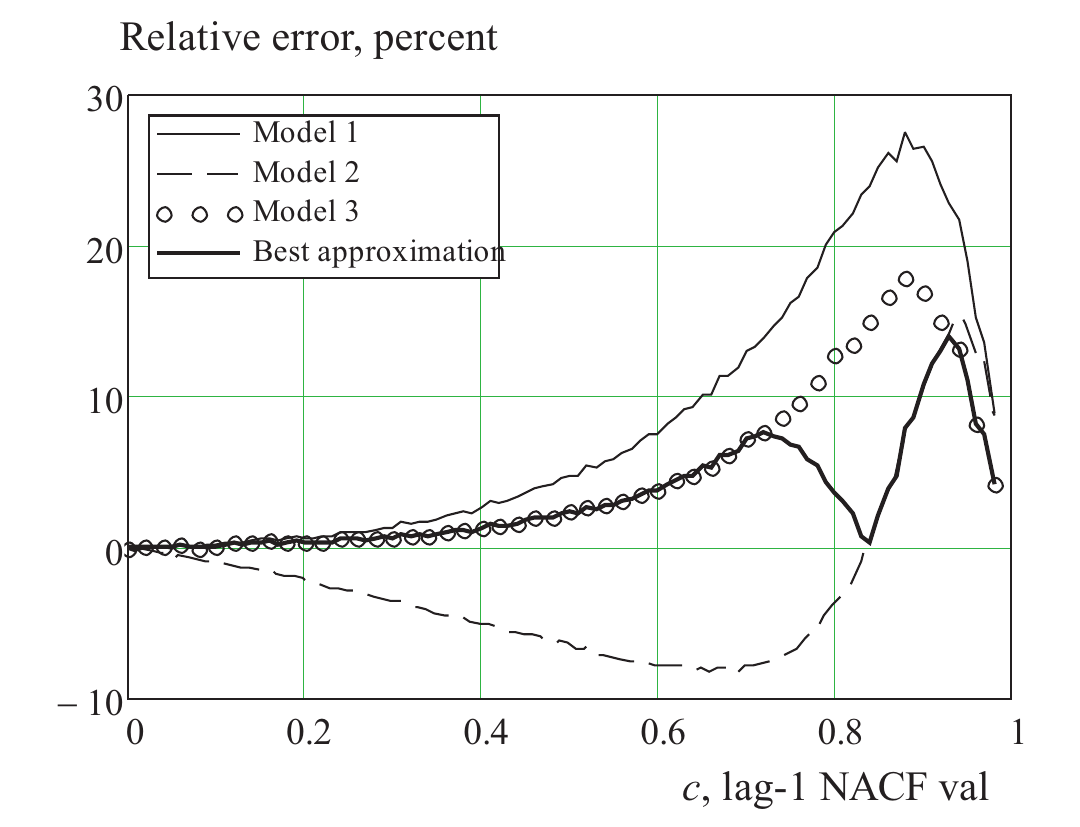}
    \label{fig:comp_corr_001_161l1}
  }
} \centerline{
  \subfigure[$(I,v)=(4,4),l=3$]{
    \includegraphics[width=0.33\textwidth]{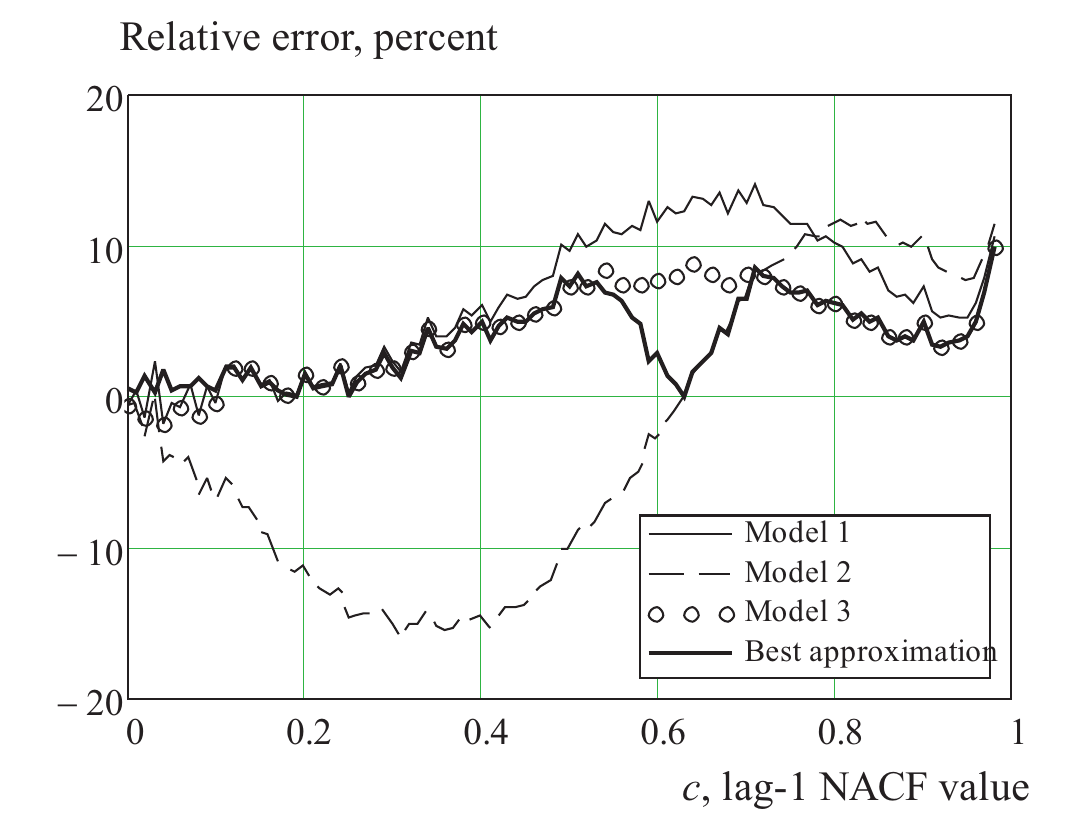}
    \label{fig:comp_corr_001_44l3}
  }
  \subfigure[$(I,v)=(8,2),l=3$]{
    \includegraphics[width=0.33\textwidth]{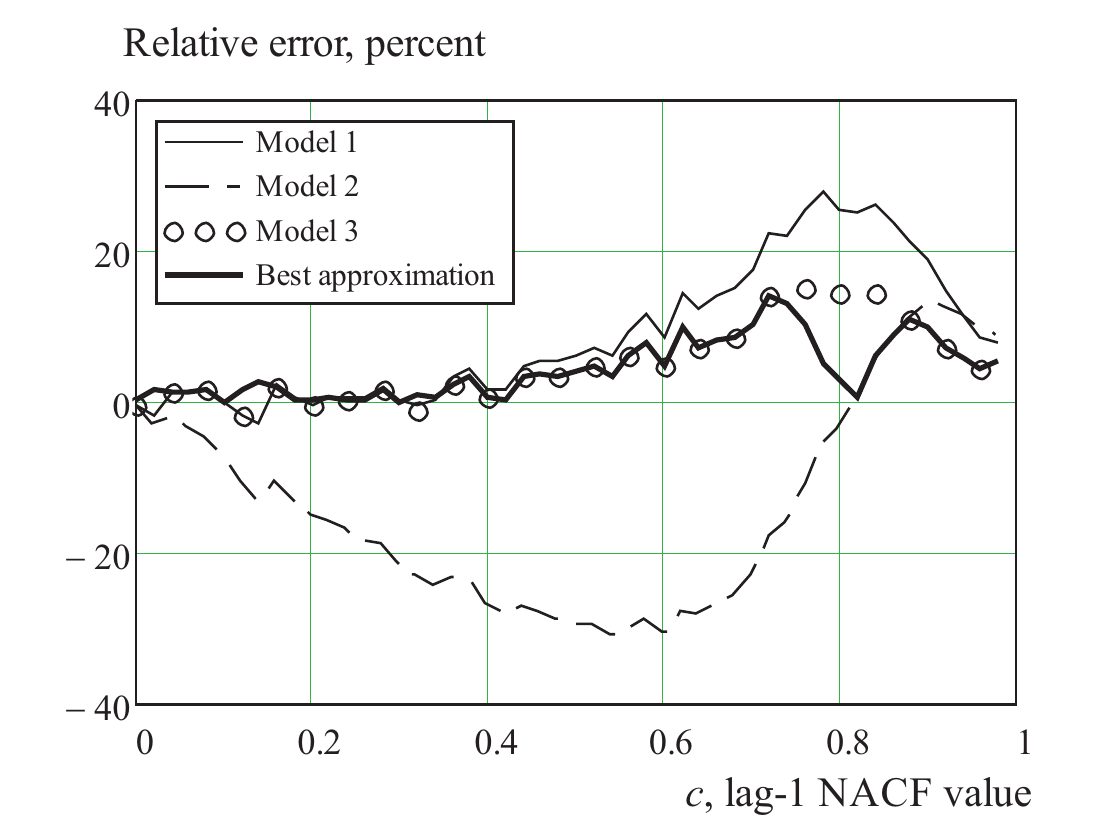}
    \label{fig:comp_corr_001_82l3}
  }
  \subfigure[$(I,v)=(16,1),l=3$]{
    \includegraphics[width=0.33\textwidth]{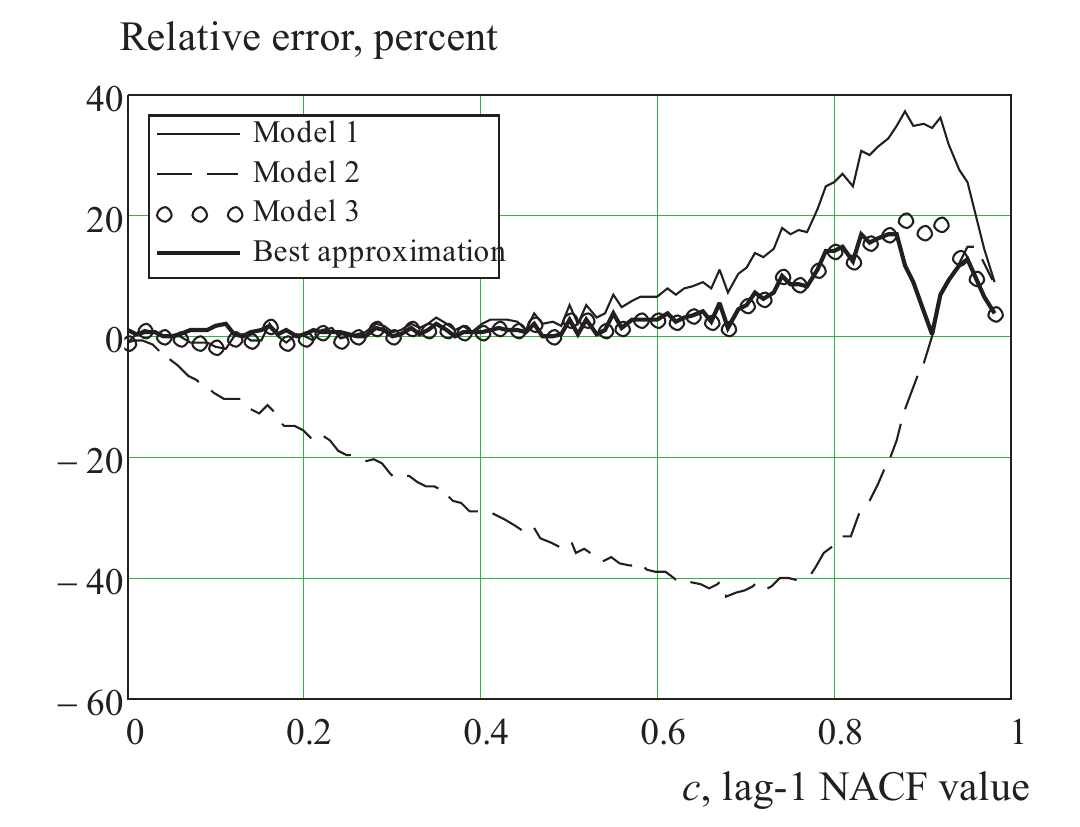}
    \label{fig:comp_corr_001_161l3}
  }
} \centerline{
  \subfigure[$(I,v)=(4,4),l=5$]{
    \includegraphics[width=0.33\textwidth]{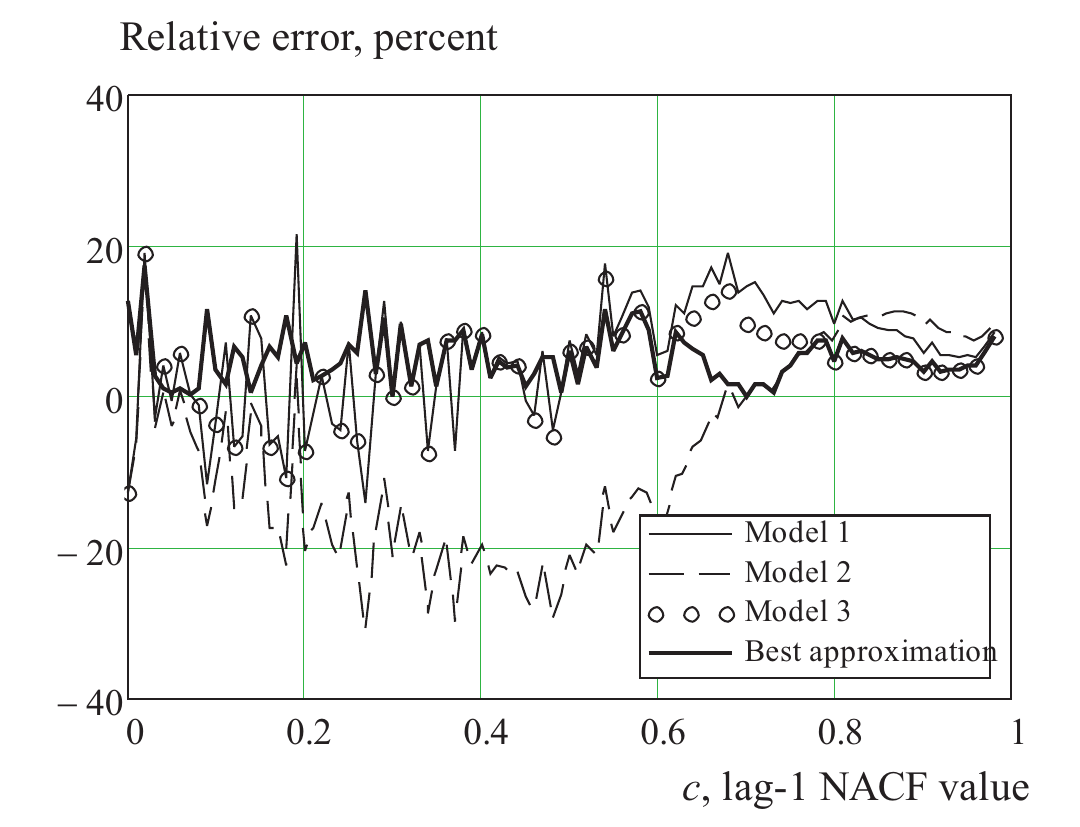}
    \label{fig:comp_corr_001_44l5}
  }
  \subfigure[$(I,v)=(8,2),l=5$]{
    \includegraphics[width=0.33\textwidth]{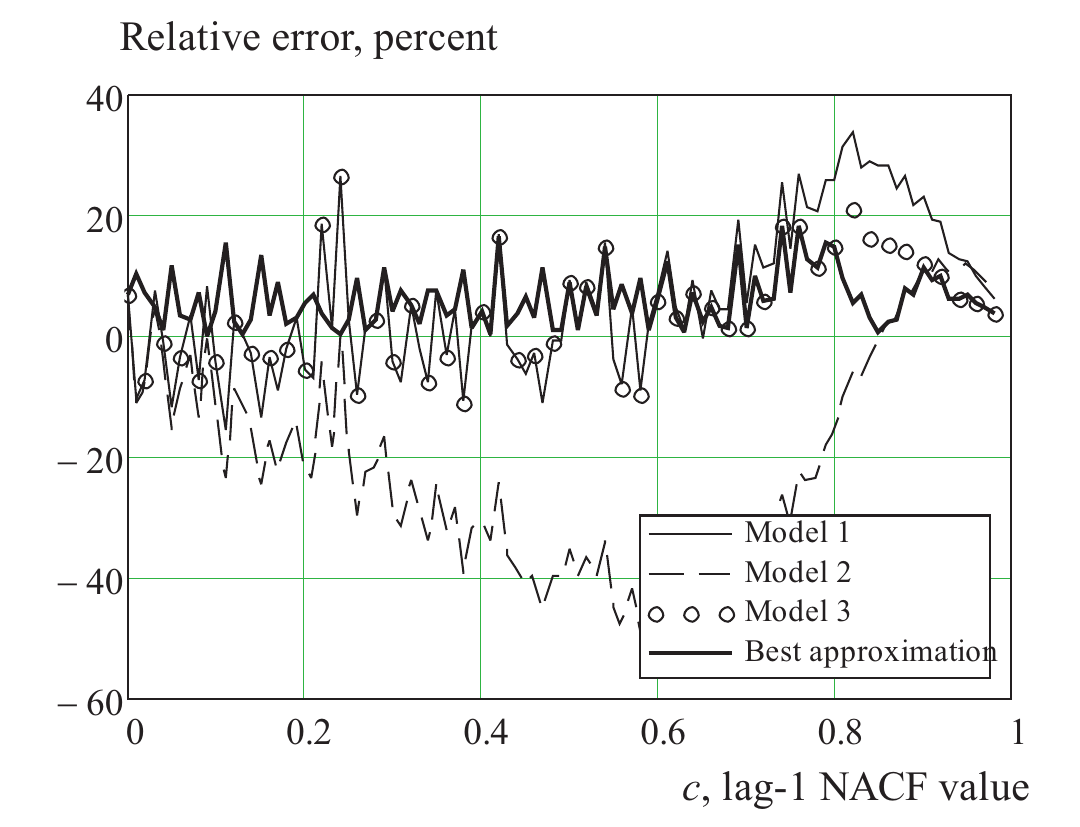}
    \label{fig:comp_corr_001_82l5}
  }
  \subfigure[$(I,v)=(16,1),l=5$]{
    \includegraphics[width=0.33\textwidth]{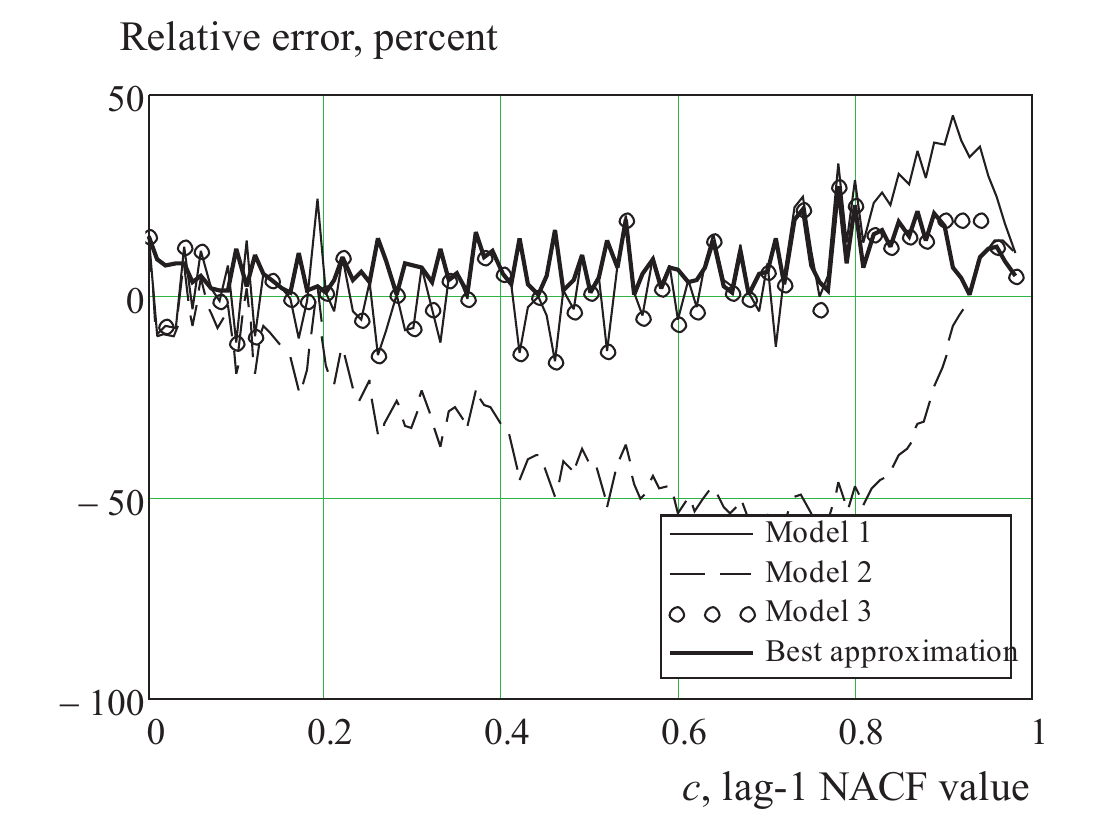}
    \label{fig:comp_corr_001_161l5}
  }
} \caption{Relative difference as a function of $c$ for $p_{E}=0.01$.}
  \label{fig:comp_corr_001}
\end{figure*}

%\subsubsection{Detailed dependence on correlation}

So far we observed that all the models are highly sensitive to the value of the lag-1
NACF and the choice of the best model depends on it. Here, we will study the effect of
correlation on accuracy of our models. The relative accuracy of all the models for
different values of $l$ and $(I,v)$ and fixed value of BER ($p=0.01$) is shown in Fig.
\ref{fig:comp_corr_001}. First of all, notice that the absolute error is not greater than $40\%$ in the worst possible case. Further, it is clear that model 2 differs from the other two. For small and medium values of $c$ this model underestimates the packet error probability. Moreover, its absolute deviation in most cases is greater compared to other models. On the other
hand, models 1 and 3 always overestimate the actual performance demonstrating
qualitatively similar behavior. The important difference is that for large values of $c$
($c>0.9$) the model 3 clearly outperforms the model 1. Recalling that both these models
are computationally intensive incorporating numerical estimation of two-dimensional
distribution we see that from the practical point of view model 3 is always preferable.
Finally, we would like to note that the accuracy of all the models for finite values of $c$
is almost a convex function of $c$ having a peak at a certain $c$. The value of $c$
maximizing the error is a function of the interleaving depth $I$ and is independent of
$l$. However, these values do not coincide for models 2 and 3. As one may observe the
accuracy of model 2 is always better when model 3 provides the worst possible prediction.
The best approximation out of these two models is shown by thick lines in Fig.
\ref{fig:comp_corr_001}.

%\subsubsection{Conclusions}

Concluding this section we claim that model 3 outperforms the other two
models for all the considered values of BER, $l$, and $c$ when accuracy is the only
metric of interest. For most input values the results of this model deviate by at most
$10\%$ and does not allow for extreme bias for small values of BER. However, if
computational complexity comes into play, for all values of input parameters there is an
alternative model (either model 1 or model 2) performing as good as model 3, while
requiring way less computational efforts. For extremely high values of lag-1 NACF model
and high values of $l$ the model 2 performs comparably to the model 3. In other cases
model 1 provides provides similar performance.

\begin{figure*}[t!]
\centerline{
  \subfigure[$l=1,c=0.3$]{
    \includegraphics[width=0.33\textwidth]{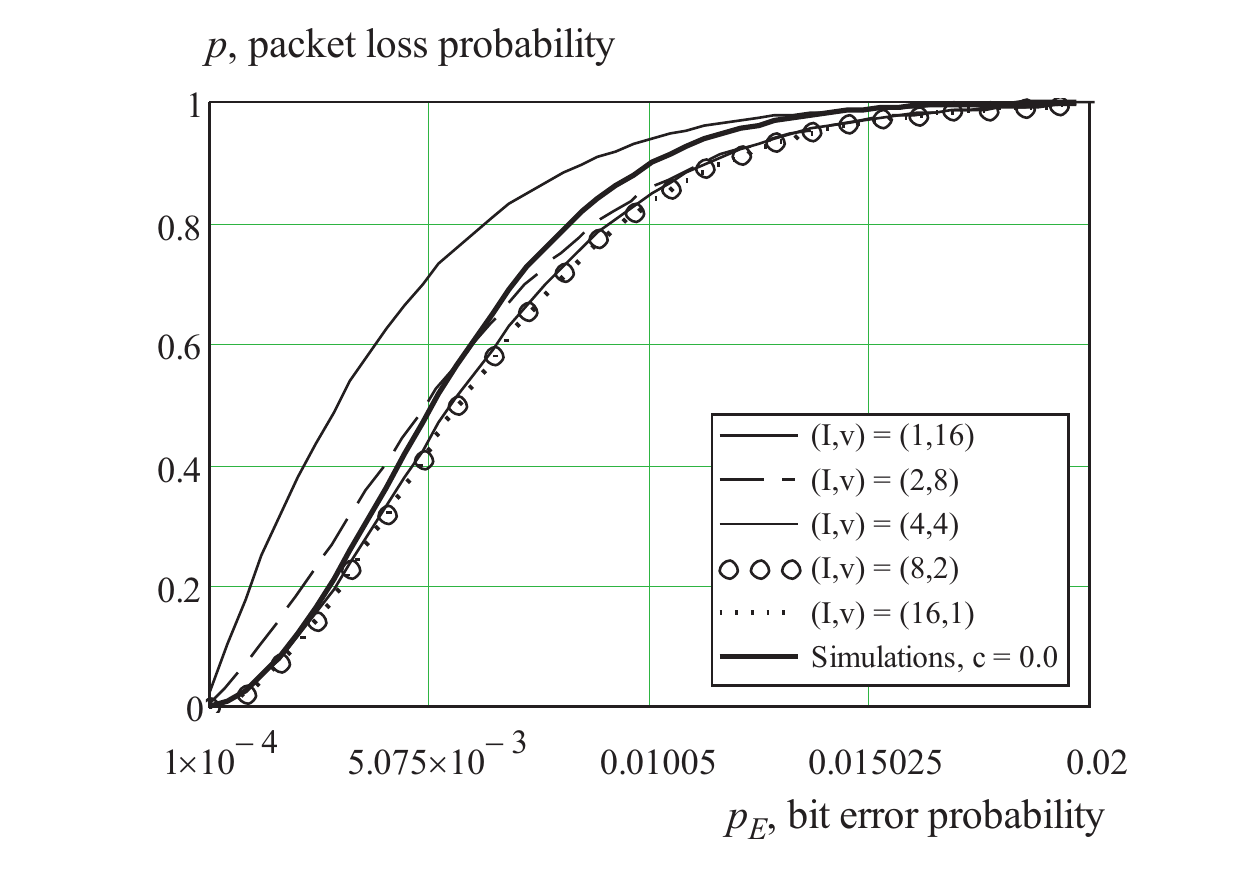}
    \label{fig:eff_iv_lc_103}
  }
  \subfigure[$l=1,c=0.6$]{
    \includegraphics[width=0.33\textwidth]{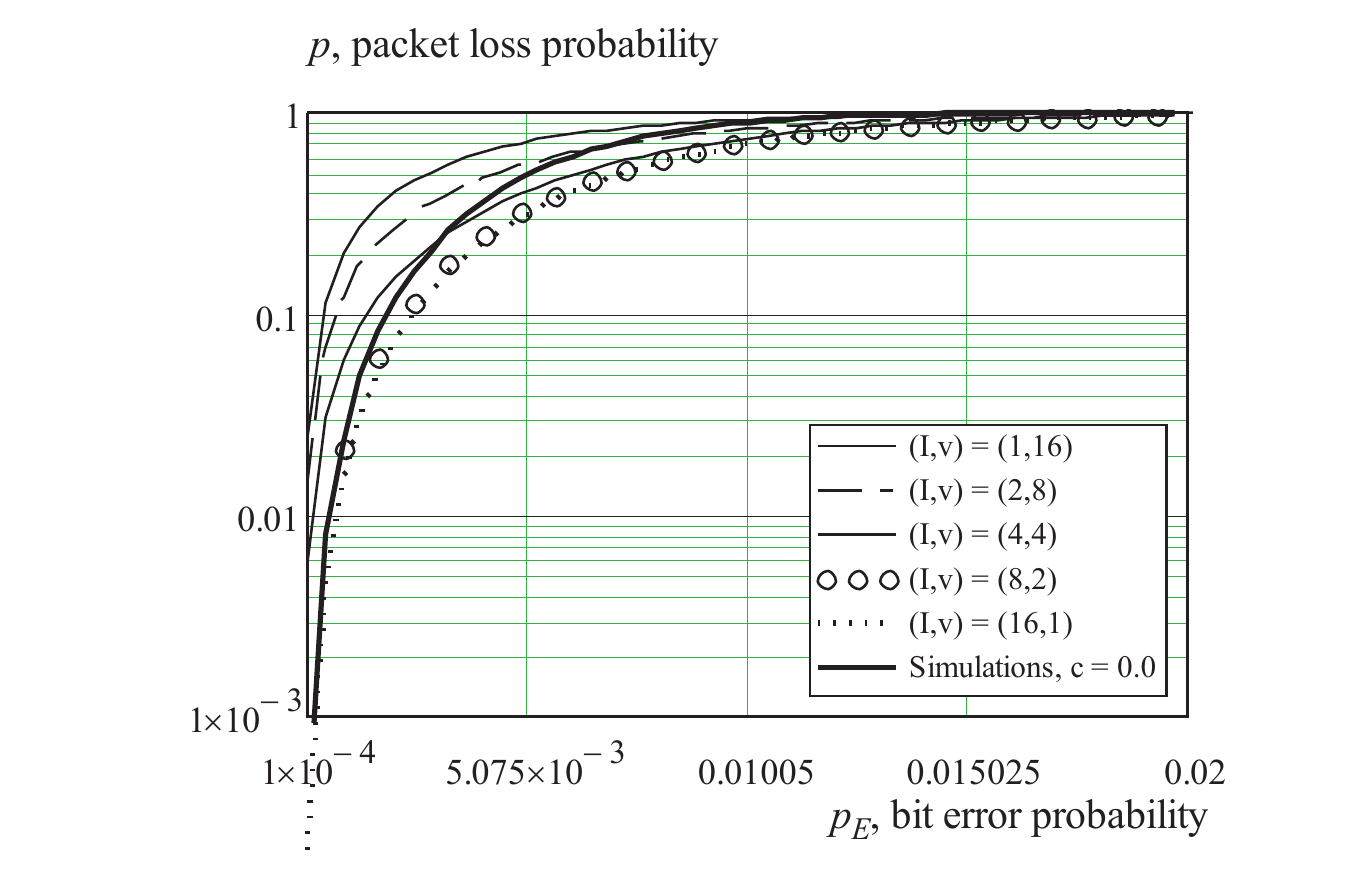}
    \label{fig:eff_iv_lc_106}
  }
  \subfigure[$l=1,c=0.9$]{
    \includegraphics[width=0.33\textwidth]{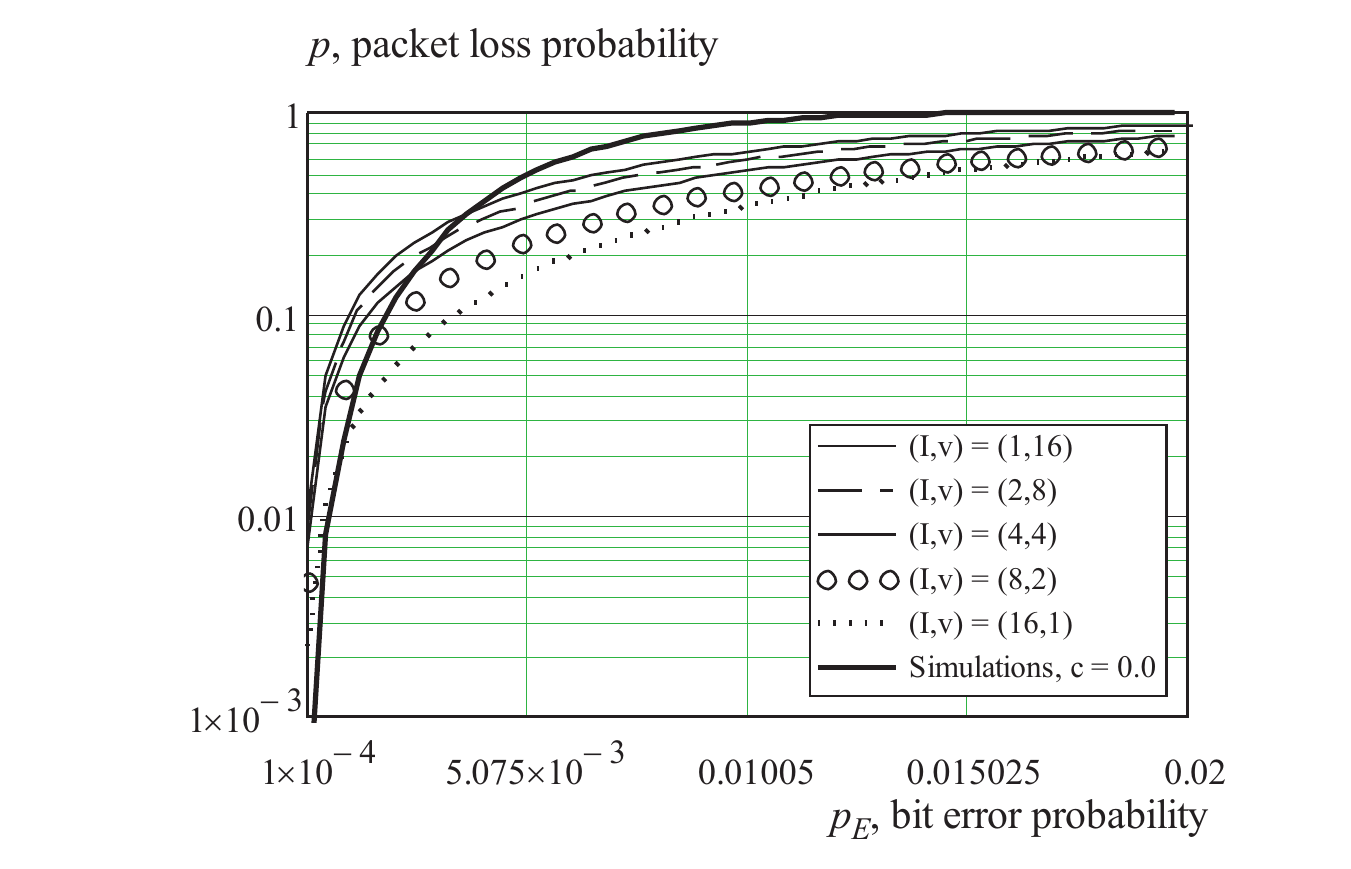}
    \label{fig:eff_iv_lc_109}
  }
} \centerline{
  \subfigure[$l=3,c=0.3$]{
    \includegraphics[width=0.33\textwidth]{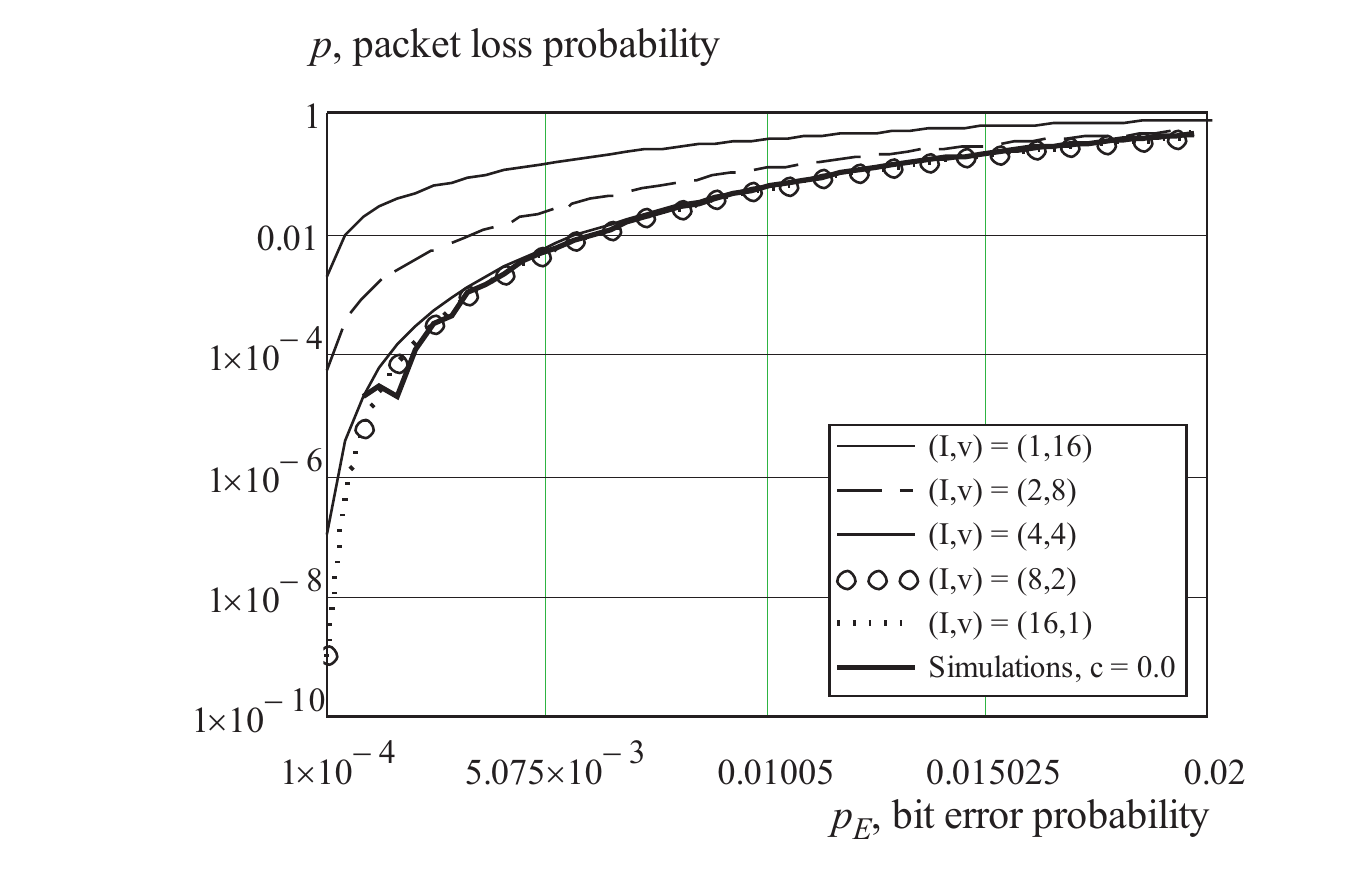}
    \label{fig:eff_iv_lc_303}
  }
  \subfigure[$l=3,c=0.6$]{
    \includegraphics[width=0.33\textwidth]{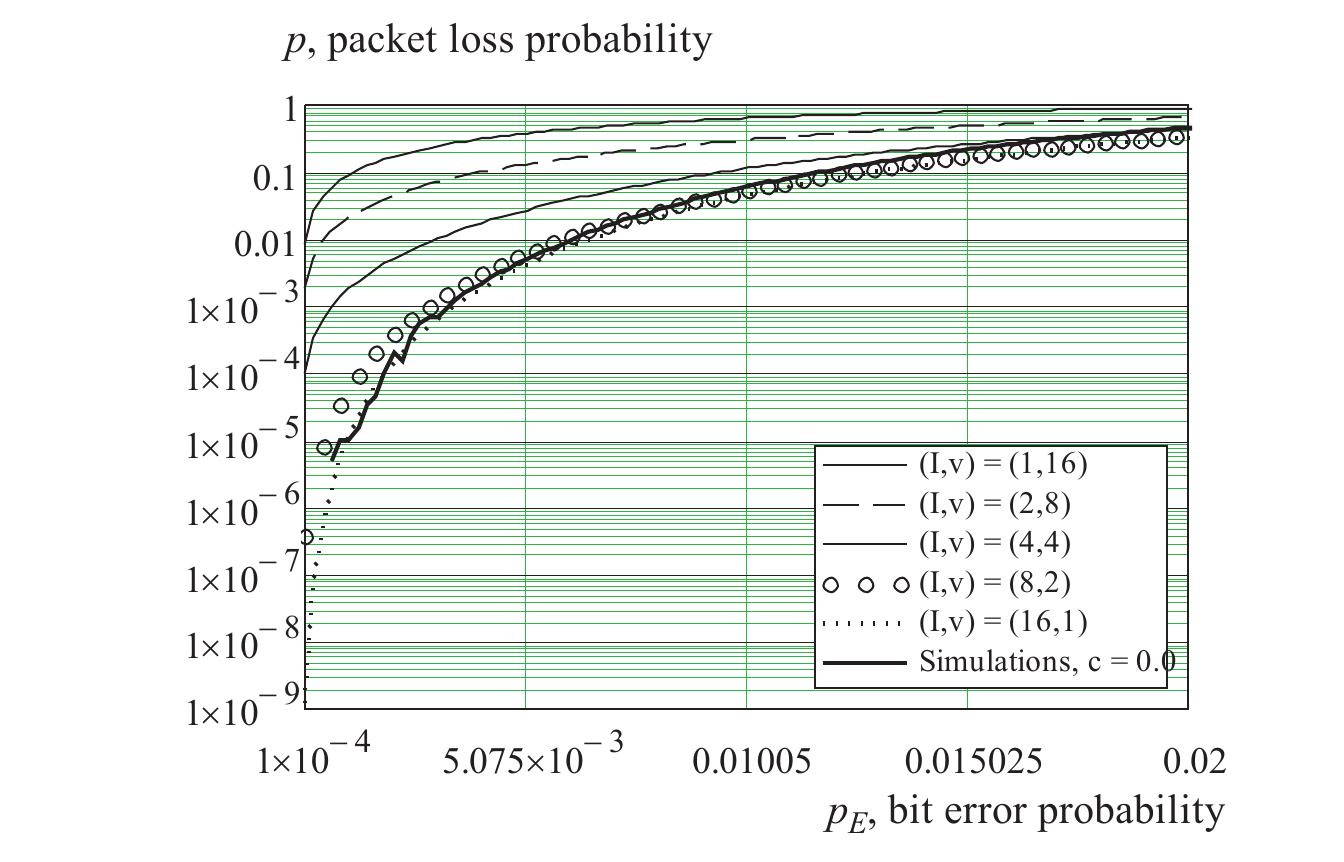}
    \label{fig:eff_iv_lc_306}
  }
  \subfigure[$l=3,c=0.9$]{
    \includegraphics[width=0.33\textwidth]{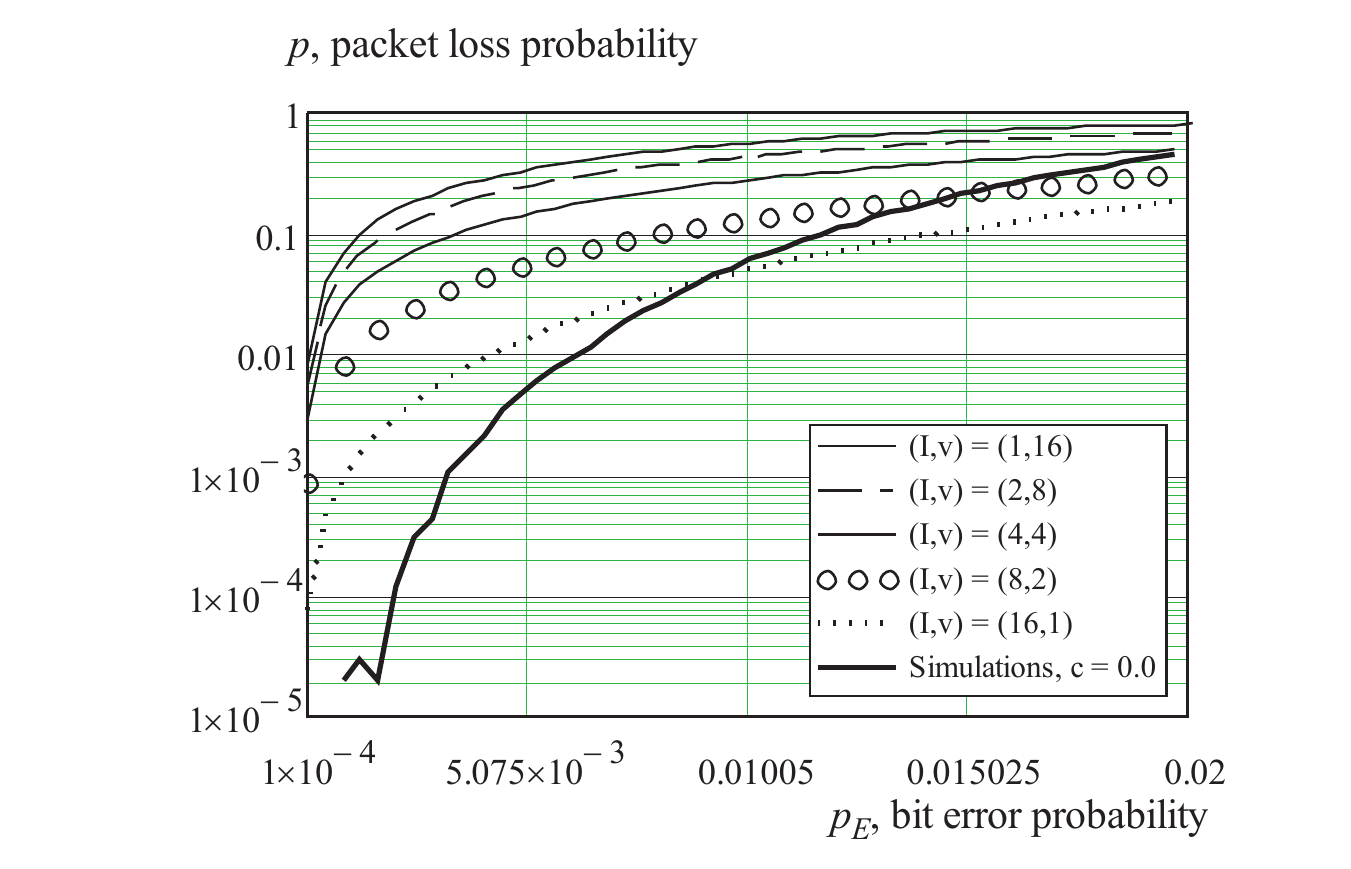}
    \label{fig:eff_iv_lc_309}
  }
} \caption{The effect of interleaving depth $I$.}
  \label{fig:eff_iv_lc}
\end{figure*}

\section{Numerical study of interleaving}\label{sect_05}

In this section we study the effect of interleaving on the packet loss performance of a
channel. In what follows, all the presented results are based on model 3 that is found to
provide the most accurate results across all ranges of BER, $c$, and $l$. We are
interested in both qualitative and quantitative effects.

\subsection{The effect of interleaving}

First, consider the effect of interleaving qualitatively, i.e. studying what
happens when we increase $I$ from $1$ to $16$ keeping the packet size constant. Fig.
\ref{fig:eff_iv_lc} demonstrates the packet error probability for all the interleaving
depths and several values of $c$ and $l$. We limit presentation to only two
values of $l$ as results for higher $l$ are similar. The best performance in terms of the
packet error probability is provided by the maximum possible value of the interleaving
depth; in our case this is $(16,1)$ scheme. As the correlation increases the difference
between the interleaving schemes grows. The reason is that the residual correlation left
between successive bits in a codeword $(R=c^{I})$ is higher for bigger values of $c$ and
it becomes tougher to conceal these errors due to their grouping in a single codeword.
For example, for $c=0.3$, $l=1$, $(8,2)$, interleaving scheme performs as good as
$(16,1)$ scheme. The reason is that the residual correlation is negligible for $I=8$ and
$I=16$ ($0.3^{8}\approx{}6.5E-5$ and $0.3^{16}\approx{}4.3E-9$). For $c=0.6$ the
difference between these two schemes is also almost unnoticeable as the residual
correlations are still fairly close to zero, $0.6^{8}\approx{}0.02$,
$0.6^{16}\approx{}2.8E-4$. For $c=0.9$ it the difference is noticeable as
$0.9^{8}\approx{}0.43$ which is significantly bigger than $0.9^{16}\approx{}0.18$.
Similar behavior is observed for $l=3$. However, as the correction capability increases
the difference becomes bigger.

The aim of interleaving is to remove the autocorrelation from the
bit error process of a wireless channel. We clearly see it observing the results provided
in Fig. \ref{fig:eff_iv_lc}. However, we also see that even high values of $I$ cannot make
the channel completely uncorrelated. This is best exemplified observing Fig.
\ref{fig:eff_iv_lc_109} and Fig. \ref{fig:eff_iv_lc_309} showing the packet loss
probability for $c=0.9$. We see that for small values of BER completely uncorrelated
channel outperforms a correlated one even for $I=16$. The reason is again the residual
correlation which is non-negligible even for $I=16$, $0.9^{16}\approx{}0.18$. The
difference is as large as one order of magnitude, see Fig. \ref{fig:eff_iv_lc_309}.

\begin{figure*}[t!]
\centerline{
  \subfigure[$l=1,p=0.005$]{
    \includegraphics[width=0.33\textwidth]{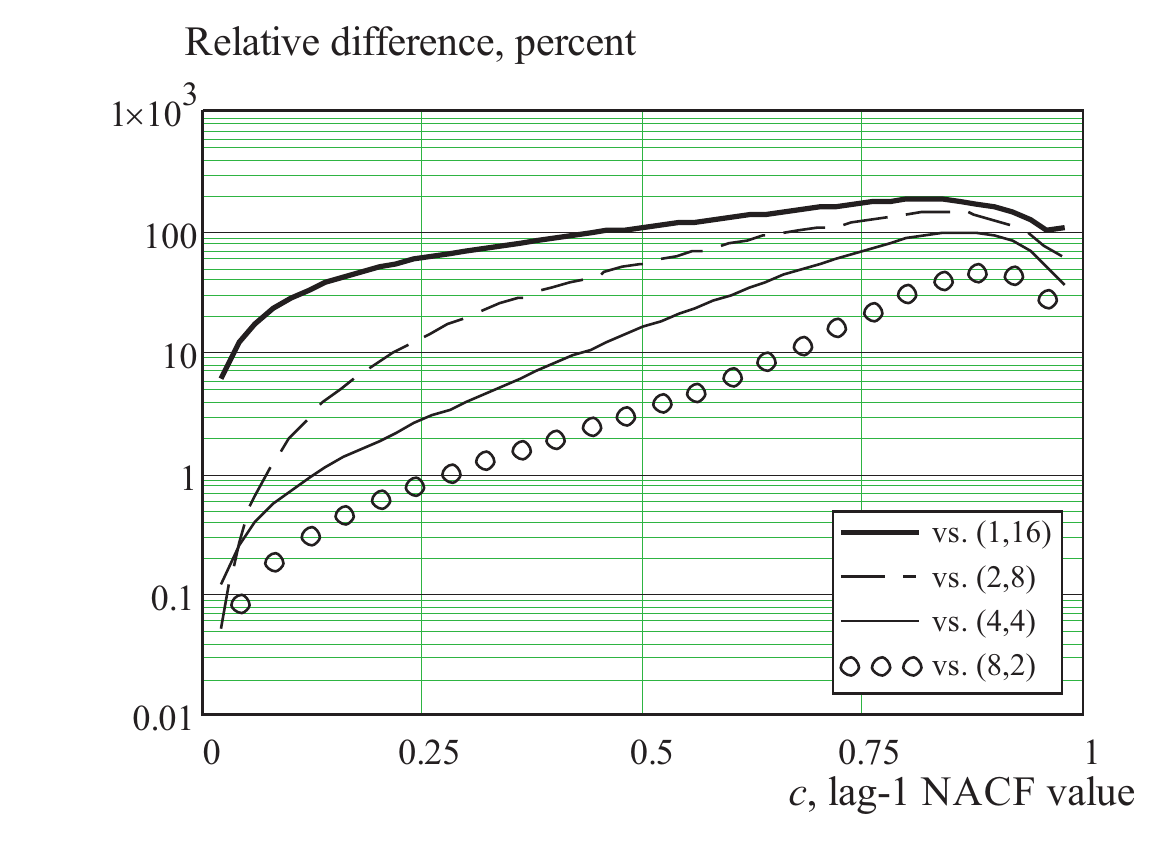}
    \label{fig:eff_iv_lc_30005}
  }
  \subfigure[$l=1,p=0.010$]{
    \includegraphics[width=0.33\textwidth]{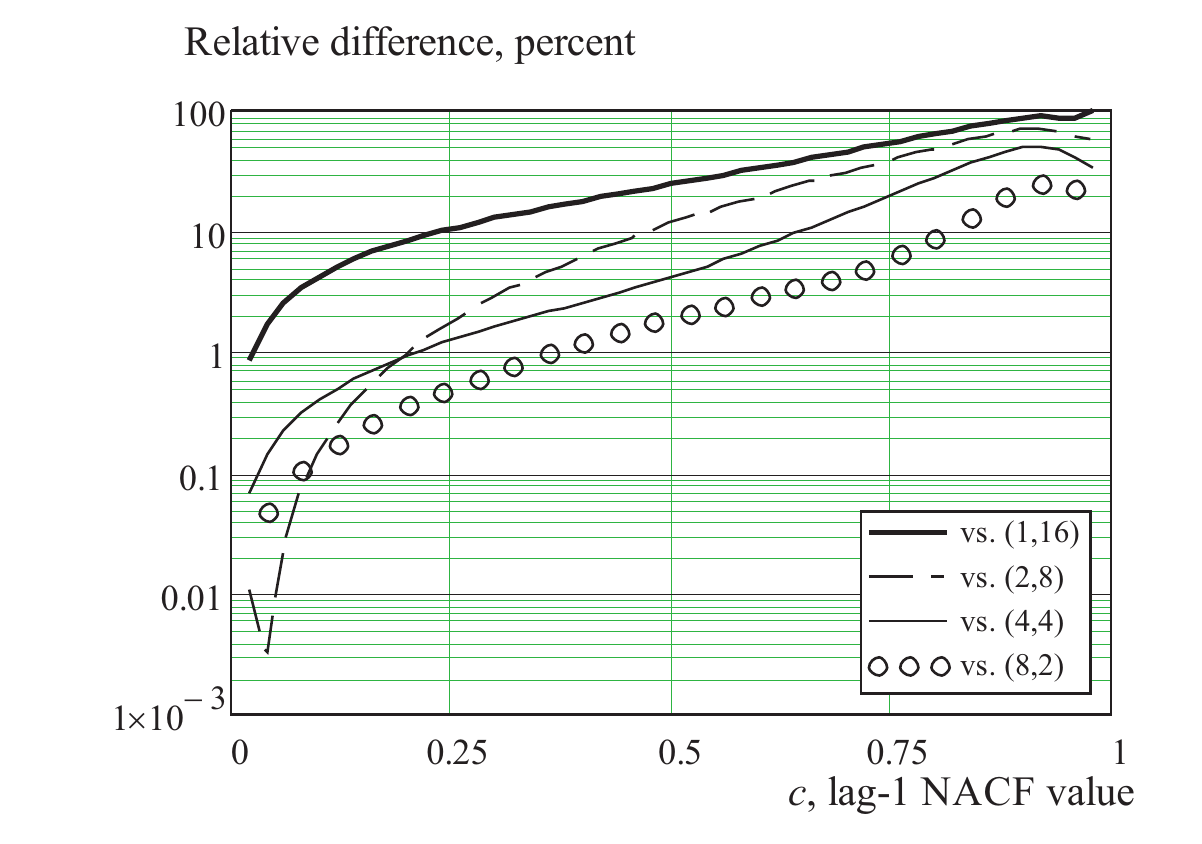}
    \label{fig:eff_iv_lc_30010}
  }
  \subfigure[$l=1,p=0.015$]{
    \includegraphics[width=0.33\textwidth]{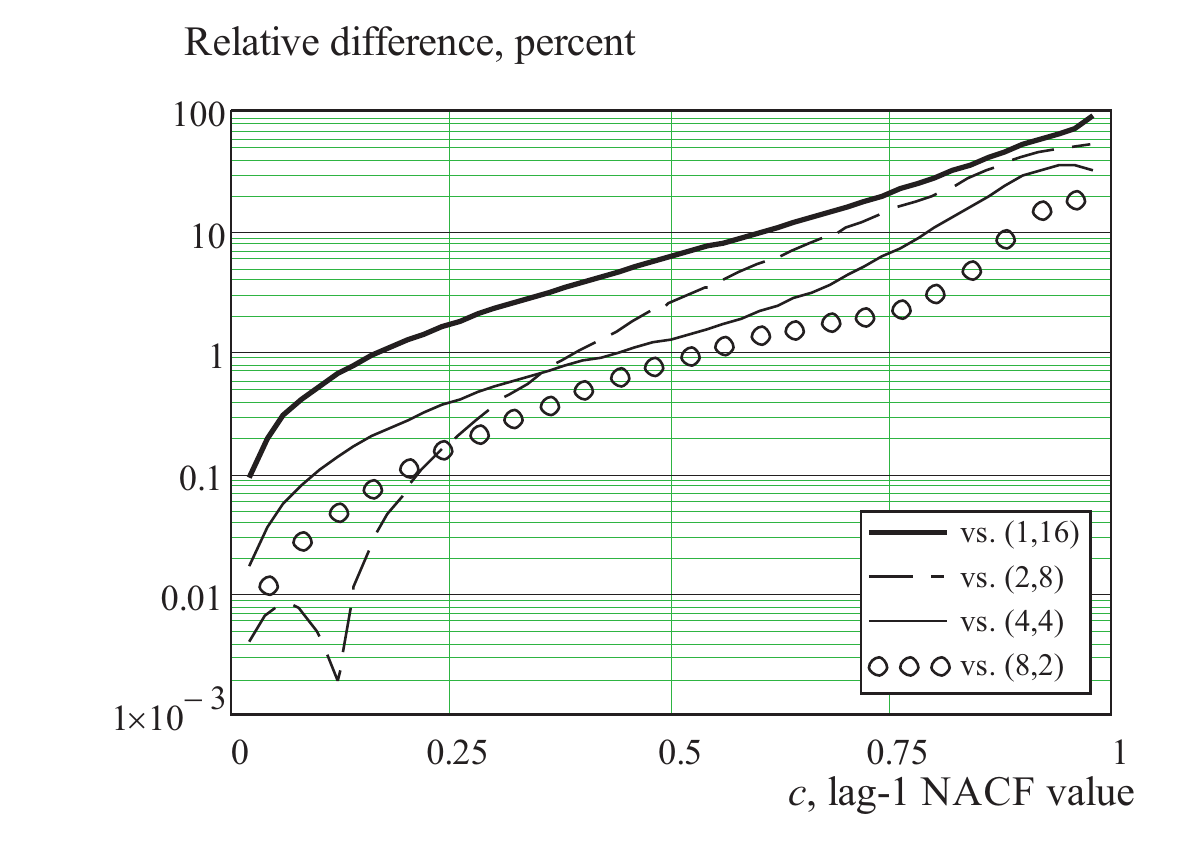}
    \label{fig:eff_iv_lc_30015}
  }
} \centerline{
  \subfigure[$l=3,p=0.005$]{
    \includegraphics[width=0.33\textwidth]{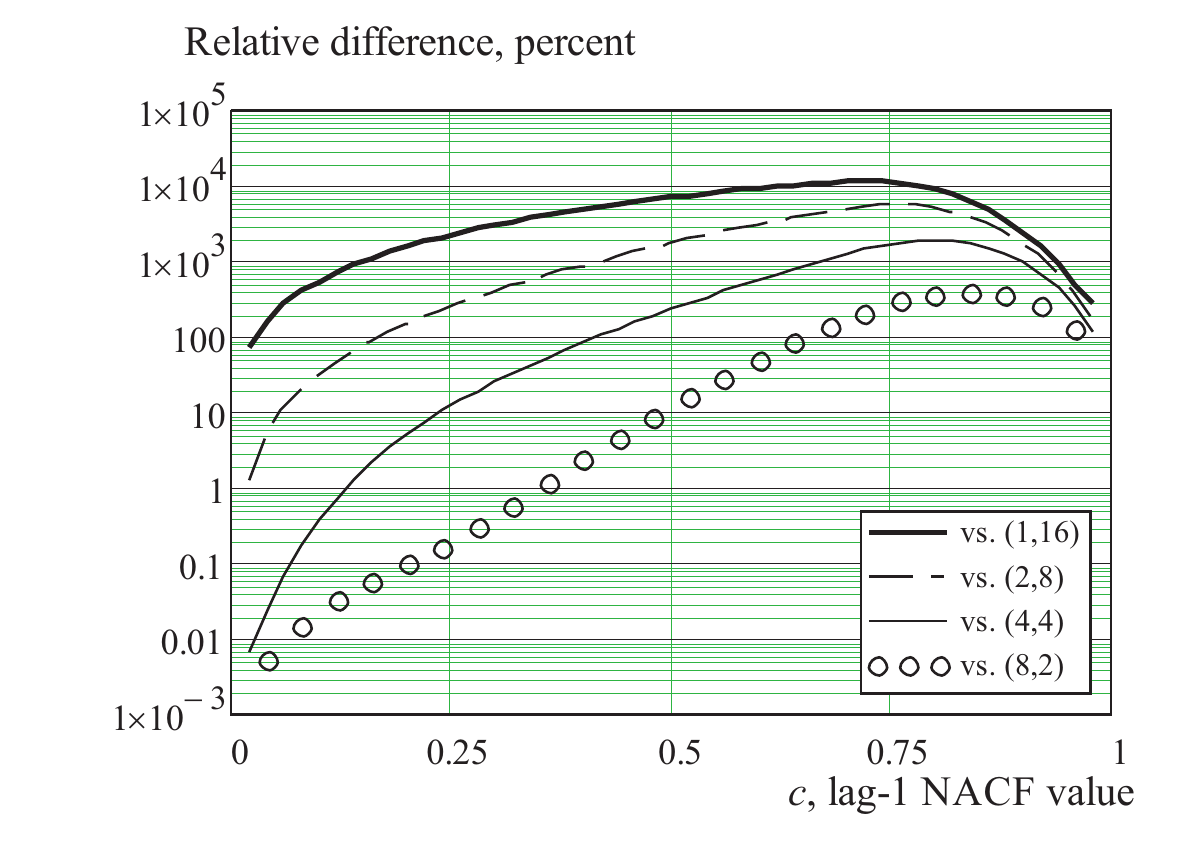}
    \label{fig:eff_iv_lc_30005}
  }
  \subfigure[$l=3,p=0.010$]{
    \includegraphics[width=0.33\textwidth]{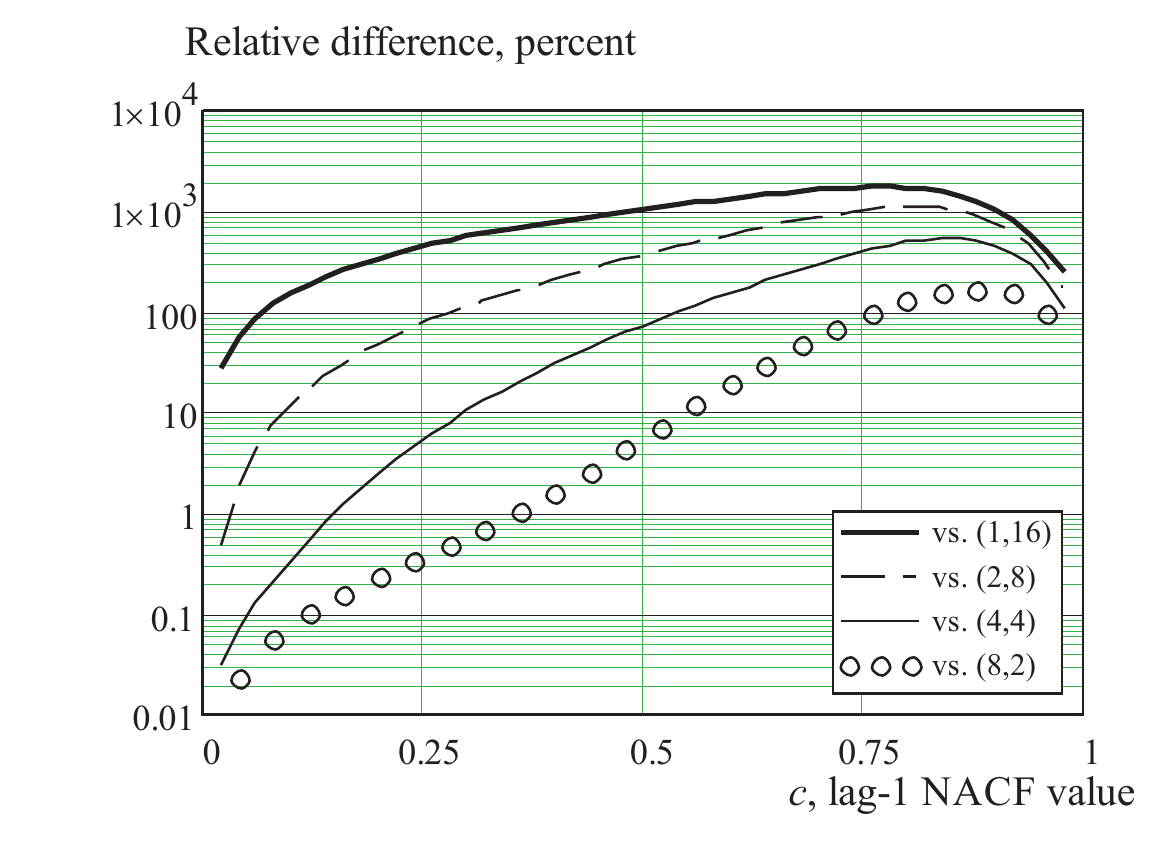}
    \label{fig:eff_iv_lc_30010}
  }
  \subfigure[$l=3,p=0.015$]{
    \includegraphics[width=0.33\textwidth]{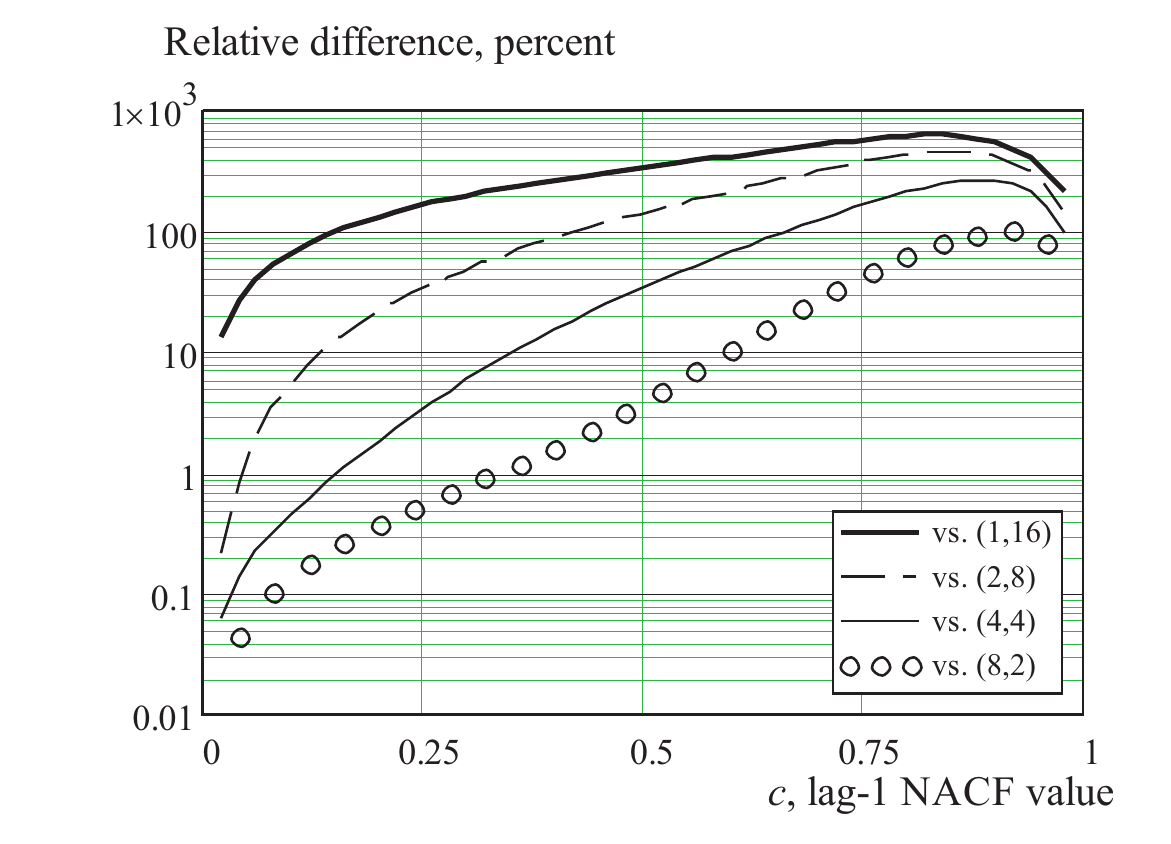}
    \label{fig:eff_iv_lc_30015}
  }
} \centerline{
  \subfigure[$l=5,p=0.005$]{
    \includegraphics[width=0.33\textwidth]{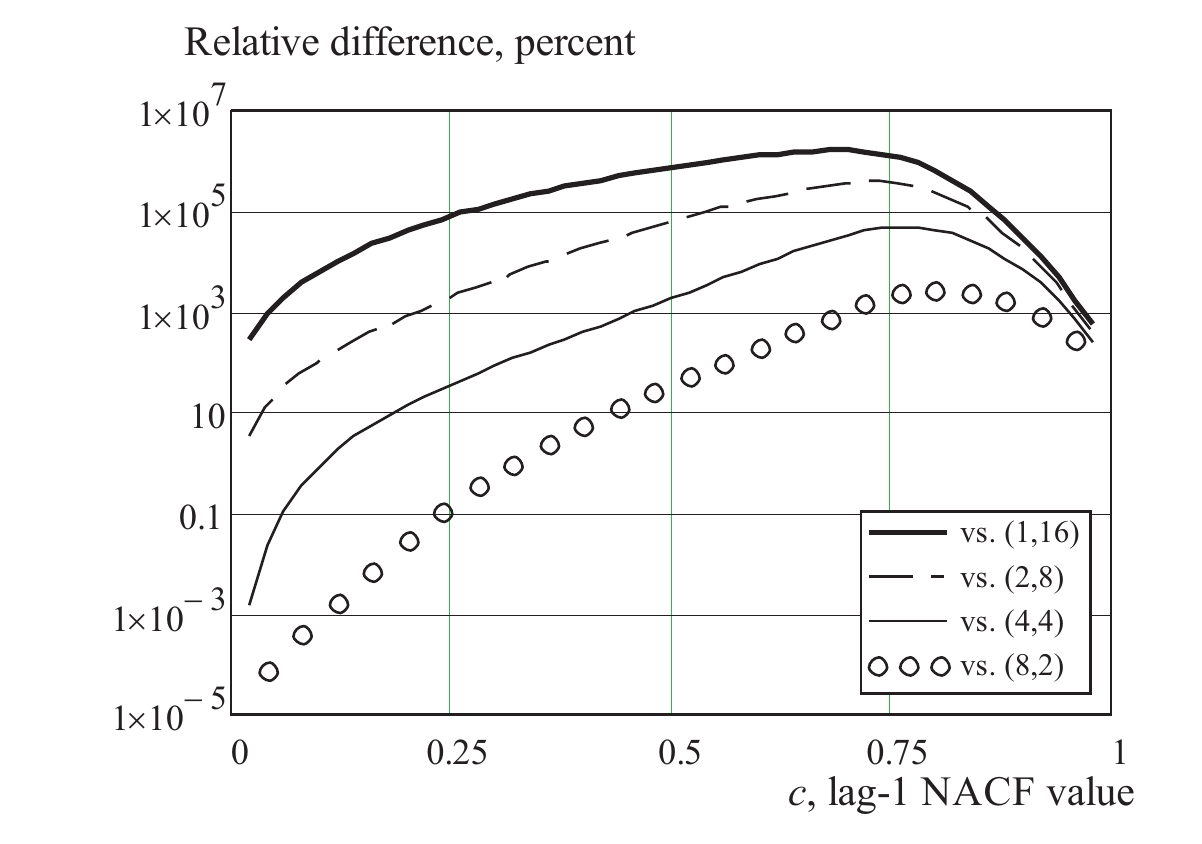}
    \label{fig:eff_iv_lc_50005}
  }
  \subfigure[$l=5,p=0.010$]{
    \includegraphics[width=0.33\textwidth]{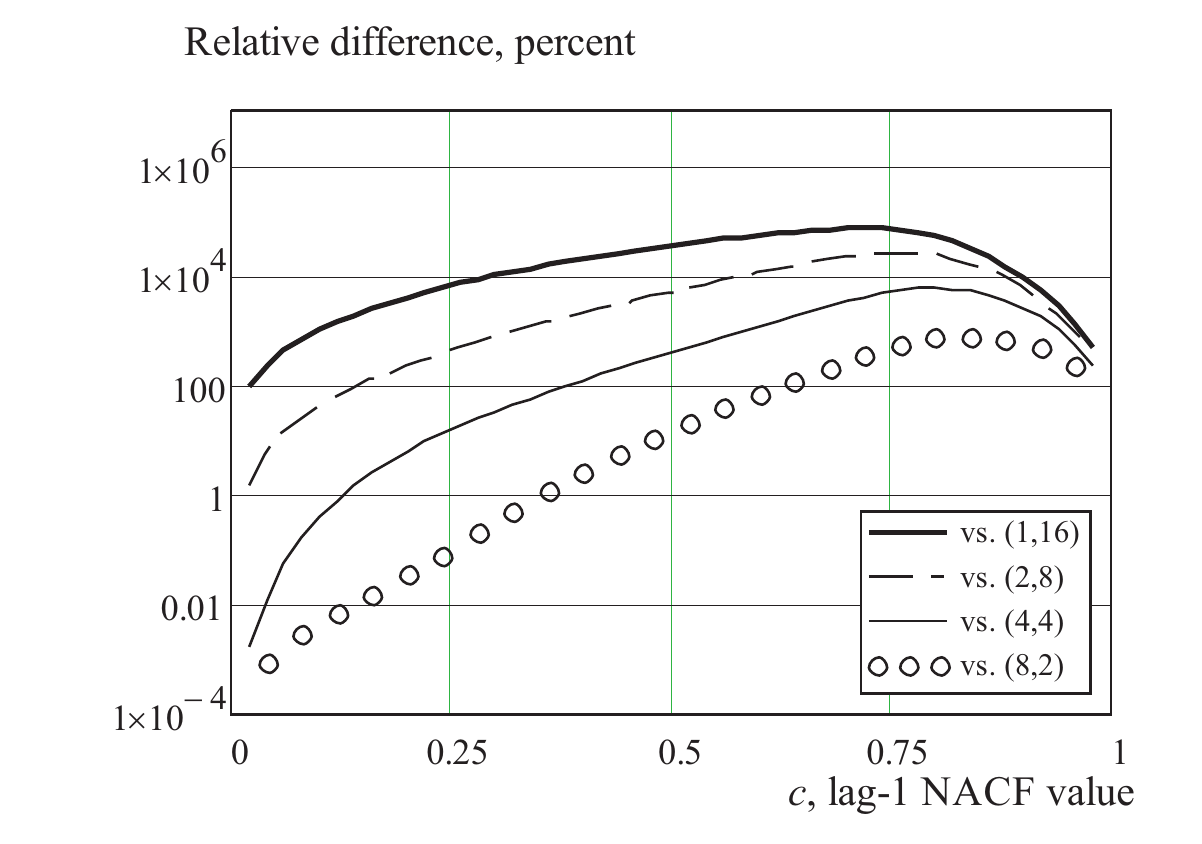}
    \label{fig:eff_iv_lc_50010}
  }
  \subfigure[$l=5,p=0.015$]{
    \includegraphics[width=0.33\textwidth]{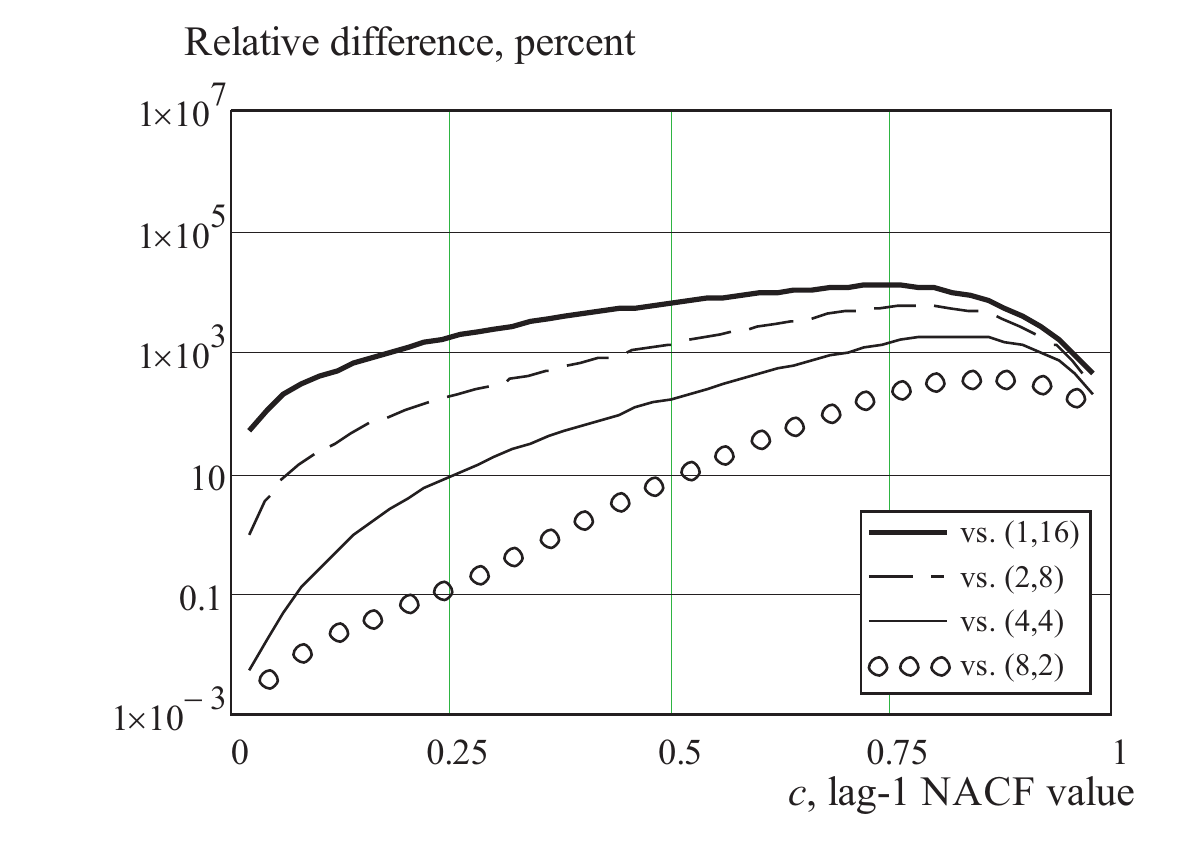}
    \label{fig:eff_iv_lc_50015}
  }
} \caption{Relative difference between $(I,v)$ and $(16,1)$.}
  \label{fig:eff_iv_lc_corr}
\end{figure*}

For large values of BER the correlated channel with $I=16$ is better than that with
$c=0.0$. The correlated channel also performs better in the region of high BER for other
values of $I$ as well. The abovementioned behavior is related to the way how correlation is removed by interleaving or, precisely, to the insufficient interleaving depth. The reason
is that the correlation manifests itself not only in grouping of bit errors but in
grouping of correctly received bits as well. In fact, for large values of $c$, $I=16$ is
insufficient to completely remove the correlation between successive bits of a single
codeword. Since $0.9^{16}\approx{}0.18$, bit errors in a codeword still tend to group
rather that be completely uncorrelated. Thus, for small values of BER the probability
that there will be more than $l$ errors in a codeword is bigger compared to the
completely uncorrelated case. On the other hand, when BER is high the correction
capability of a code may not be sufficient to correct all the errors in completely
uncorrelated case. However, when memory is high these errors tend to group in a single
codeword implying that correctly received bits also happen in batches. Since it does not
matter how many bits are received in error as long as the codeword is incorrectly
received, correlated channel with high lag-1 NACF performs better for large values of
BER. When $l$ increases this special effect fades away. In practice, it rarely happens
that NACF is strictly exponential, thus, allowing us to use simple approximation for the
residual correlation $R=c^{I}$. If the NACF decays slowly than that the residual
correlation is bigger and the correlated channel starts to perform better than completely
uncorrelated one sooner compared to what is shown in Fig. \ref{fig:eff_iv_lc}. This
effect is of theoretical interest as the absolute values of the packet error probability
in this regime are unacceptable.

Note that we cannot increase the interleaving depth arbitrarily for a given packet size
and a codeword length. For our choice of $n$ and $v$ we are limited to $I=16$. However, choosing shorter codewords with comparable code rate $k/n$ we may increase the interleaving depth and get closer to completely uncorrelated channel behavior. For
example, choosing a BCH code out of $n=31$ family we may increase the maximum
interleaving to approximately $32$ instead of $16$. Applying the rule of thumb $R=c^{I}$ we see that the residual correlation between successive bits of the same codeword will be approximately $0.9^{32}=0.034$ implying that the channel becomes almost uncorrelated and approximately five times less correlated ($0.18/0.034\approx{}5$) compared to $n=63$ and $I=16$.

\begin{figure*}[t!]
\centerline{
  \subfigure[$(I,v)=(2,8),c=0.6$]{
    \includegraphics[width=0.33\textwidth]{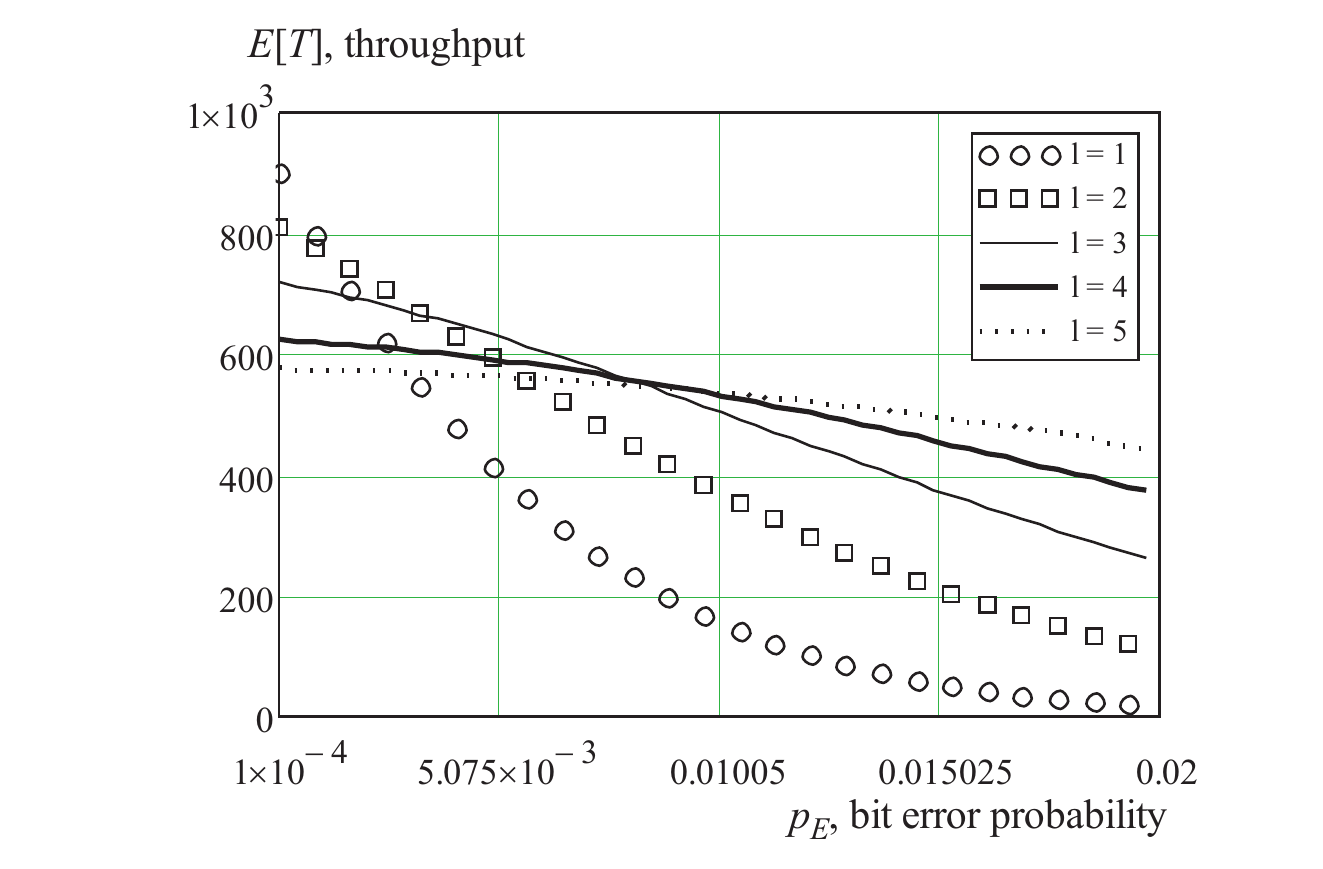}
    \label{fig:eff_l_ber_2806}
  }
  \subfigure[$(I,v)=(4,4),c=0.6$]{
    \includegraphics[width=0.33\textwidth]{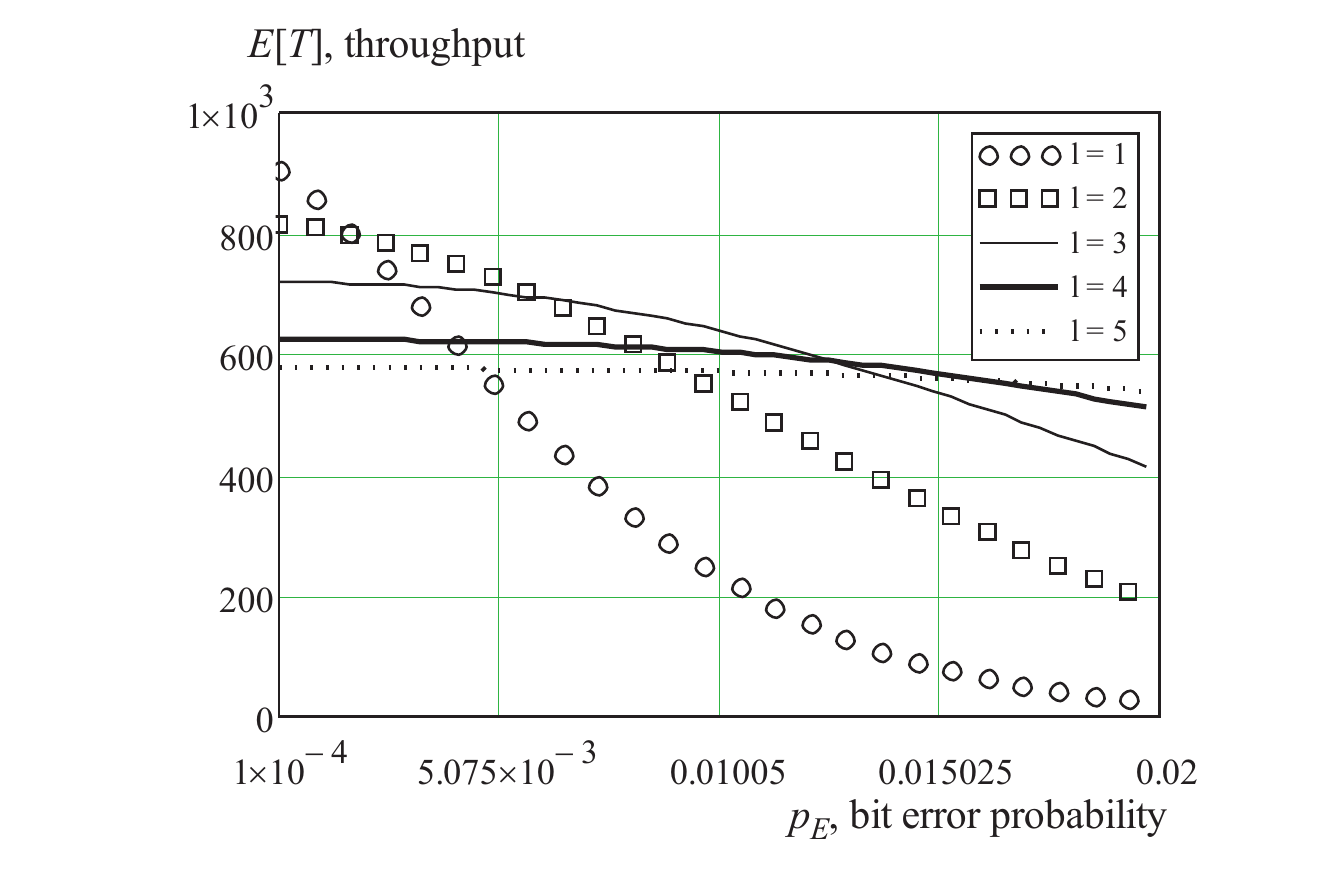}
    \label{fig:eff_l_ber_4406}
  }
  \subfigure[$(I,v)=(16,1),c=0.6$]{
    \includegraphics[width=0.33\textwidth]{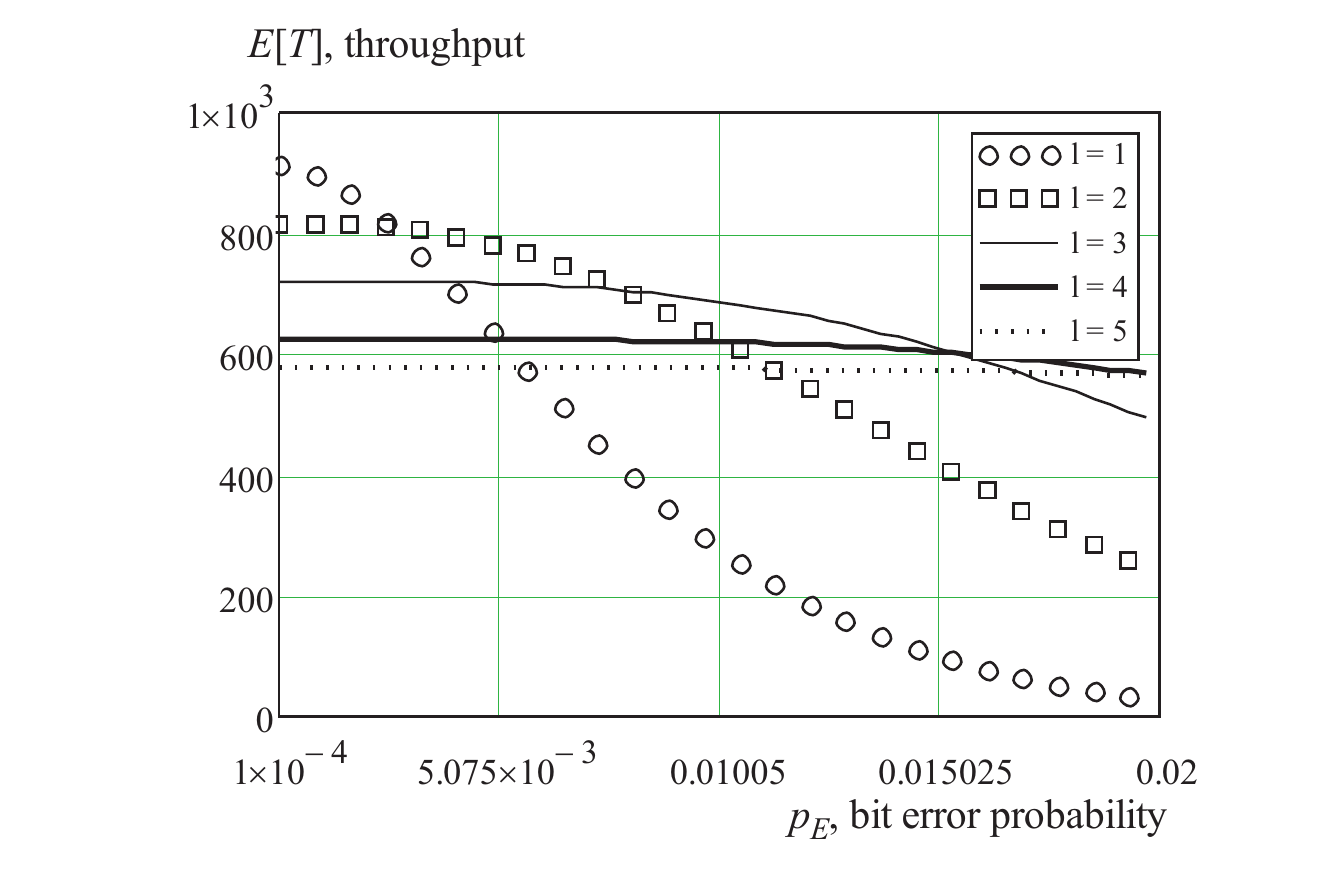}
    \label{fig:eff_l_ber_16106}
  }
} \centerline{
  \subfigure[$(I,v)=(2,8),c=0.9$]{
    \includegraphics[width=0.33\textwidth]{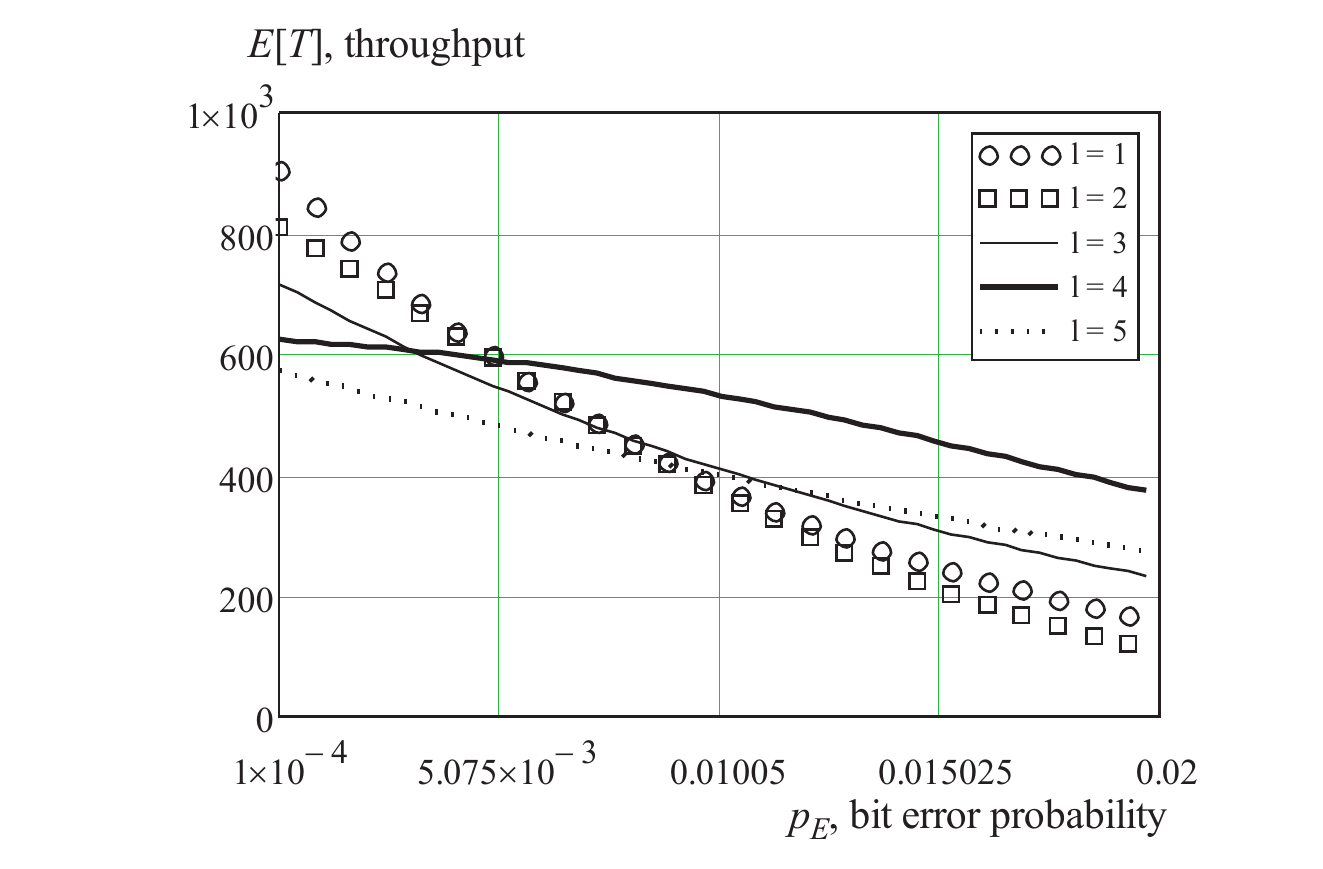}
    \label{fig:eff_l_ber_2809}
  }
  \subfigure[$(I,v)=(4,4),c=0.9$]{
    \includegraphics[width=0.33\textwidth]{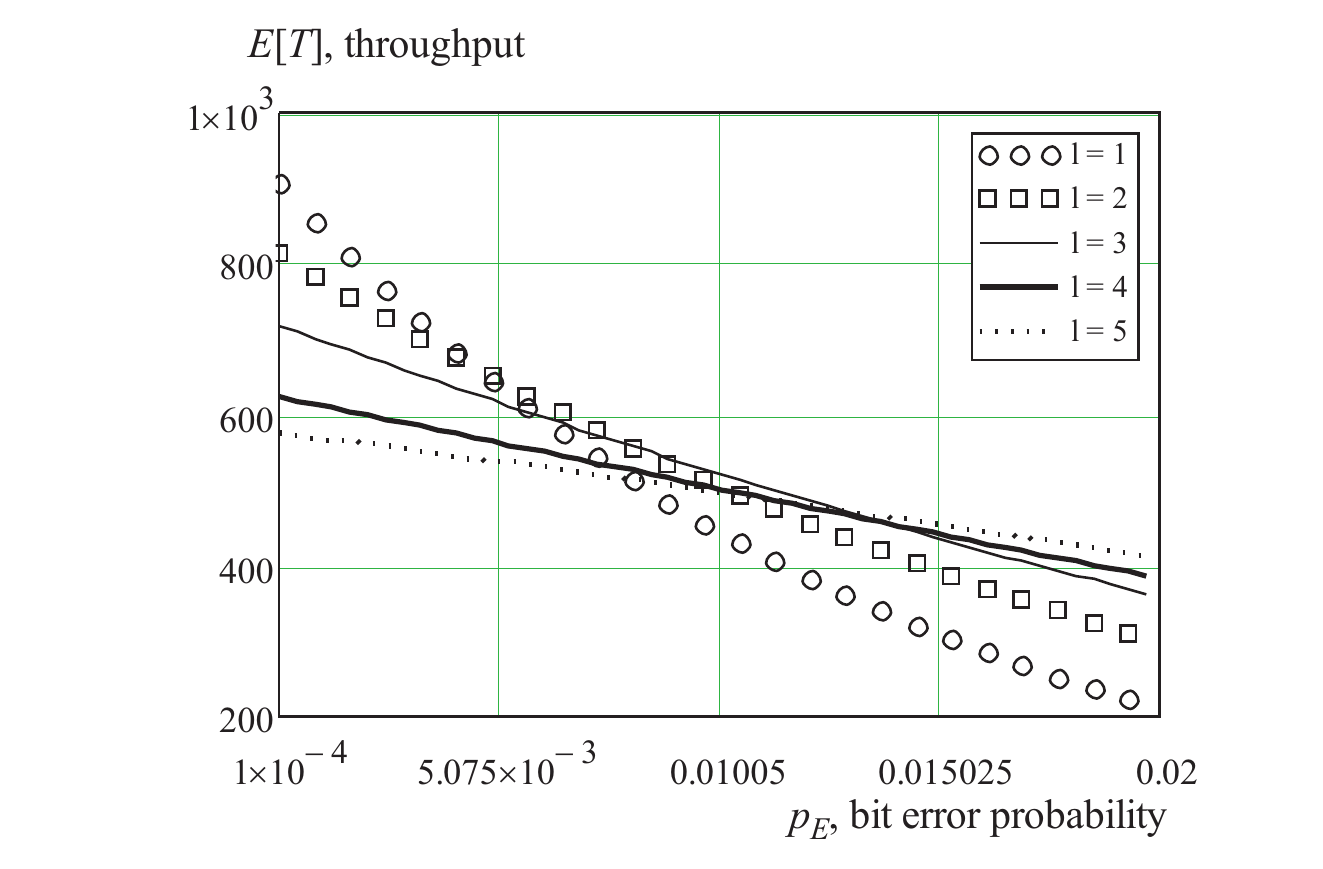}
    \label{fig:eff_l_ber_4409}
  }
  \subfigure[$(I,v)=(16,1),c=0.9$]{
    \includegraphics[width=0.33\textwidth]{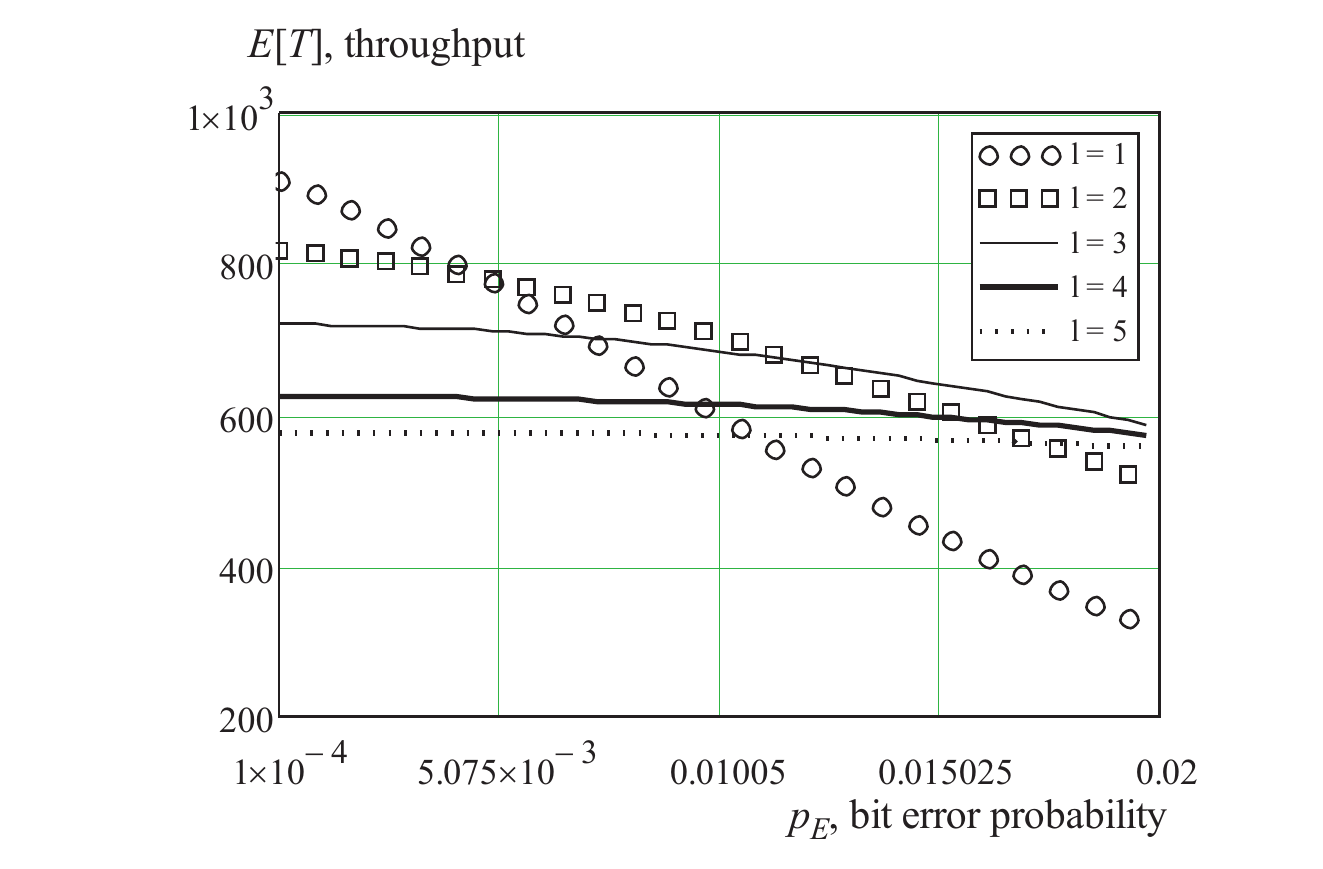}
    \label{fig:eff_l_ber_16109}
  }
} \caption{Throughput as a function of BER for different $l$ and $c$.}
  \label{fig:eff_l_ber}
\end{figure*}

\subsection{The detailed response to lag-1 NACF}

So far we considered the detailed response of our models to a wide range of BER values
for three values of lag-1 NACF. We see that $c$ and $l$ are the two factors affecting the
behavior of interleaving schemes $(I,v)$. In previous section we noticed that the
increase of the lag-1 NACF value leads to worse performance in terms of the packet loss
probability due to grouping of bit errors in codewords. However, this may or may not hold for extremely high values of $c$ as such behavior may actually lead to better performance due to extreme grouping of correctly received bits in the whole packet. Here, we will study the detailed response of a channel to lag-1 NACF.

The relative difference between different $(I,v)$ interleaving schemes and $(16,1)$
scheme as a function of $c$ is shown in Fig. \ref{fig:eff_iv_lc_corr}. The working
expression is $(a-b)/b$, where $a$ is the value corresponding to $(16,1)$ scheme. Simply put, these figures show by how many percents $(I,v)$ interleaving schemes are worse compared to the maximum possible interleaving depth $I=16$. As one may observe the maximum difference sometimes approaches four orders of magnitude ($l=5$, $p=0.005$). Another interesting observation is that going from $(1,16)$ through $(2,8)$, $(4,4)$, and $(8,2)$ to $(16,1)$ scheme increases performance by the same amount, i.e. increasing interleaving depth twice roughly doubles performance of a wireless channel. Further, we see that the gain is different for various values of $c$. The biggest gains are experienced at moderate values of $c$ and the gain decreases as lag-1 NACF values approaches $1$. It is understandable as for extremely high values of $c$ we either do not experience errors at all or they tend to happen inside a single frame eventually leading to the loss of the whole packet. It is also interesting to observe that the effect of correlation is significant even for very small values of $c$. All the performance curves demonstrated in Fig. \ref{fig:eff_iv_lc_corr} start in the same point corresponding to $c=0.0$ and then quickly start to deviate from each other. Already for $c=0.02$ they may deviate by as much as one order of magnitude. 

%This is an extremely important observation supporting wide usage of interleaving in modern networks. Indeed, other techniques implemented in modern wireless access technologies (i.e., RS codes) only partially remove correlation leaving some memory of the channel intact. Here, we see that even slight correlation may drastically change the channel response in terms of the packet loss probability.

\begin{figure*}[t!]
\centerline{
  \subfigure[$p$ for $l=1$ and $l=2$]{
    \includegraphics[width=0.5\textwidth]{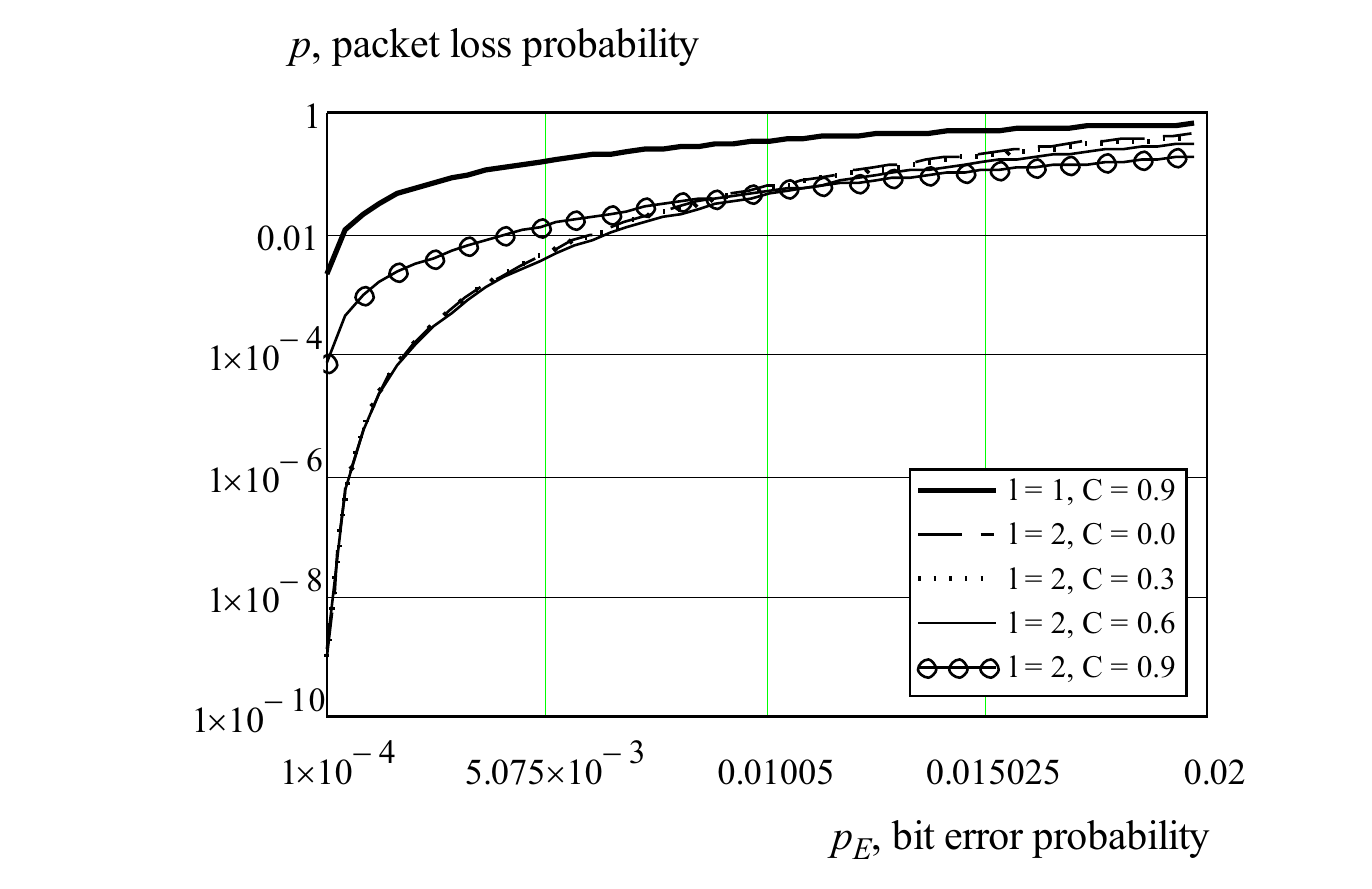}
    \label{fig:ex_1_plr}
  }
  \subfigure[Throughput for $l=1$ and $l=2$]{
    \includegraphics[width=0.5\textwidth]{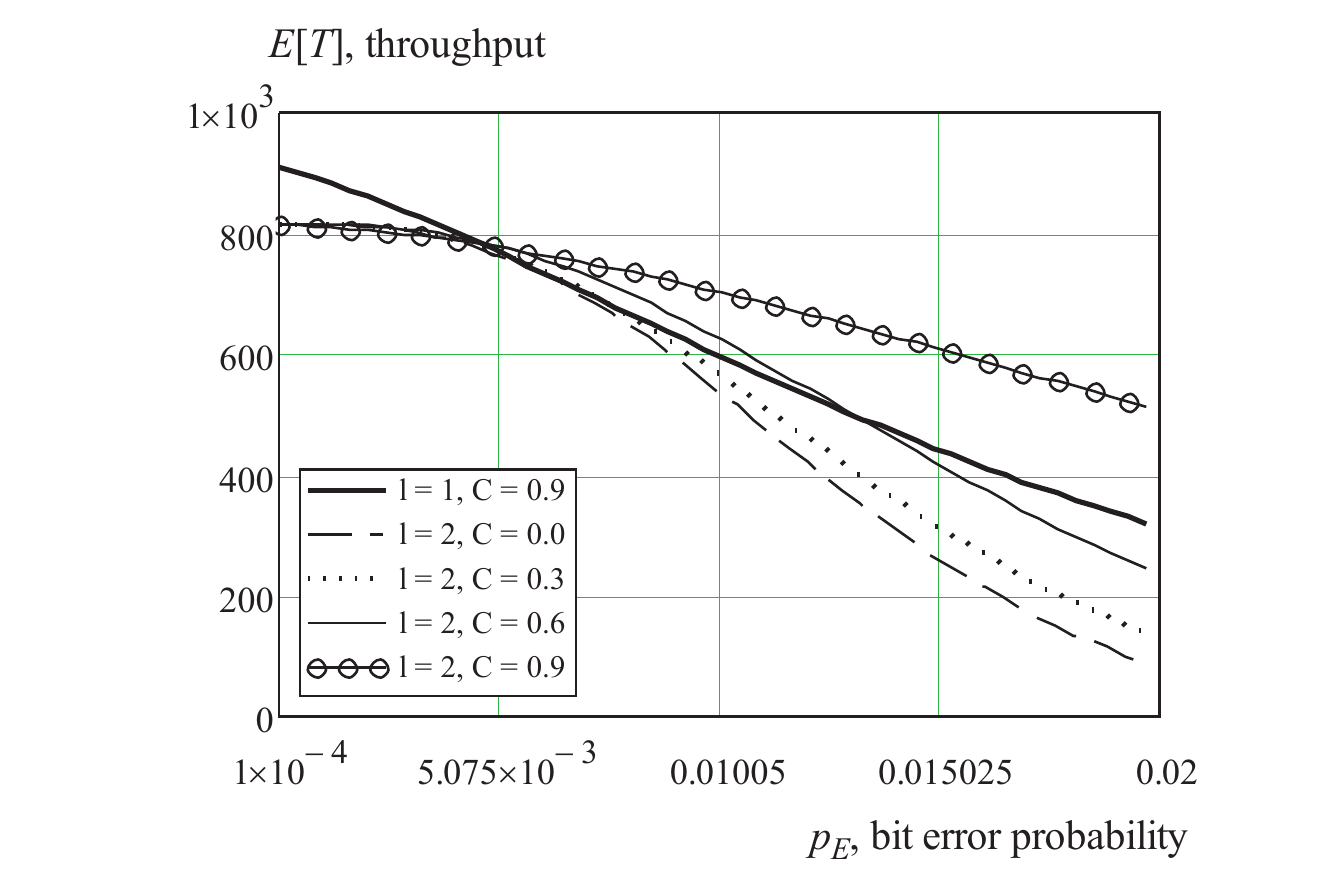}
    \label{fig:ex_1_thr}
  }
} \centerline{
  \subfigure[$p$ for $l=2$ and $l=3$]{
    \includegraphics[width=0.5\textwidth]{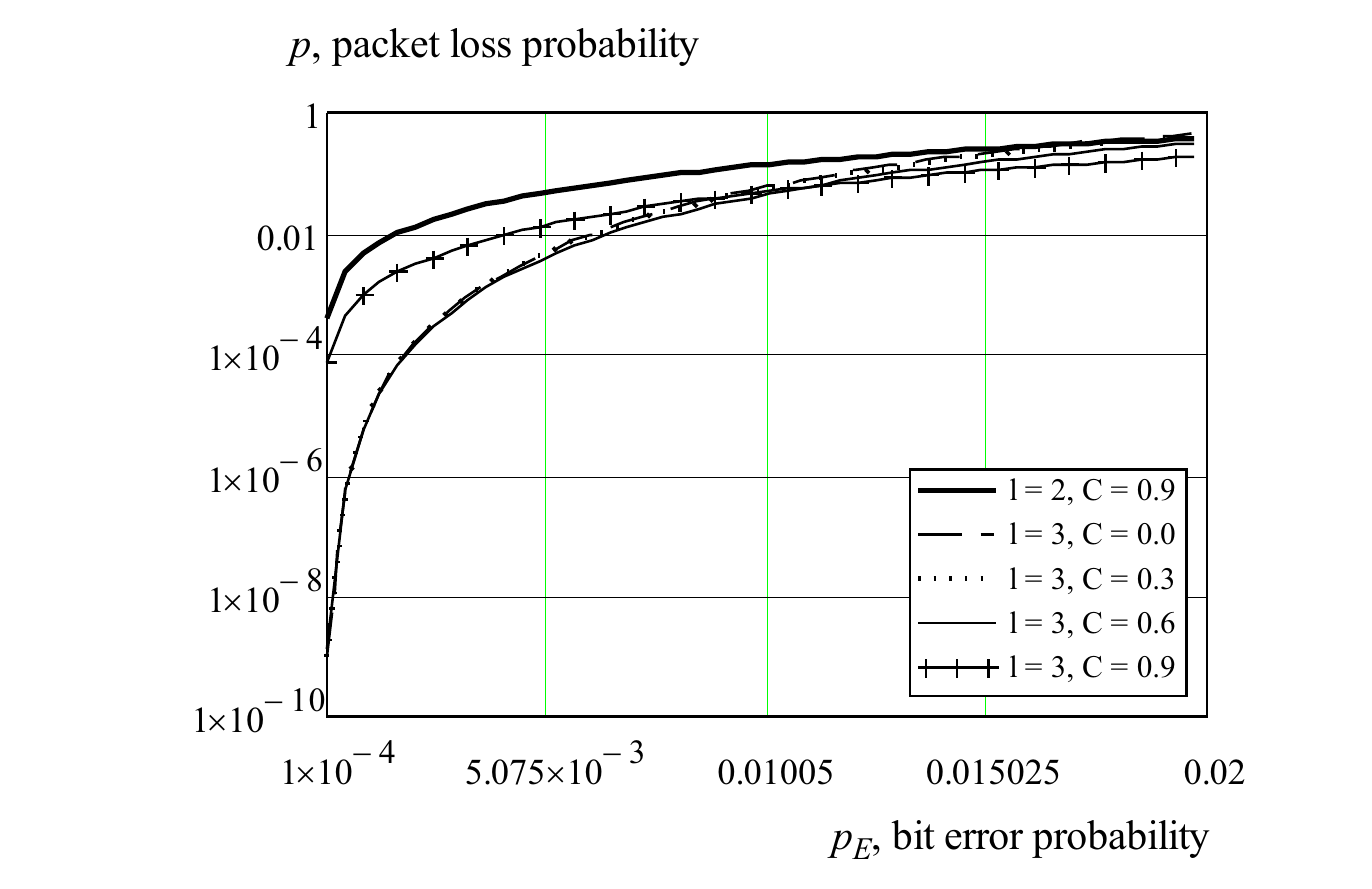}
    \label{fig:ex_2_plr}
  }
  \subfigure[Throughput for $l=2$ and $l=3$]{
    \includegraphics[width=0.5\textwidth]{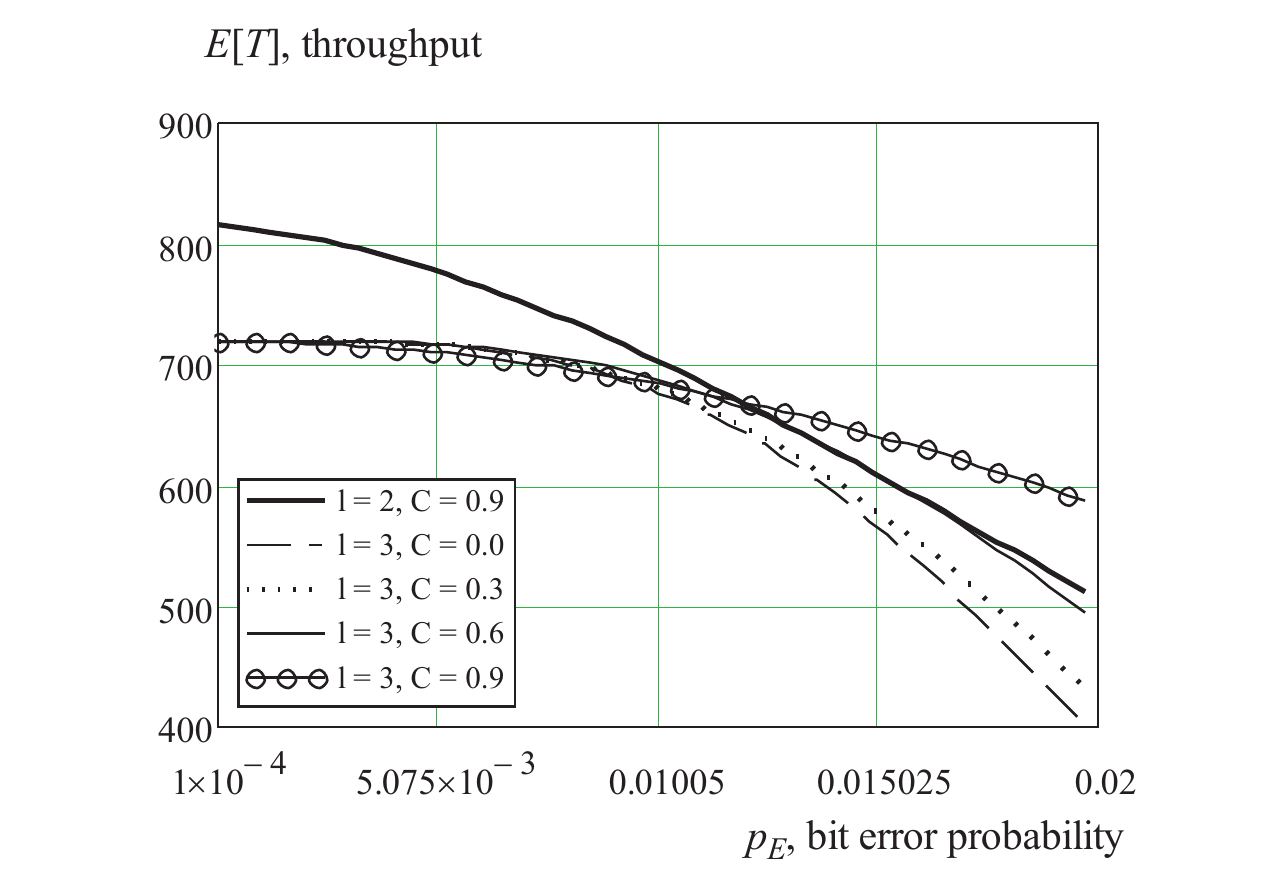}
    \label{fig:ex_2_thr}
  }
} \caption{Packet error probability and throughput for $(I,v)=(16,1)$.}
  \label{fig:exam}
\end{figure*}

\subsection{Throughput}

The packet error probability we concentrated on in previous sections is indeed the most
important metric when deciding upon the choice of the interleaving depth for a code with a given error correction capability $l$. However, the code rate $k/n$ usually decreases faster than a linear function as the correction capability increases. We also saw that the packet error probability decreases faster than a linear function as $l$ increases.
Thus, when choosing an optimal code rate we need to rely on the achievable throughput not on the packet error probability alone. In this subsection we highlight how the throughput changes when codes with different $l$ are used. The throughput, $E[T]$, is defined as the number of delivered bits averaged over $N$ packet transmissions as $N\rightarrow\infty$ and can be computed as $1-Ivkp$, where $k$ is the number of data bits per codeword, $p$ is the packet error probability, $I$ is the interleaving depth, and $v$ is the number of interleaving codeblocks per packet. 

%Note that the effect the interleaving scheme $(I,v)$ on throughput (as a function of BER) is rather trivial. For all values of the lag-1 NACF and for all the interleaving schemes $(I,v)$ the throughput decreases exponentially fast as BER increases. The performance of interleaving schemes with greater $I$ is always better for non-zero lag-1 NACF.

Fig. \ref{fig:eff_l_ber} compares the throughput of several codes with increasing
correction capabilities for few values of $(I,v)$ and $c$. Depending
on the lag-1 NACF value different codes need to be used across the considered range of
BER. For example, for $(2,8)$ scheme and $c=0.6$, going from $p_{E}=0.0001$ to
$p_{E}=0.02$ we need to change five difference codes. When $c$ goes up to $0.9$ the choice
of the code becomes simpler as $(63,56,1)$ is the best over the ($0.0001-0.01$) range
of BER while $(63,36,5)$ is the best for ($0.01-0.02$) range. The choice is obviously
different as we increase the interleaving depth to $(4,4)$ and then to $(16,1)$.

The analysis carried out in this section highlights several special effects. One of them
is that for a certain value of $c$ stronger FEC code may perform worse compared to those
having smaller $l$ even though it may result in smaller packet error probability. Few such
examples for $(I,v)=(16,1)$ interleaving scheme are shown in Fig. \ref{fig:exam}. As one
may observe FEC code $(63,57,1)$ operating over the channel with $c=0.9$ is vastly
outperformed by $(63,51,2)$ code operating over the channel with $c=0.0$ over the
($0.0001-0.01$) range of BER. In terms of throughput the former code is better
for the whole range of BER.

\section{Conclusions}\label{sect_06}

Motivated by both the lack of analytical models for interleaving we proposed three models characterized by different degree of complexity. Another motivation behind this work was to enable interleaving in cross-layer performance evaluation and optimization studies of modern and future wireless access technologies. Although we used BCH code to specify the model we also discussed extensions of the model to the case of other code including both block codes and convolutional ones.

We study performance of the all proposed models in details. The model based on estimation of two-dimensional distribution of the number of errors in adjacent codewords and successive application of absorbing Markov chain provide the best possible results for all ranges of lag-1 NACF values, BER, error correction capabilities and interleaving depths. The deviation of this model from the statistical data has a very special form with two peaks. One of these peaks depends on the interleaving depth $I$ and and happens when the values of lag-1 NACF are close to $1$. For all the considered parameters the worst observed deviation was $20\%$. However, we also observed that that less complex models such as those based on two-state Markov chains perform only slightly worse than the abovementioned one. What is more important is that the simplest model based on one-dimensional distribution shows the closest results. This simplest model also allows for closed form expression for the packet error probability making it attractive for analytical performance optimization studies. Finally, given a computational package with arbitrary precision arithmetic (e.g. GNU Multiple Precision Arithmetic Library, GNP, used in Mathematica, \cite{8}) our model allows to compute packet error probabilities for any desirable arbitrarily small value of BER which is hardly possible using simulation studies.

We also studied the effect of interleaving in detail using the most accurate model based
on absorbing Markov chain. Aside from the common belief that small values of interleaving depth (i.e. up to $5$, say) are sufficient for correlated channel we saw that the best possible choice of $I$ depends on the channel memory and should be chosen carefully. In fact, increasing the interleaving depth makes the channel more and more random. For the working range of BER and big values of the lag-1 NACF the advise would be to use as big value of $I$ as possible. Such big values of $I$ do not affect the complexity of the system significantly but allow for complete removal of channel correlation. However, one needs to recall that, the value of interleaving depth is often limited by the size of a packet to be transmitted. For extremely large packet sizes, when we can use rather big values of $I$, the simple rule of thumb would be to estimate the residual value of autocorrelation $R=c^{I}$, where $c$ is the lag-1 NACF. The first allowed valued of $I$ for which $R$ is sufficiently close to zero can be used.

\section*{Acknowledgement}
 
The publication was supported by the Ministry of Education and Science of the Russian Federation (project No. 2.3397.2017).

\balance
\section*{References}
\bibliographystyle{IEEEtran}

%\bibliography{inter}

\begin{thebibliography}{10}
\providecommand{\url}[1]{#1}
\csname url@samestyle\endcsname
\providecommand{\newblock}{\relax}
\providecommand{\bibinfo}[2]{#2}
\providecommand{\BIBentrySTDinterwordspacing}{\spaceskip=0pt\relax}
\providecommand{\BIBentryALTinterwordstretchfactor}{4}
\providecommand{\BIBentryALTinterwordspacing}{\spaceskip=\fontdimen2\font plus
\BIBentryALTinterwordstretchfactor\fontdimen3\font minus
  \fontdimen4\font\relax}
\providecommand{\BIBforeignlanguage}[2]{{%
\expandafter\ifx\csname l@#1\endcsname\relax
\typeout{** WARNING: IEEEtran.bst: No hyphenation pattern has been}%
\typeout{** loaded for the language `#1'. Using the pattern for}%
\typeout{** the default language instead.}%
\else
\language=\csname l@#1\endcsname
\fi
#2}}
\providecommand{\BIBdecl}{\relax}
\BIBdecl

\bibitem{1}
D.~Moltchanov, ``Source code of interleaving models,'' Department of
  Electronics and Communications, Available at
  http://www.cs.tut.fi/˜\textasciitilde{}moltchan/bchInterleavingModels.zip.

\bibitem{3}
D.~Moltchanov, Y.~Koucheryavy, and J.~Harju, ``Simple, accurate and
  computationally efficient wireless channel modeling algorithm,'' in
  \emph{Proc. {WWIC}}, Xanthi, Greece, May 2005, pp. 234--245.

\bibitem{4}
Q.~Zhang and S.~Kassam, ``Finite-state markov model for rayleigh fading
  channels,'' \emph{{P}erf. {E}val.}, vol.~47, no.~11, pp. 1688–--1692, Nov.
  1999.

\bibitem{5}
A.~Konrad, B.~Zhao, and R.~Ludwig, ``Markov-based channel model algorithm for
  wireless networks,'' \emph{{W}ir. {N}etw.}, vol.~9, no.~3, pp. 189–--199,
  2003.

\bibitem{6}
D.~Moltchanov, ``State description of wireless channels using change-point
  statistical tests,'' in \emph{Proc. {WWIC}}, Bern, Switzerland, May 2006, pp.
  275--286.

\bibitem{dunscombe1989optimal}
E.~Dunscombe and F.~Piper, ``Optimal interleaving scheme for convolutional
  coding,'' \emph{Electronics Letters}, vol.~25, no.~22, pp. 1517--1518, 1989.

\bibitem{chen2004improving}
L.-J. Chen, T.~Sun, M.~Sanadidi, and M.~Gerla, ``Improving wireless link
  throughput via interleaved fec,'' in \emph{Computers and Communications,
  2004. Proceedings. ISCC 2004. Ninth International Symposium on},
  vol.~1.\hskip 1em plus 0.5em minus 0.4em\relax IEEE, 2004, pp. 539--544.

\bibitem{el2012performance}
M.~A. M.~M. El-Bendary, A.~E. Abou-El-azm, N.~A. El-Fishawy, F.~Shawki, F.~E.
  Abd-ElSamie, M.~A.~R. El-Tokhy, and H.~B. Kazemian, ``Performance of the
  audio signals transmission over wireless networks with the channel
  interleaving considerations,'' \emph{EURASIP Journal on Audio, Speech, and
  Music Processing}, vol. 2012, no.~1, pp. 1--14, 2012.

\bibitem{kiyani2008performance}
N.~F. Kiyani, J.~H. Weber, A.~G. Zaji{\'c}, and G.~L. St{\"u}ber, ``Performance
  analysis of a system using coordinate interleaving and constellation rotation
  in rayleigh fading channels,'' in \emph{Vehicular Technology Conference,
  2008. VTC 2008-Fall. IEEE 68th}.\hskip 1em plus 0.5em minus 0.4em\relax IEEE,
  2008, pp. 1--5.

\bibitem{kang2008probabilistic}
K.~Kang, ``Probabilistic analysis of data interleaving for reed-solomon coding
  in bcmcs,'' \emph{Wireless Communications, IEEE Transactions on}, vol.~7,
  no.~10, pp. 3878--3888, 2008.

\bibitem{kang2010hybrid}
K.~Kang, C.~Kim, and K.-J. Park, ``A hybrid architecture for delay analysis of
  interleaved fec on mobile platforms,'' \emph{Vehicular Technology, IEEE
  Transactions on}, vol.~59, no.~4, pp. 2087--2092, 2010.

\bibitem{kemeny1960finite}
J.~G. Kemeny, J.~L. Snell \emph{et~al.}, \emph{Finite markov chains}.\hskip 1em
  plus 0.5em minus 0.4em\relax van Nostrand Princeton, NJ, 1960, vol. 356.

\bibitem{o1999phase}
C.~A. O'cinneide, ``Phase-type distributions: open problems and a few
  properties,'' \emph{Stochastic Models}, vol.~15, no.~4, pp. 731--757, 1999.

\bibitem{7}
S.~Bhattacharya and A.~Gupta, ``Occupation times for two-state markov chains,''
  \emph{{D}isc. {A}ppl. {M}ath.}, vol.~2, no.~3, pp. 249–--250, March 1980.

\bibitem{8}
C.~library, ``Gnu multiple precision arithmetic (gmp),'' GNU General Public
  License, Available at http://gmplib.org/, Acessed on 16.07.2013.

\end{thebibliography}

\end{document}